\documentclass[11pt,letterpaper]{JHEP3}

  \usepackage{latexsym,bm,amsmath,amssymb,amsfonts}
  \usepackage{epsfig,graphics,graphicx}
  \usepackage{slashed}
  \usepackage[latin1]{inputenc}
  \usepackage{mathrsfs}
\usepackage{mathcomp}
\usepackage{yfonts}

\usepackage{bm,mathrsfs,slashed}

\long\def\comment#1{ }
\newcommand{\eqnum}[1]{Eq.~\eqref{#1}}
\newcommand{\eqn}[1]{Eq.~\eqref{#1}}

\newcommand{\beq}{\begin{eqnarray}}
\newcommand{\eeq}{\end{eqnarray}}
\newcommand{\nn}{\nonumber\\}

\newcommand{\rmd}{{\rm d}}
\newcommand{\rme}{{\rm e}}

\newcommand{\qq}{{q \bar q}}
\newcommand{\bq}{{\bar q}}
\newcommand{\bu}{{\bm u}_\perp}
\newcommand{\bbu}{{\bar{\bm u}}_\perp}
\newcommand{\bv}{{\bm v}_\perp}
\newcommand{\bk}{{\bm k}_\perp}
\newcommand{\tqq}{\theta_{q\bar q}}
\newcommand{\tq}{\theta_{q}}
\newcommand{\tbq}{\theta_{\bar q}}
\newcommand{\tf}{\theta_{f}}
\newcommand{\ts}{\theta_{s}}

\newcommand{\del}{\partial}
\newcommand{\grad}{\nabla_{\perp}}

\newcommand{\order}[1]{\mcal{O}{(#1)}}
\newcommand{\mcal}{\mathcal}

\newcommand{\calA}{{\mathcal A}}

\newcommand{\bxt}{\bm{x}_{\perp}}
\newcommand{\br}{\bm{r}}

\def\Prf{{2 g^2 C_F}}
\def\Ptotin{{\mathcal P}^{\mbox{(in)}}_{tot}}

\def\Intx{{\int_0^{L^+} \rmd x^+}}
\def\Inty{{\int_0^{x^+} \rmd y^+}}
\def\bperp{{\bm b}_\perp}
\def\Intb{{\int \rmd^2 \bperp}}

\def\bqv{{\bm q}_\perp}

\def\Sqqgen{{S_{F, L}(y^+, {\bf 0})}}
\def\Igen{{I_{u_{L}, u_{F}}( x^+, y^+, \bperp)}}
\def\Igenxy{{I_{u_{L}, u_{F}}( x^+, y^+, \bperp)}}
\def\Igenq{{I_{u_{L}, u_{F}}( x^+, y^+, \bqv)}}

\def\ulperp{{{\bm u}_{L}}}
\def\ufperp{{{\bm u}_{F}}}

\def\tint{{\tau_{int}}}
\def\tlambda{\tau_{\lambda}}

\def\Fgen{{F_{u_{L}, u_{F}}( x^+, y^+, \bk)}}

\let\Oldcdot\cdot
\renewcommand{\cdot}{\mspace{-2mu}\Oldcdot\mspace{-2mu}}
\let\Oldtimes\times
\renewcommand{\times}{\mspace{-2mu}\Oldtimes\mspace{-2mu}}

\title{\Large Interference effects in medium--induced gluon
radiation}

\author{J.~Casalderrey--Solana,$^{\,a}$\ \ E. Iancu,$^{\,a,b}$ \\
${}^a$CERN, Theory Division, CH-1211 Geneva, Switzerland\\
 ${}^b$Institut de Physique Th\'eorique,
CEA Saclay,
 F-91191 Gif-sur-Yvette, France\\ \\
        {\it E-mail:} \email{jorge.casalderrey@cern.ch}\,,
        \email{edmond.iancu@cea.fr}}

  \abstract{As a step towards understanding the in--medium
evolution of a hard jet, we consider the interference
pattern for the medium--induced gluon radiation produced by a color
singlet quark--antiquark antenna embedded in a QCD medium with size $L$.
We focus on the typical kinematics for
medium--induced gluon radiation in the BDMPS--Z regime, that is,
short formation times $\tau_f \ll L$ and relatively large emission
angles $\theta \gg\theta_c\equiv {2}/{\sqrt{\hat q L^3}}$,
with $\hat q$ the `jet quenching' parameter.
We demonstrate that, for a dipole opening angle $\theta_{q\bar q}$
larger than $\theta_c$, the interference between the
medium--induced gluon emissions by the quark and the antiquark is
parametrically suppressed with respect to the corresponding
direct emissions. Physically, this is so since the
direct emissions can be delocalized anywhere throughout the medium
and thus yield contributions proportional to $L$. On the contrary,
the interference occurs only between gluons
emitted at very early times, within the characteristic time
scales for quantum and color coherence between the two emitters,
which in this regime are much smaller than $L$. This implies that,
for $\tqq \gg\theta_c$, the medium--induced radiation
by the dipole is simply the sum of the two BDMPS--Z spectra
individually produced by the quark and
the antiquark, without coherence effects like angular ordering.
For $\tqq \ll\theta_c$, the medium--induced radiation by the dipole
vanishes.
}

\begin{document}

\section{Introduction}
\label{intro}

The phenomenon of {\em jet quenching} globally denotes the modifications
in the properties of a jet which occur when the jet propagates through
the dense QCD matter created in the intermediate stages of a
ultrarelativistic heavy ion collision. One of the most striking effects
of this kind is the large di--jet asymmetry observed in Pb+Pb collisions
at the LHC, as reported by the ATLAS \cite{Aad:2010bu} and CMS
\cite{Chatrchyan:2011sx} collaborations (see also \cite{Ploskon:2009zd}
for related results at RHIC). These data imply that, as a consequence of
the interactions between the jet and the medium, the jet energy is
transported to larger angles and redistributed into softer fragments as
compared to the p+p baseline. Understanding this phenomenon of strong jet
broadening and also the strong suppression of particle production at high
$p_T$ in nucleus--nucleus collisions as compared to p+p, as observed at
RHIC \cite{Arsene:2004fa,Back:2004je,Adams:2005dq,Adcox:2004mh} and the
LHC \cite{Aamodt:2010jd}, is essential for using jet probes as a
diagnosis tool of hot and dense QCD matter.

From a microscopic point of view, the dominant mechanism for jet
quenching at weak coupling and high energy is radiative energy loss
associated with medium--induced gluon radiation
\cite{Baier:1996kr,Baier:1996sk,Baier:1998yf,Zakharov:1996fv,Gyulassy:2000er,Wiedemann:2000za,Salgado:2003rv,Wang:2002ri,Arnold:2002ja}
(see also the review papers \cite{Baier:2000mf,CasalderreySolana:2007zz}
for more references). If the medium is sufficiently dense, both the
parton that initiates the jet and its descendants undergo multiple
scattering leading to additional radiation which is described by the
BDMPS--Z  (from Baier, Dokshitzer, Mueller, Peign\'e, Schiff, and
Zakharov) formalism. While this mechanism for jet quenching has been
quite successful in describing the suppression of single particle spectra
observed at RHIC (see, for example \cite{d'Enterria:2009am,
Majumder:2010qh}), it has been realized for long that the respective data
refer to inclusive measurements which are quite limited in constraining
the underlaying dynamics. By contrast, the differential jet measurements
that are performed at the LHC provide more detailed informations, in
particular, on the spectrum of the medium--induced radiation which could
help us to better pinpoint the physical mechanisms at work.

At this point, one should stress that the BDMPS--Z mechanism predicts
that gluons are emitted at relatively large angles  --- the softer the
gluon, the larger its emission angle --- and thus it has the potential to
explain the di--jet asymmetry measured at the LHC (see the recent
publications
\cite{CasalderreySolana:2010eh,Qin:2010mn,Lokhtin:2011qq,Young:2011qx}
for related studies). However, from the experience with jet evolution in
the vacuum, one knows that large angle radiation can be prohibited by
coherence effects leading to {\em angular ordering} : within the partonic
cascade produced via jet fragmentation in the vacuum, the emission angles
are bound to decrease from one emission to the next one. So far very
little is known about the corresponding property for the medium--induced
gluon radiation. The only analyses in that sense so far
\cite{MehtarTani:2010ma,MehtarTani:2011tz} are either restricted to the
single--scattering approximation \cite{MehtarTani:2010ma} or concerned
with a different mechanism for medium--induced radiation
\cite{MehtarTani:2011tz}, which applies to relatively soft and collinear
emissions which are less effective in broadening the transverse energy
distribution of a jet.

It is therefore crucial to clarify whether interference effects can
frustrate medium--induced radiation at large angles, in the interesting
regime where the medium is relatively opaque and the multiple scattering
is important. This is the main objective in this paper. To that aim, we
shall study the interference between the medium--induced gluon emissions
by two sources immersed into the medium: a quark ($q$) and an antiquark
($\bq$). More precisely, we shall address the problem of the in--medium
`dipole antenna pattern', that is, the radiation produced by a $q\bq$
pair in a color singlet state (a `color dipole') where the two particles
separate from each other at constant velocities which make a relative
angle $\tqq$ --- the dipole opening angle. This `dipole antenna' is a
familiar set--up for studies of interference and angular ordering for
radiation in the vacuum \cite{Dokshitzer:1991wu,Ellis:1991qj} and has
been generalized in Refs. \cite{MehtarTani:2010ma,MehtarTani:2011tz} to
corresponding studies in a medium. As usual in the related literature, we
shall work in the `multiple soft scattering approximation' which assumes
that successive scattering centers are independent from each other.
Formally, the results of Ref.~\cite{MehtarTani:2010ma} can be recovered
from this  formalism as the lowest order term in the `opacity expansion',
that is, the perturbative expansion of the medium effects. But this is
only formal, since the effects of multiple scattering are
non--perturbative and the final results cannot be expanded out
anymore\footnote{This is similar to the failure of the twist expansion
for high--energy scattering in the vicinity of the unitarity limit, or in
the gluon saturation region.}. In that sense, we expect our conclusions
to differ from those in Ref.~\cite{MehtarTani:2010ma} at {\em
qualitative} level, and not only quantitatively.

The main conclusion which emerges from our analysis is that the
interference effects for medium--induced gluon radiation are
parametrically small and hence irrelevant for all values of the dipole
angle $\tqq$ except for very small values $\tqq\lesssim \theta_c$, where
direct emissions and interference terms become comparable with each
other, and even cancel each other when $\tqq\ll\theta_c$.

In order to explain this conclusion and in particular the special angle
$\theta_c$, we need to first recall some basic features of the BDMPDS--Z
mechanism (see Sect.~\ref{heur} for a physical discussion). The
corresponding phase--space is characterized by two limiting values, a
maximal frequency $\omega_c$ and a minimum angle $\theta_c$, which are
expressed in terms of the medium properties as $\omega_c =\hat q L^2/2$
and $\theta_c={2}/{\sqrt{\hat q L^3}}$. ($\hat q$ is the jet quenching
parameter and $L$ is the longitudinal extent of the slice of the medium
which is crossed by the dipole.) The energy loss by the leading particle
is dominated by the emission of relative hard gluons with
$\omega\simeq\omega_c$, but such gluons make a small angle
$\theta\simeq\theta_c$ with respect to their source and thus are not
effective in broadening the jet energy distribution in the transverse
plane. Rather, the dominant transverse broadening comes from softer
gluons with $\omega\ll \omega_c$, which are emitted at relatively large
angles $\theta\gtrsim \tf(\omega)\gg\theta_c$. Here, $\tf(\omega)$ is the
minimal emission angle for a gluon with frequency $\omega$ (the
`formation angle') and increases when decreasing $\omega$ below
$\omega_c$.

An important property of these soft gluons, which is favorable too for
the physics of jet broadening, is the fact that they are {\em promptly}
emitted: the corresponding formation time is much smaller than the medium
length $L$. Hence, such gluons can be emitted at any point inside the
medium. Accordingly, the longitudinal phase--space for {\em direct}
emissions by the quark or the antiquark is proportional to $L$. By
contrast, the {\em interference} between the two partonic sources occurs
only for the gluons emitted at sufficiently early times $t<\tau_{min}$,
when the quark and the antiquark are still close enough to each other to
ensure {\em color} and {\em quantum coherence}. The precise mechanism
which determines $\tau_{min}$ depends upon the value of the dipole angle
$\tqq$~: \texttt{(i)} when $\theta_c\ll \tqq\ll \tf(\omega)$, the
interference is limited by the {\em color decoherence} of the $q\bq$ pair
(the two sources suffer different color precessions in the medium);
\texttt{(ii)} when $\tqq\gtrsim \tf(\omega)$, $\tau_{min}$ is rather
determined by the condition of {\em quantum coherence} (the radiated
gluon must overlap with both sources). But in both cases, {\em i.e.} so
long as $\tqq\gg\theta_c$, this upper limit $\tau_{min}$ is much smaller
than $L$, meaning that the phase--space for interference, which is
proportional to $\tau_{min}$, is parametrically suppressed relative to
that for direct emissions. Then the interference effects are negligible.
On the other hand, when $\tqq\lesssim\theta_q$, $\tau_{min}$ becomes as
large as $L$, so the interference is not suppressed anymore. But then the
total medium--induced radiation vanishes, since a gluon emitted at an
angle $\theta\gtrsim \tf(\omega)\gg\theta_c$ `sees' the total color
charge of the $q\bq$ pair, which is zero.

To summarize, the medium--induced radiation by the dipole is non--zero
only when $\tqq\gg\theta_c$ and in that case it is simply the sum of the
two BDMPS--Z spectra separately produced by the two emitters, without any
coherence effect like angular ordering. In order to substantiate this
conclusion and the above physical picture, we shall explicitly estimate
the contribution of the interference effects to the spectrum of the
medium--induced radiation by the dipole and compare the result with the
corresponding contribution due to direct emissions. Our main results in
that sense, namely \eqn{Ispec} for the contribution of the interference
terms and \eqn{BDMPS} for that of the direct emissions, are confirmed by
two different calculations (one exposed in the main text, the other one
in the Appendix), which involve approximation schemes with different
degrees of rigor, but which agree with each other to parametric accuracy.

Our paper is organized as follows: In Sect.~\ref{heur} we present a
qualitative discussion of the medium--induced radiation, including the
BDMPS--Z mechanism (for completeness and pedagogy), but focusing on our
original results on interferences. Our purpose there is to motivate our
conclusions via physical considerations, which hopefully will provide the
guidelines for the subsequent, more formal, developments. In
Sect.~\ref{general}, we give a streamlined presentation of the BDMPS--Z
formalism adapted to the problem at hand and also make contact with the
analysis in Ref.~\cite{MehtarTani:2011tz}. Sects.~\ref{direct} and
\ref{Int} are the main sections of this paper. They present detailed
calculations of the medium--induced contributions to direct emissions
(Sect.~\ref{direct}) and to the interference terms (Sect.~\ref{Int}). To
keep the presentation as fluent as possible, in these sections we resort
on analytic approximations which are correct at parametric level. (More
refined versions of these calculations are deferred to Appendix
\ref{App}.) This allows us to provide an explicit expression for the
BDMPS--Z spectrum (consistent with the respective results in the
literature) and to deduce an equally explicit result for the interference
contribution to the spectrum. Finally, in Sect.~\ref{outlook} we discuss
the implications of our results for the in--medium evolution of a hard
jet and we mention some open problems.

\subsection*{Note added}
When this work was already finished, a preprint appeared,
Ref.~\cite{MehtarTani:2011jw}, in which the general formula for the
interference contribution to the  medium--induced radiation by the dipole
(our \eqn{Iinfin}) was also derived. However, the physical consequences
of this formula were not explicitly worked out.  In particular
Ref.~\cite{MehtarTani:2011jw} did not identify the physical mechanism
responsible for the suppression of the in--medium interference terms,
which is  the reduction in the corresponding longitudinal phase--space.
The conclusions drawn there by inspection of the general formula turned
out to be incorrect in some cases. In the mean time, such issues have
been clarified via private communications.

\section{Physical discussion and summary}
\label{heur}

We start our presentation with a section which summarizes, at a
qualitative level, the physical picture and the main conclusions that we
shall eventually reach through our analysis. This strategy of
presentation is motivated by the fact that the physical problems that we
shall address are both complex and subtle, as they involve several (time
and momentum) scales and also non--perturbative phenomena (like high
density effects). The theoretical treatment of these phenomena will
therefore be quite involved and lengthy. Notwithstanding, the ultimate
physical picture which will emerge from our calculations is quite simple
and can be transparently structured in terms of a hierarchy of (time and
angular) scales, which for the benefit of the reader are summarized in
Table~\ref{tab:dict}. So we feel that the reading of the paper will gain
in clarity if the relevant scales are {\em a priori} identified, via
physical considerations, and if the final conclusions are exposed on the
basis on these scales alone. This will motivate the subsequent, more
formal, manipulations and hopefully provide the guidelines for the
approximations to come. The discussion in this section is intended to be
self--contained (provided, of course, the reader will be ready to accept
some arguments to be rigorously demonstrated later) and could be read as
a summary of our present work, independently of the subsequent, more
formal developments.

\subsection{A primer on BDMPS--Z physics}
\label{bdmps0}

The propagation of a high energy parton through a dense QCD medium leads
to energy loss and transverse momentum broadening via {\em
medium--induced radiation}, that is, the emission of gluons stimulated by
the interactions between the quark, or the radiated gluon, and the
medium. The radiation process requires a characteristic {\em formation
time} which can be understood as the time for the (virtual) gluon to
separate enough from its parent quark for the quantum (in particular,
color) coherence between the two quanta to be lost. This formation time
$\tau_q$ can be estimated from the condition that the transverse
separation $b_\perp=\tau_q v_\perp$ between the quark and the gluon at
the formation time be of the order of the gluon transverse wavelength
$\lambda_\perp=1/k_\perp$. Here `transverse' refers to the direction
orthogonal to the trajectory of the quark, $k_\perp$ is the gluon
transverse momentum and $v_\perp=k_\perp/\omega$ is its transverse
velocity. We have also introduced the gluon energy $\omega$, assumed to
be large compared to $k_\perp$. Accordingly, the gluon emission angle is
small, $\theta_q\simeq k_\perp/\omega \ll 1$. The previous discussion
implies\footnote{The factor of 2 in \eqn{tau} is conventional; this
result is to be seen as a parametric estimate, valid within the limits of
the uncertainty principle. }
 \beq\label{tau}
 \tau_q\,\frac{k_\perp}{\omega}\,\simeq\,\frac{2}{k_\perp}\quad
 \Longrightarrow\quad \tau_q\,\simeq\,\frac{2\omega}{k_\perp^2}
 \,\simeq\,\frac{2}{\omega\tq^2}\,.
 \eeq
The above argument is completely general: it holds for gluon emissions in
either the medium or the vacuum. What is different, however, is the
typical value of $k_\perp$  in the two cases.

For emissions in the vacuum, $k_\perp$ is {\em a priori} arbitrary.
However, the associated, bremsstrahlung, spectrum
 \beq\label{brems}
\omega \,\frac{\rmd N^{\rm vac}}{\rmd \omega\rmd k^2_\perp}
 \,\simeq\,
 \frac{\alpha_s C_F}{k_\perp^2}\,\simeq\,\alpha_s C_F
 \tq^2\tau_q^2\,,\eeq
is such that large values of $k_\perp$ are strongly suppressed, so most
of the radiation is quasi--collinear with its source ($\theta_q\to 0$).
The second equality in \eqn{brems}, which is clearly true in view of
\eqn{tau}, has a simple physical interpretation. After the quark is
created at $t_0=0$, a gluon with energy $\omega$ and transverse momentum
$k_\perp$ is emitted by a time $t\sim \tau_q$ and not much later. Then
the integral over time which enters the calculation of the emission
amplitude (see Sect.~\ref{amplit} below) yields a factor $\tau_q$.
Correspondingly, there will be a factor $\tau_q^2$ in the emission
probability which represents the temporal (or longitudinal) phase--space
for gluon emissions in the vacuum. The factor $\tq^2$  is the square of
the transverse velocity $v_\perp\!\simeq\tq$ of the emitted gluon, which
is the emission vertex in a vectorial, gauge, theory.

\begin{figure}[tb]
\begin{center}
\includegraphics[width=0.55\textwidth]{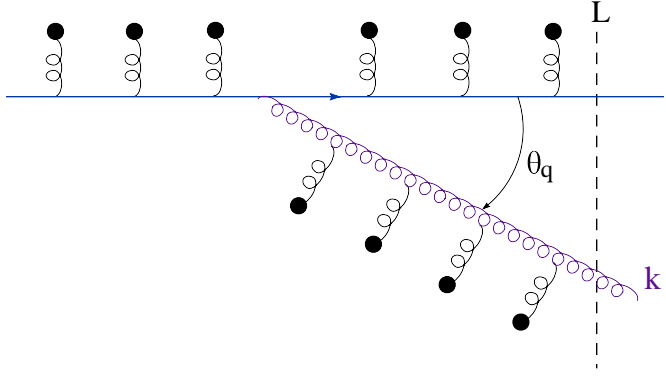}
\end{center}
\caption{\label{fig:medium}\sl The standard representation of the Feynman
graph for medium--induced gluon radiation: both the quark and the
emitted gluon undergo multiple scattering off the medium constituents.}
\end{figure}

Within a medium, on the other hand, the gluon can acquire an additional
transverse momentum via scattering off the medium constituents, both
during the formation time and after (see the graphical representation in
Fig.~\ref{fig:medium}). If the medium is sufficiently dense, the momentum
acquired in this way can be large and then the gluon spectrum is  shifted
towards a non--zero central value. At weak coupling, one can assume that
successive collisions in the medium proceed independently from each
other, even when the medium is dense: the gluon mean free path $\ell$
scales like $1/\alpha_s$ and for sufficiently small $\alpha_s=g^2/(4\pi)$
it becomes much larger than the screening length $\mu_D^{-1} \propto 1/g$
for charge correlations in the medium ($\mu_D$ is the Debye mass). The
gluon receives random kicks from the medium constituents, with each kick
transferring a momentum squared $\sim \mu_D^2$, so its {\em average}
transverse momentum squared grows at a rate
 \beq\label{dkdt}
 \frac{\rmd \langle k_\perp^2 \rangle}{\rmd t}\,\simeq\,
 \frac{\mu_D^2}{\ell}\,\equiv\,\hat q\,.
 \eeq
The quantity $\hat q$ is a local transport coefficient known as the {\em
jet quenching parameter}.

Returning to the process of gluon emission, we see that the gluon can be
set free by its interactions with the medium, which then determine the
{\em in--medium formation time} $\tau_f$. Specifically, within a time
$\tau_f$, the gluon acquires a transverse momentum squared $k_{f}^2
\simeq \hat q \tau_f$ with $\tau_f$ related to $k_{f}^2$ as shown in
\eqn{tau}. We thus deduce
 \beq\label{tauf}
  k_{f}^2 \,\simeq\,
(2\omega\hat q)^{1/2}\, \qquad\mbox{and}\qquad \tau_f\,\simeq
\,\sqrt{\frac{2\omega}{\hat q}}\,.\eeq
 % \,\simeq\,L \sqrt{\frac{\omega}{\omega_c}}\,.\eeq
We see that the medium rescattering controls the formation time $\tau_f$,
the gluon average transverse momentum at the time of formation, $k_{f}$,
and hence also the {\em formation angle} $\tf\simeq k_{f}/\omega$. In
order for the gluon to be formed in the medium, one needs, clearly,
$\tau_f\le L$, with $L$ the longitudinal extent of the slice of the
medium which is crossed by the quark. Hence, the maximal possible value
for $k_{f}$, known as the {\em saturation momentum} $Q_s$, is given by
$Q_s^2= \hat q L$ and is achieved for a gluon with a frequency $\omega_c$
such that $\tau_f(\omega_c)=L$. These relations imply
 \beq\label{omegac}
\omega_c\, =\, \frac{1}{2}\, \hat q L^2\,,\qquad Q_s^2\,=\, \hat q L
\,,\qquad\theta_c=\,\frac{Q_s}{\omega_c}\,=\,\frac{2}{Q_s L}\,=
 \,\frac{2}{\sqrt{\hat q L^3}}
 \,.\eeq
$\theta_c$ is the formation angle for a gluon with frequency $\omega_c$
and is the minimal angle in the problem: gluons with larger frequencies
$\omega> \omega_c$ and smaller angles $\theta_q<\theta_c$ cannot be
emitted via this mechanism. The medium is {\em dense} provided the
transverse momentum $k_f$ acquired by the gluon via multiple scattering
during the process of formation is much larger than the Debye mass
$\mu_D$. In view of \eqn{dkdt}, this requires the in--medium formation
time $\tau_f$ to be substantially larger than the mean free path $\ell$,
and hence even more so as compared to the Debye length:
$\tau_f\gg\ell\gg\mu_D^{-1}$. Since, moreover, $Q_s>k_f$ and $L>\tau_f$,
these relations imply that the limiting angle $\theta_c\sim 1/(Q_sL)$ is
truly small: $\theta_c\ll 1$. (Some typical values for heavy ion
collisions at RHIC and the LHC are $L=6$~fm and $\hat q=2\div 10\,{\rm
GeV}^2/{\rm fm}$, yielding $\theta_c = 0.005\div 0.01$.)

More generally, for a given energy $\omega < \omega_c$, the quantities
$k_{f}$ and $\tf$ introduced above --- the average transverse momentum
and emission angle {\em at the formation time} --- represent lower limits
on the respective kinematical variables in the BDMPS--Z spectrum. For
what follows, it is useful to express the {\em quasi--local} quantities
$k_{f}$ and $\tf$, which are controlled by the physics at the scale
$\tau_f$, in terms of the {\em global} ($L$--dependent) quantities in
\eqn{omegac}, which represent their absolute limits attained when
$\tau_f=L$:
  \beq\label{formation}
\tau_f\,=\,
 L \sqrt{\frac{\omega}{\omega_c}}\,,\qquad k_{f}\,=\,
 Q_s\left(\frac{\omega}{\omega_c}\right)^{1/4}\,,\qquad
 \theta_f\equiv\,\frac{k_{f}}{\omega} =\,
\theta_c\left(\frac{\omega_c}{\omega}\right)^{3/4}
 \,.\eeq
These formul\ae{} make clear that the {\em relatively soft} gluons with
$\omega \ll \omega_c$ are emitted {\em very fast} ($\tau_f\ll L$) and at
relatively {\em large angles} ($\tf\gg\theta_c$). Such gluons are very
efficient in broadening the jet energy in the transverse plane.

While the quark propagates though the medium it receives kicks from the
medium constituents and it can radiate after any of those kicks. The
typical distance $\ell$ between two consecutive kicks is generally much
shorter than the formation time of the gluon $\tau_f$; accordingly, a
large number of scattering centers $N_{coh}\simeq \tau_f/\ell\gg 1$ will
coherently act as a single source of radiation. This subset of $N_{coh}$
constituents can be located anywhere inside the medium, meaning that the
time $t_1$ at which a particular emission is initiated is delocalized
within the interval $0\le t_1\le L$. Thus, unlike vacuum emissions which
start right away after a hard scattering, the medium--induced emissions
can be initiated at any point inside the medium. Accordingly, the
longitudinal phase--space for medium--induced gluon radiation is
$(L-\tau_f)\tau_f\sim L\tau_f$, which for $\omega\ll\omega_c$ is {\em
parametrically larger} than the corresponding bremsstrahlung phase--space
$\tau_q^2$ (for the same kinematics). We shall later derive the gluon
spectrum {\em at the formation time} and thus find a Gaussian centered at
$k_{f}$:
 \beq\label{spectrumf}
 \omega \,\frac{\rmd N}{\rmd \omega\rmd k_\perp^2}\bigg|_{\rm form}
 \,\propto\,\alpha_s C_F
\tq^2\,\tau_f L\exp\biggl\{
 -\frac{k_\perp^2}{k_{f}^2}\biggr\}\,.
 \eeq
The prefactor in this expression is similar to that in \eqn{brems}~: the
only difference refers to the replacement $\tau_q^2\to\tau_f L$ for the
longitudinal phase--space. Since $\theta_q\propto k_{\perp}$, it is clear
that this spectrum is strongly peaked at $k_\perp=k_{f}$. Hence, for
parametric estimates, one can replace $\theta_q\to\tf$ in the prefactor.
For $k_\perp\sim k_f$ and $\omega\ll\omega_c$, one has
$\tau_q\sim\tau_f\ll L$, hence \eqn{spectrumf} is indeed enhanced w.r.t.
the vacuum spectrum \eqref{brems}, by the large factor $L/\tau_f\gg 1$.
This factor counts the number of times that a medium--induced gluon can
be formed inside the medium.

\begin{figure}[tb]
\begin{center}
\includegraphics[width=0.65\textwidth]{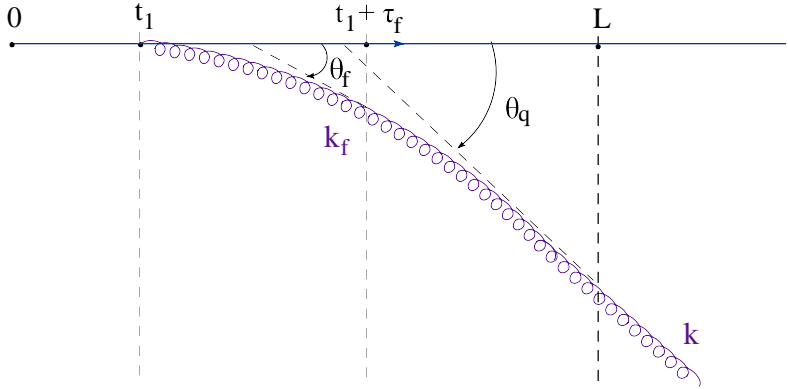}
\end{center}
\caption{\label{fig:angles}\sl A cartoon illustrating the space--time
picture of a medium--induced gluon radiation. The gluon formation
is initiated at time $t_1$ and terminated at time $t_1+\tau_f$.
The gluon leaves the medium with its final momentum ${\bm k}$ at time $L$.
The interactions with the medium are not explicitly shown.}
\end{figure}

\eqn{spectrumf} is not yet the final BDMPS--Z spectrum since after being
formed, the gluon will still propagate inside the medium over a distance
$L-\tau_f-t_1$, with $0<t_1<L-\tau_f$, and thus acquire an additional
momentum broadening $\Delta k_{\perp}^2\simeq \hat q (L-\tau_f-t_1)$.
Accordingly, its final momentum $k_{\perp}^2=k_{f}^2+\Delta k_{\perp}^2$
can take any value from $k_{f}^2=\hat q\tau_f$ to $Q_s^2= \hat q L$. This
results in the following kinematics domain for the final gluon, as
measured by a detector:
 \beq\label{range}
\hat q^{1/3}\,\lesssim\,\omega\,\lesssim\,\omega_c\,,\qquad
 k_{f}\,\lesssim\,k_{\perp}\,
 \lesssim\,Q_s\,.\eeq
The lower limit $\hat q^{1/3}$ on $\omega$ comes from the condition that
$\omega>k_{\perp}>k_{f}$. Within this range in $k_\perp$, the BDMPS--Z
distribution is roughly flat (see \eqn{BDMPSZ} below).

In discussing interference phenomena in what follows, it will be more
convenient to use angular variables instead of transverse momenta. Using
\eqn{range}, one immediately finds the following range for the {\em
final} gluon angle $\tq\simeq k_\perp/\omega$ :
\beq\label{theta}
\tf\,\equiv \theta_c\left(\frac{\omega_c}{\omega}\right)^{3/4}
\,\lesssim\ \tq\,\lesssim\,\ts\equiv\, \theta_c\,\frac{\omega_c}{\omega}
 \,.\eeq
Clearly, $\ts$ (the `saturation angle') is the same as $\ts=Q_s/\omega$.
For $\omega \ll\omega_c$, the final angle $\tq$ is much larger than
$\theta_c$ --- at least as large as $\tf$. The space--time picture and
the main scales for medium--induced radiation are illustrated in
Fig.~\ref{fig:angles}.

\subsection{Qualitative discussion of interference}
\label{intphys}

To study interference effects, we shall replace the quark probe
considered in the previous subsection with a color dipole, that is a
quark ($q$) and antiquark ($\bq$) in a color singlet state which separate
from each other at constant velocities which make a relative angle $\tqq$
--- the dipole opening angle. The quark and the antiquark are massless,
so they propagate at the speed of light. The dipole is created at time
$t_0=0$ and for later times its transverse size grows like
$r_\perp(t)\simeq\tqq \,t$. (We assume the dipole angle to be relatively
small, $\tqq\ll 1$. For the case of medium--induced emission, this angle
should be correlated with the respective emission angles, as we shall
later discuss.) By `transverse' we here mean the direction perpendicular
to the common direction of motion of the quark and the antiquark (the
`longitudinal axis'), as defined by the trajectory of their center of
mass. Interference occurs if the transverse wavelength of the gluon which
is about to be emitted is large enough for the gluon to have an overlap
with both sources. When this happens, the gluon `sees' the overall color
charge of the $q\bq$ pair, which is zero, so it is not
emitted anymore (destructive interference). %

Let us first recall the respective argument in the case of the vacuum,
where it is well known to lead to {\em angular ordering}. The distance
between the quark and the antiquark at the time of emission is, roughly,
$\tqq\,\tau_q$ with $\tau_q=2\omega/k_\perp^2$. (We assume the gluon to
be emitted by the quark.) Interference occurs if this distance is smaller
than the transverse wavelength $\lambda_\perp \simeq {1}/{k_\perp}$ of
the gluon. Using $k_\perp\simeq \omega\tq$, with $\tq$ the angle of
emission, we deduce
 \beq\label{interfvac}
 \tqq\ \frac{1}{\omega\tq^2}\
 \lesssim\ \frac{1}{\omega\tq}\qquad\Longrightarrow\qquad
 \tq\,\gtrsim\,\tqq\,.\eeq
That is, the interference is important only for gluon emissions at
relatively large angles, outside of the dipole cone.

\begin{figure}[tb]
\centerline{%\hspace*{-1.cm}
\includegraphics[width=0.45\textwidth]{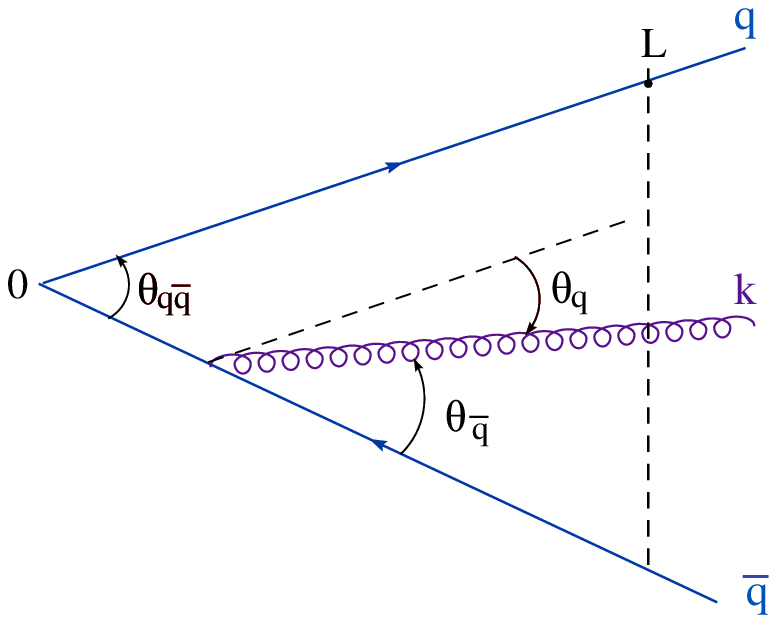}
\hskip 3mm
\includegraphics[width=0.48\textwidth]{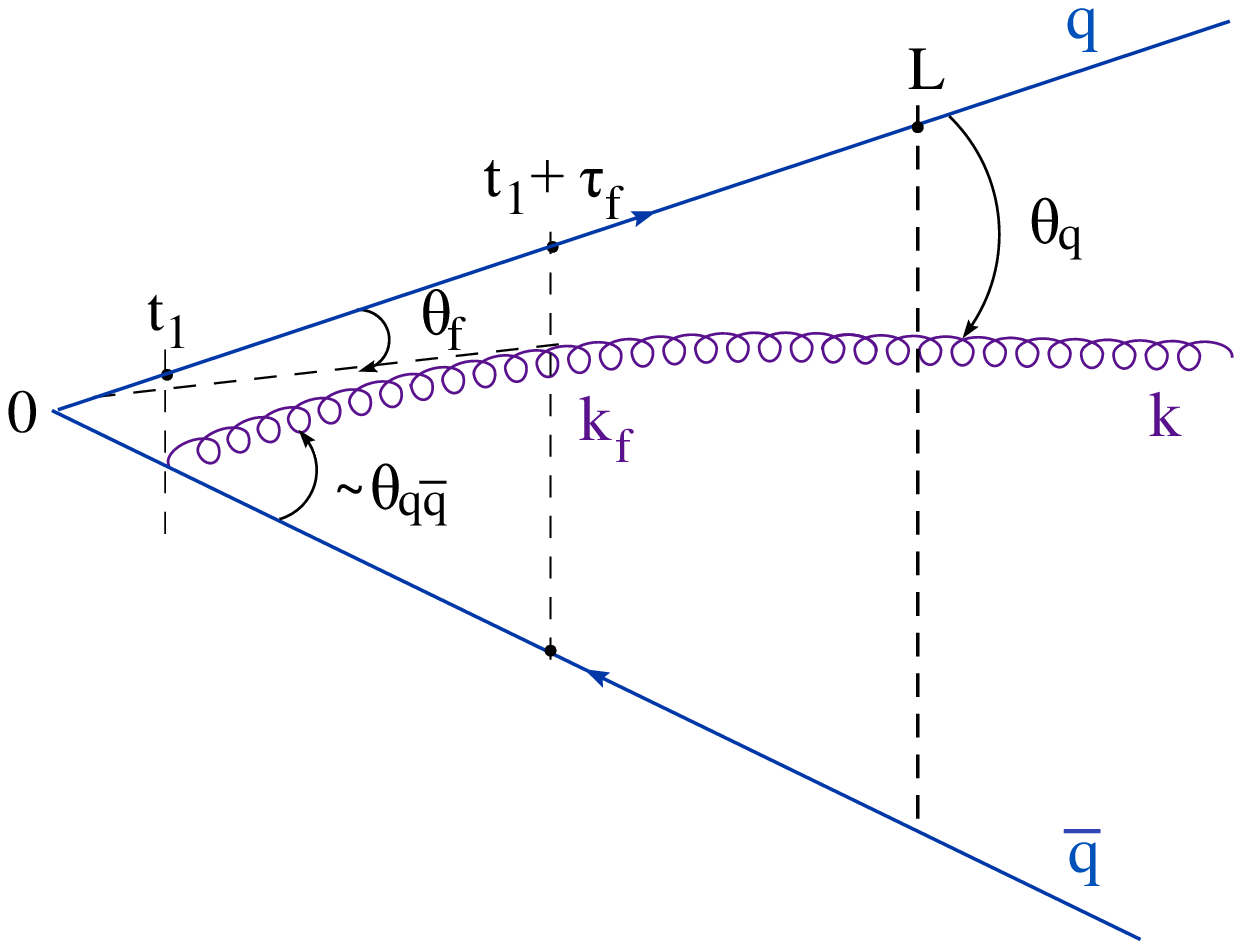}}
\caption{\label{fig:interf}\sl Gluon emission by the dipole. Left:
the geometry of the final state. Right: The space--time picture of a
typical emission contributing to interference in the case $\tqq>\tf$.
When the emission is initiated, at time $t_1\sim \tau_{int}$ by the
antiquark, the virtual gluon is co--moving with the quark.
When the emission is completed,  at time $t_1+\tau_f$,
the gluon makes an angle $\tf$ with the quark.}
\end{figure}

The interference effects can be also discussed at the level of the
bremsstrahlung spectrum produced by the dipole. This is given by the
following generalization of \eqn{brems}
 \beq\label{bremsd}
\omega \,\frac{\rmd N^{\rm vac}_{\rm dip}}{\rmd^3 {\bm k}}
 \,\simeq\,\alpha_s C_F \big(\tq\tau_q-\tbq\tau_\bq\big)^2\,,\eeq
with $k^\mu=(\omega, {\bm k})$ the 4--vector of the emitted gluon,
$\tau_q\simeq 2/(\omega\tq^2)$ and $\tau_\bq\simeq 2/(\omega\tbq^2)$.
(Note that our sign conventions for the emission angles are such that
$\tq <0$ and $\tbq>0$  for emissions inside the dipole and $\tq \,\tbq>0$
for emissions outside the dipole; see also Fig.~\ref{fig:interf} left.
With this convention, one has $\tqq=\tbq-\tq > 0$. Also, when using an
angle within a parametric estimate or an inequality, we always mean its
absolute value.) By expanding the square in the r.h.s. one generates the
{\em direct emission} terms, from the quark ($\tq^2\tau_q^2$) and
respectively the antiquark ($\tbq^2\tau_\bq^2$), together with the {\em
interference term} $-2\tq\tbq\tau_q\tau_\bq$, which has an overall minus
sign because the two sources have opposite charges. Inside the dipole
cone, where the two emission angles have opposite signs, the interference
term is relatively unimportant: the radiation is strongly peaked around
the direction of the quark ($\tq\approx 0$) or of the antiquark
($\tbq\approx 0$), where it is dominated by the respective direct
emission. However for large emission angles $\tq,\,\tbq\gg\tqq$ one has
$\tq\simeq\tbq$ and $\tau_q\simeq\tau_\bq$ and then the total radiation
vanishes --- the direct emissions are compensated by the interference
term.

We now turn to the description of the medium--induced gluon radiation
from an energetic $q\bq$ dipole created in the medium. It is intuitively
clear that, if the dipole angle is sufficiently large (larger than the
maximal emission angle $\ts$ introduced in the previous subsection, cf.
\eqn{theta}), the radiation patterns produced by the quark and the
antiquark via interactions in the medium have no overlap with each other
and thus they are independent. The question we would like to address is
what happens when the dipole opening angle is not that large. In that
case, and in view of the experience with radiation in the vacuum, one may
expect the dipole antenna pattern to be affected by interference effects
between the emissions by the quark and the antiquark. However, as we
shall now argue, this expectation is generally incorrect: for a
sufficiently dense medium and a dipole angle $\tqq$ which is not very
small (see below for the precise condition), only those gluons which are
emitted very close to the $q\bq$ vertex can be coherent with both
emitters and thus lead to interference. Accordingly, the interference
effects are parametrically suppressed as compared to the direct emissions
from each of the quarks.

Our subsequent arguments will heavily rely on the three main properties
of the medium--induced radiation (by a single source), as described in
Section~\ref{bdmps0} and that we summarize here for further convenience:
\begin{enumerate}
\item  The formation of a medium--induced gluon takes a time
    $\tau_f$. When the gluon is formed it is emitted at a typical
    angle $\tf$ from the parent quark, that is, it carries a typical
    transverse momentum $k_f\simeq\omega \tf$.
\item After formation,  multiple scattering leads to an additional
    broadening of the gluon spectrum. The final gluon distribution is
    concentrated within a typical  angle $\ts>\tf$ around the parent
    parton.
\item As long as $\tau_f\ll L$, gluons are emitted all along the
    medium length $L$.  As a consequence, the medium--induced gluon
    spectrum is proportional to $L$.
\end{enumerate}

The concept of {\em coherence} will play an important role in the
subsequent discussion of interference effects in the medium--induced
gluon radiation. But as we shall see, several types of coherence come
into the play and, moreover, some of them are not specific to the
interference phenomena, but also occur for the direct emissions.
Specifically, the process of in--medium gluon formation is by itself a
phenomenon of {\em decoherence}~: it proceeds via the gradual loss (due
to interactions in the medium) of the coherence between the virtual gluon
and the partonic sources which compose the dipole. This applies to all
the emissions of the BDMPS--Z type --- direct emissions and
interferences. In addition, the interference phenomena require the two
partonic sources ($q$ and $\bq$) to be {\em coherent} with each other
during the gluon formation. In turn, this last requirement has {\em two}
aspects: \texttt{(i)} {\em quantum} coherence, namely the emitted gluon
needs to overlap with both sources, and \texttt{(ii)} {\em color}
coherence, that is, the $q\bq$ pair needs to preserve its original,
overall, color charge
--- here, to remain in a color singlet state. Each of these coherence
phenomena introduces a characteristic time (or length) scale, that we
shall shortly describe.

\texttt{(i) Quantum coherence.} The emission process preserves the
quantum coherence of the $q\bq$ system so long as the virtual gluon
overlaps with both sources in the course of formation. (After formation,
the multiple scattering of the gluon in the medium can change the gluon
momentum but it cannot affect the interference process since the gluon is
already decorrelated from both partonic sources.) For that to be
possible, the transverse resolution of the gluon at the time of
formation, as measured by the respective transverse wavelength
$\lambda_f= 1/k_f$, should be larger than the typical distance between
the quark and the antiquark around that time. For emissions in the
vacuum, this condition immediately leads to angular ordering, as
discussed around \eqn{interfvac}. But for the medium--induced emissions,
the situation is more subtle: during the gluon formation, the transverse
size of the $q\bq$ system increases from $r_{min}\simeq\tqq\, t_1$ to
$r_{max}\simeq\tqq\, t_2$~; here, $t_1$ is the time when the emission is
initiated and can be delocalized anywhere within the medium, while $t_2=
t_1+\tau_f$ is the time when the emission is completed. Then it is not
{\em a priori} clear which one among these scales, $r_{min}$ or
$r_{max}$, should be compared to $\lambda_f$. Yet, our calculations, to
be detailed in Sect.~\ref{Int} and in Appendix \ref{App}, yield an
unambiguous answer to this question: the $q\bq$ size to be compared with
$\lambda_f$ is the {\em geometric average} of these two extreme scales.
That is, the condition of quantum coherence amounts to
$\sqrt{r_{min}r_{max}}\lesssim\lambda_f$, or
 \beq\label{tlambdadef}
 \sqrt{t_1(t_1+\tau_f)}\ \lesssim\ \frac{\lambda_f}{\tqq}\,\equiv\,
 \tau_\lambda\,.\eeq
Most likely, the appearance of the geometrical average in the l.h.s.
reflects the fact that the gluon undergoes transverse diffusion during
the formation process (see Sects.~\ref{direct} and \ref{Int} below). The
r.h.s. of \eqn{tlambdadef} defines the {\em transverse resolution time}
$\tlambda$, which is the scale controlling quantum coherence in the
interference of the BDMPS--Z gluons. Using $\lambda_f=1/k_f$ with $k_f$
given by \eqn{tauf}, it is easily to see that
 \beq \label{tlambda}
 \tau_\lambda\,=\,\frac{1}{\tqq\,(\hat q\omega)^{1/4}}
 \,=\,
\tau_f\, \frac{\tf}{\tqq}\,,
 \eeq
where the second estimate follows since $k_f\simeq \omega \theta_f$ and
$\tau_f\simeq 2/(\omega\tf^2)$.

\eqn{tlambdadef} implies an upper limit on $y^+$ that we shall now
explicitly work out. The second estimate in \eqn{tlambda}, which compares
$\tlambda$ to $\tau_f$, makes clear that there are two limiting regimes,
depending upon the ratio $\tqq/\tf$ :

\texttt{(i.a)} For relatively small dipole angles $\tqq\ll\tf$, one has
$\tlambda\gg\tau_f$ and then \eqn{tlambdadef} implies $y^+\lesssim
\tlambda$. This upper limit is relatively large showing that, in this
case, quantum coherence is relatively easy to ensure. This is natural:
during the formation process, the BDMPS--Z spectra produced by the two
emitters are confined to angles $\lesssim \tf$ around their respective
sources. So, if the dipole angle is much smaller than $\tf$, these two
spectra have a strong overlap with each other and can interfere.

\texttt{(i.b)} For larger dipole angles $\tqq\gg\tf$, one has
$\tlambda\ll\tau_f$ and therefore $t_1\ll\tau_f$ as well. (Indeed, $t_1$
is is strictly smaller than $\tlambda$, as obvious from
\eqn{tlambdadef}.) Using this information, the condition
\eqref{tlambdadef} simplifies to
 \beq\label{tintdef}
 t_1\,\lesssim\, \frac{\tlambda^2}{\tau_f}\,\equiv\,\tau_{int}
  \,,\eeq
which introduces a new scale $\tau_{int}$, the {\it interference time}.
This scale can be rewritten as
 \beq \label{tint1} \tau_{int}=\frac{2}{\omega \tqq^2}=\tau_f
\left(\frac{\tf}{\tqq}\right)^2\,,
 \eeq
where the first expression is recognized as the {\em vacuum--like}
formation time for a gluon emitted at an angle $\sim\tqq$. This situation
is reminiscent of the interference phenomena in the vacuum  (cf.
\eqn{interfvac}) : for the interference to be possible, the process must
begin with the emission of a gluon at a relatively large angle, of order
$\tqq$, with respect to its parent quark. Of course, this gluon could
then interfere with a similar, large--angle, emission by the other quark;
if so, it would contribute to the interference piece of the
bremsstrahlung spectrum in \eqn{bremsd}. Alternatively --- and this is,
of course, the situation that we are currently interested in ---, it can
interfere with a medium--induced emission by the other quark. Indeed,
among the gluons emitted by one parton (say, the antiquark) at an angle
$\gtrsim\tqq$, there is a non--trivial fraction which are co--moving with
the {\em other} parton (the quark). These gluons overlap with the quark
and are co--moving with it, so they cannot be distinguished from the
typical gluons from the {\em quark} wavefunction. Accordingly, they will
follow the formation process for in--medium radiation by the {\em quark}
and eventually emerge (at time $t_1+\tau_f$) at an angle $\tf$ w.r.t. the
latter. This process, which is qualitatively illustrated in
Fig.~\ref{fig:interf} right, will yield an interference contribution to
the BDMPS--Z spectrum of the dipole, whose phase--space however is
relatively small, since restricted by $\tau_{int}$.

\texttt{(ii) Color coherence.} In addition to quantum coherence, the
existence of interference effects between the two partonic emitters
require the preservation of the color coherence between the quark and the
antiquark. In the vacuum, the color state of the dipole is conserved
until a gluon emission takes place and the interference pattern  is
governed solely by quantum coherence. In the medium, on the contrary, the
interactions with the medium constituents change the color of each of the
propagating parton (via `color rotation'). For a very energetic parton,
this rotation amounts to multiplying the respective wavefunction by a
SU$(N_c)$ matrix--valued phase --- a Wilson line --- which involves the
random color field generated by the charged constituents of the medium
evaluated along the trajectory of the particle.

For the $q\bq$ pair we have two such Wilson lines which diverge from each
other (since so do the quark and the antiquark) at constant angle $\tqq$.
The color coherence is measured by the 2--point correlation function of
these Wilson lines, as obtained after averaging over the fluctuations of
the background field.  Within the `multiple soft scattering
approximation', this 2--point function can be computed to all orders in
the medium effects. The results of these calculations, to be detailed in
Sects.~\ref{out} and  \ref{Int}, show that the quark and the antiquark
loose any trace of their original color correlation after the {\it
decoherence time}
 \beq\label{tdecoh}
 \tau_{coh}\,=\,\frac{2}{
 (\hat q\tqq^2)^{1/3}}\,
 =\tau_f \, \left(\frac{\tf}{\tqq}\right)^{2/3}
 .\eeq
Accordingly, this scale too acts as an upper limit on $t_1$ : there is no
interference between the two partonic sources for the gluons whose
emission is initiated at a time $t_1$ larger than $\tau_{coh}$. It will
be later useful to have a direct comparison between this scale
$\tau_{coh}$ and the medium length $L$. This is obtained from
\eqn{tdecoh} and \eqn{formation} as
\beq\label{cohL} \tau_{coh}\simeq
\left(\frac{\theta_c}{\tqq}\right)^{2/3} L\,.
 \eeq
Quite remarkably, this estimate involves the limiting angle $\theta_c$ of
the BDMPS--Z spectrum, although the present physical context is quite
different: the scale $\tau_{coh}$ refers to the color coherence between
the two emitters {\em independently} of their radiation.

From the previous discussion, we see that the analysis of interference
effects in the medium--induce dipole radiation is a multi--scale problem.
While the details of the in--medium dipole antenna pattern are expected
to depend upon all these scales, the phase--space for interference is
controlled by the {\em smallest} of them,
$\tau_{min}=\min(\tau_\lambda,\tau_{int},\tau_{coh})$. Indeed, the
coherence of the $q\bq$ dipole is only preserved at times which are
simultaneously shorter than any of these scales. As a consequence, the
interference contribution to the gluon spectrum (that is, the
contribution of diagrams in which the gluon is emitted by the quark in
the amplitude and by the antiquark in the complex conjugate amplitude, or
{\it vice versa}), does not scale with the medium length, as the emission
from each of the sources does, but with $\tau_{min}$.

Remarkably, Eqs.~\eqref{tlambda}, \eqref{tint1}  and \eqref{tdecoh},
which can be summarized as
\beq\label{taucomp}
 \tau_f\,\sim\,\tau_{coh}\, \left(\frac{
 \tqq}{\theta_f}\right)^{2/3}\sim\, \tau_{\lambda}\
 \frac{\tqq}{\theta_f}
 \,\sim\, \tau_{int}
 \left(\frac{\tqq}{\theta_f}\right)^{2}\sim\,L
 \sqrt{\frac{\omega}{\omega_c}}\,
 \,,\eeq
show that, for a given in--medium formation time $\tau_f$, all the three
time scales relevant for coherence depend solely upon the ratio
$\tqq/\tf$. Physically, this is a consequence of the fact that, at
formation, the medium--induced gluon distribution \eqref{spectrumf} is
characterized by the formation angle $\tf$ alone (for a given $\tau_f$).
Hence, when discussing interference effects, it is natural to distinguish
between the values of $\tqq$ which are larger than $\tau_f$ and those
which are smaller. In addition, as we shall see, there is a change of
regime when $\tau_{coh}$ becomes as large as the medium size $L$, which
according to \eqn{cohL} happens when $\tqq\simeq \theta_c$. We are thus
led to consider the  three following ranges for $\tqq$:

\texttt{1. Relatively large dipole angles, $\tf\lesssim \tqq \lesssim
\ts$.} In this regime Eqs.~\eqref{tlambda}, \eqref{tint1} and
\eqref{tdecoh} imply the following hierarchy of scales
 \beq   \tau_{int}\lesssim \tau_\lambda \lesssim  \tau_{coh}
 \lesssim  \tau_f\qquad\mbox{when}\qquad \tqq \gtrsim \tf\,,
 \eeq
which shows that, for such large dipole angles, the condition
\eqref{tintdef} of quantum coherence is the most restrictive one.
Accordingly, in this regime, the longitudinal phase--space for
interferences is of order $\sim\tau_{int}\tau_f$ and thus is suppressed
with respect to the corresponding phase--space $\sim\tau_f L$ for direct
emissions by a factor
 \beq\label{Rint}  \mathcal{R}=\frac{\tau_{int}}{L}
  \sim \sqrt{\frac{\omega}{\omega_c}}
 \left(\frac{\tf}{\tqq}\right)^2  \ll 1\,.
 \eeq
The range of values spanned by $ \mathcal{R}$ within this regime will be
displayed in \eqn{IDratio1}.

%\vspace*{0.2cm} \noindent

\texttt{2. Relatively small dipole angles $\theta_c\ll \tqq\ll\tf$.} In
this case, the strongest limitation on the phase--space for interference
comes from the requirement of color coherence, as clear from the fact
that the ordering of time scales is now reverted:
\beq \tau_f \ll \tau_{coh} \ll \tau_\lambda \ll \tau_{int}
\qquad\mbox{when}\qquad \tqq\ll\tf\,.
 \eeq
So, the longitudinal phase--space for interference is now of order
$\tau_{coh}\tau_f$. So long as $\tqq\gg\theta_c$, this is still strongly
suppressed as compared to the phase--space for direct emissions, as
manifest from \eqn{cohL}. Hence, in this regime too, the interference
contribution to the spectrum is parametrically small  (see also
\eqn{IDratio2} for the corresponding range) :
 \beq  {\mathcal R}\,=\,\frac{\tau_{coh}}{L}\,=
 \left(\frac{\theta_c}{\tqq}\right)^{2/3}\ll\,1\,.
 \eeq

Note that, in this case, the medium--induced radiation by the dipole (the
incoherent sum of the two corresponding spectra by the quark and the
antiquark) is distributed at large angles $\tq\simeq
\tbq\gtrsim\tf\gg\tqq$, that is, well outside the dipole cone. One may
wonder why the total radiation in that case is not simply zero (as it
would be for the large angle radiation by a color--singlet dipole in the
vacuum). The reason is that, so long as $\tqq\gg\theta_c$, a $q\bq$ pair
immersed in the medium is {\em not} a `color singlet' anymore, except for
a very brief period of time $\sim \tau_{coh}$.
%\bigskip
 %\vspace*{0.2cm}\noindent

\texttt{3.  Very small dipoles angles $\tqq\lesssim \theta_c$.} When the
dipole angle is even smaller, $\tqq\lesssim \theta_c$, the color
coherence time $\tau_{coh}$ becomes as large as the medium size $L$, as
clear from \eqn{cohL}. In that case, the $q\bar q$ pair preserves its
color and quantum coherence throughout the medium, so the interference
effects are not suppressed anymore and they act towards reducing the
medium--induced radiation by the dipole. For sufficiently small angles
$\tqq\ll\theta_c$, the color decoherence is parametrically small and the
total (in--medium) radiation becomes negligible: the interference effects
and the direct emissions nearly compensate each other\footnote{The net
result should be of order $(\tqq/\theta_c)^2$, as clear by inspection of
\eqn{Sqq1} below.}. This conclusion is in agreement\footnote{We would
like to thank Carlos Salgado for useful discussions on this point.} with
the results in \cite{MehtarTani:2010ma}, obtained by working to leading
order in the `opacity expansion' \cite{Gyulassy:2000er} --- {\em i.e.} in
the single scattering approximation --- which is a legitimate
approximation when $\tqq\ll\theta_c$.

 %\vspace*{0.2cm}

In summary, we have argued that for sufficiently large dipole angles
$\tqq\gg\theta_c$, the interference effects for the medium--induced
radiation are negligible, so the total BDMPS--Z spectrum by the dipole is
the incoherent sum of the respective spectra produced by the quark and
the antiquark. For smaller angles $\tqq\lesssim\theta_c$, the
interference effects are not suppressed anymore and they eventually
cancel the direct emissions when $\tqq\ll\theta_c$. The transition
between the two regimes, occurring at $\tqq\simeq\theta_c$, could in
principle be studied within the formalism that we shall develop later.
However, such a study goes beyond the approximation schemes that we shall
use throughout this paper and which are adapted to the most interesting
regime at $\tqq\gg\theta_c$.

Note that, although so far we have focused on gluons with relatively soft
energies, $\omega\ll\omega_c$, our main conclusions remain valid when
$\omega$ approaches the limiting value $\omega_c$, as we now argue. When
$\omega\sim\omega_c$, one has $\tau_f\sim L$ and $\theta_f\sim\theta_c$,
so the intermediate regime of `relatively small dipole angles' ceases to
exist. Yet, \eqn{Rint} implies that, so long as
$\tqq\gg\tf(\omega_c)=\theta_c$, the interference effects are relatively
small even for $\omega\sim\omega_c$. This is so because the time scale
$\tau_{int}$ which limits quantum coherence is still much smaller than
$L$ in this regime.

We conclude our discussion by observing that the radiation of
BDMPS--Z--like gluons is not the only medium induced radiation by the
dipole. Indeed, in Ref~ \cite{MehtarTani:2011tz}, it has been shown that
in the presence of a dense medium, leading to color decoherence over a
distance of the order of the medium size $L$, the dipole produces
additional soft gluon radiation, which is emitted outside of the medium
and also outside the dipole cone ($\tq,\,\tbq>\tqq$). As we shall further
discuss in Sects.~\ref{out} and ~\ref{outlook}, this alternative
mechanism operates only for relatively small dipole angles and the
associated radiation is mostly collinear with the $q\bq$ axis, in the
sense that $\theta_q\sim \theta_\bq \sim \tqq\ll \tf$. Moreover, the
associated gluon spectrum is simply the bremsstrahlung spectrum and hence
it is independent of the medium size $L$. On the contrary, the
medium--induced radiation that we consider is emitted at larger angles,
within a range $\theta_f\lesssim \theta_q\lesssim \ts$, and it has a
strength proportional to the medium length. Hence, these two mechanisms,
which are simultaneously present in the medium, lead to gluon spectra
with very small overlap.

\begin{TABLE}[t]
{
\begin{tabular}{cccl}
    Parameter & Definition
& Parametric estimate & Physical meaning
 \\\hline   \\ %[2pt]
     $\tau_{q}$ & $\frac{2\omega}{k^2_\perp}$ & $\tau_f \left(\frac{\theta_f}{\theta_q}\right)^2$ &
    vacuum formation time
       \\[10pt]
       \hline   %\\ %[2pt]
   $\tau_f$ & $\sqrt{\frac{2 \omega}{\hat q}}$
& $\sqrt{\frac{\omega}{\omega_c}} \,L$ & in--medium formation time
   \\[10pt]
   $\theta_f$ & $\left(\frac{2 \hat q}{\omega^3}\right)^{1/4}$ &
       $\theta_c\left(\frac{\omega_c}{\omega}\right)^{3/4}$ & formation angle
    \\[10pt]
    $\theta_s$ & $\frac{\sqrt{\hat q L}}{\omega}$ & $\theta_c\, \frac{\omega_c}{\omega}$ &
    saturation angle
    \\[10pt]
           \hline  %\\%[2pt]
    $\tau_{int}$ & $\frac{2}{\omega \theta^2_{q\bar q}}$ & $\tau_f
    \left(\frac{\theta_f}{\theta_{q\bar q}}\right)^2$ &
    interference time
    \\[10pt]
                $\tlambda$ & $\frac{1}{\theta_{q \bar q}
                 \left(\omega \hat q\right)^{1/4} }$ &
                  $\tau_f \frac{\theta_f}{\theta_{q \bar q}} $&
                   transverse resolution time
    \\[10pt]
    $\tau_{coh}$ & $\frac{2}{
 (\hat q\tqq^2)^{1/3}}$ &
    $\tau_f \left(\frac{\theta_f}{\theta_{q\bar q}}\right)^{2/3} $ &
    color decoherence time
\end{tabular}
\caption
   {
 Scales relevant for medium--induced gluon radiation. The dimensionless ratios are related to the BDMPS medium parameters $\omega_c=\hat q L^2 /2 $ and $\theta^2_c=1/ \hat q L^3$
   }
\label{tab:dict} }
\end{TABLE}

\section{General set--up and formalism}
\label{general}

In this section, we shall more precisely describe our physical problem
--- a color dipole which radiates gluons while propagating through a QCD
medium (say, a quark--gluon plasma) --- and the formalism that we shall
use in order to study the dipole interactions with the medium and its
radiation. As noticed in the Introduction, a similar set--up has been
also used in Refs.~\cite{MehtarTani:2010ma,MehtarTani:2011tz}. But the
focus there was on some special physical conditions, allowing for
additional simplifications: the single scattering approximation (`dilute
medium') in \cite{MehtarTani:2010ma} and the restriction to
out--of--medium emissions (`soft and collinear gluons') in
\cite{MehtarTani:2011tz}. Here, we shall keep our discussion as general
as possible, in such a way to encompass the physics of medium--induced
gluon radiation in the multiple soft scattering regime. In the process,
we shall also make contact with the results in
Ref.~\cite{MehtarTani:2011tz} and thus clarify the precise kinematical
region for their applicability.

\subsection{The amplitude for gluon emission}
\label{amplit}

The in--medium dipole dynamics will be treated in the semi--classical
approximation, that is, by solving classical equations of motion in which
the dipole enters as a classical source of color charge. The medium
rescattering will be resummed to all orders via a background field
method. The effects of this rescattering on the quark and the antiquark
legs of the dipole will be treated in the eikonal approximation. The
corresponding effects on the emitted gluon will be treated exactly
(within the semi--classical approximation), by using an appropriate
background field propagator. The background field is assumed to be
random, with a Gaussian distribution, and the average over its
fluctuations will be performed using techniques borrowed from the color
glass condensate \cite{Iancu:2003xm}. The underlying assumptions are that
the two quarks are very energetic, with momenta much larger than any
momentum which can be transferred by the medium, whereas the emitted
gluon carries (transverse) momenta comparable to those of the medium.
Under these assumptions, our calculations are correct to lowest order in
the color charge of the dipole but to all orders in the medium effects.
This formalism has been used in Ref.~\cite{MehtarTani:2006xq} to study
the radiation by a single quark and shown to encompass the essential
BDMPS--Z physics.

The {\em color dipole} is truly a pair of classical, massless, particles
with opposite color charges (so that the pair is a color singlet) which
is produced at time $t_0=0$ by some hard process occurring inside the
medium. After being produced, the two particles separate from each other
at constant velocities,  ${\bm u}= {\bm p}_q/E_q$ for the quark ($q$) and
$\bar{\bm u}= {\bm p}_\bq/E_\bq$ for the antiquark ($\bar q$), which make
a relative angle $\tqq$ : ${\bm u}\cdot \bar{\bm u}=\cos\tqq$. Here ${\bm
p}_q$ and $E_q=|{\bm p}_q|$ represent the 3--momentum and the energy of
the quark (and similarly for the antiquark), assumed to much larger than
any other momentum scale in the problem. The only way that the medium can
act on such an energetic particle is by inducing a {\em color precession}
(see below). We choose the longitudinal axis ($x^3$) as the direction of
motion of the center--of--mass of the $q\bar q$ pair. In the medium rest
frame, the dipole has a relatively large longitudinal boost $\gamma\gg 1$
and hence a relatively small opening angle $\theta_\qq\sim 1/\gamma$.
This angle will be nevertheless assumed to be significantly larger than
the critical angle for medium--induced radiation $\theta_c\sim 1/(Q_s
L)$, which is very small ($\theta_c\ll 1$) as explained in the previous
section. The dipole propagates through the medium along a longitudinal
distance $L$ before escaping into the vacuum.

The {\em QCD medium} is described as a random color background field
$\calA_a^\mu$ with a Gaussian distribution. As well known e.g. from the
experience with the color glass condensate \cite{Iancu:2003xm}, this
description becomes simpler by working in a Lorentz frame in which the
medium is strongly boosted (an `infinite momentum frame'). For the
problem at hand, it is convenient to choose the `dipole frame' in which
the COM of the $q\bar q$ pair is nearly at rest, meaning that the plasma
is boosted (essentially, by the dipole $\gamma$ factor introduced above)
in the negative $x^3$ direction. In this new frame and in light--cone
(LC) gauge $A^+_a=0$, the background field has only one non--trivial
component, $\calA_a^\mu=\delta^{\mu-}\calA^-_a$, which is moreover
independent of the LC `time' $x^-$, by Lorentz time dilation. We have
introduced here the LC components of the 4--vector $\calA_a^\mu$, defined
in the standard way; e.g. $x^\mu=(x^+,x^-,\bxt)$, with
 \beq \label{LC}
 x^+=\frac{1}{\sqrt{2}}\,(x^0+x^3)\,,\quad
 x^-=\frac{1}{\sqrt{2}}\,(x^0-x^3)\,,\quad\bxt=(x^1,\,x^2)\,.
 \eeq
In the dipole frame, both the dipole angle $\theta_\qq$ and the
characteristic medium angle $\theta_c$ are enhanced by a factor $\gamma$
(so, in particular, $\theta_\qq\sim \order{1}$), but the inequality
$\theta_{q\bar q}\gg \theta_c$ remains of course true. In view of that,
and in order to avoid a proliferation of symbols, we shall use the same
notations for quantities in the plasma rest frame and in the dipole frame
--- the difference should be clear from the context. Moreover, we shall
often use the small--angle version of a parametric estimate (e.g.,
$\tau_q\simeq{2}/{\omega\tq^2}$) even when working in the boosted frame,
where the angles are not necessarily small; what we truly mean by such a
writing, is an estimate which becomes true after boosting back to the
plasma rest frame.

Another advantage of using the dipole frame refers to the correlations
between the charged constituents of the medium (say, quark and gluon
quasiparticles in the case of a weakly--coupled QGP): the longitudinal
($x^+$) range of the correlations, which was $1/\mu_D$ in the original
frame, is now Lorentz--contracted to $1/(\gamma\mu_D)$. When probing this
distribution over relatively large longitudinal separations $\Delta
x^+\gg 1/(\gamma\mu_D)$, one can describe the medium constituents as
independent color charges with a current density $J^{\mu,a}_{\rm
med}(x)=\delta^{\mu-}\rho^a(x^+,\bxt)$ and a local 2--point correlation
 \beq\label{2prho}
\langle\rho^a(x^+,\bxt)\rho^b(y^+,{\bm y}_\perp)\rangle \,=\,\delta^{ab}
 \delta(x^+-y^+)\,\delta^{(2)}(\bxt-{\bm y}_\perp)\,n_0\,,\eeq
where $n_0$ is the average color charge squared per unit volume, assumed
to be homogeneous. (For a longitudinally expanding medium, this would be
a function of $x^+$.) If the medium is a QGP, then $n_0\propto \gamma
T\mu_D^2$ in the dipole frame. Such a color charge distribution gives
rise to the following distribution for the background field $\calA^-_a$ :
 \beq\label{2pA}
\langle \calA^-_a(x^+,{\bm q}_\perp)\calA^-_b(y^+,{\bm k}_\perp)\rangle
\,=\,\delta_{ab}
 \delta(x^+-y^+)\,(2\pi)^2 \delta^{(2)}({\bm q}_\perp-{\bm k}_\perp)\,
 \frac{ n_0}{(q_\perp^2+\mu_D^2)^2}\,.\eeq
Note that the Debye screening has been heuristically implemented as a
`gluon mass' $\mu_D$, although the actual mechanisms at work are
generally more complicated.

The {\em gluon radiation} by the dipole will be described in the
classical approximation, as the additional color field (on top of the
background field) generated by the $\qq$ pair. The classical
approximation is correct when the gluon is soft relative to its sources,
meaning that $\omega=|{\bm k}|\ll E_q,\,E_\bq$. The color field $a^\mu_a$
describing the radiation is a small perturbation of the background field
and will be obtained by solving the linearized version of the Yang--Mills
equation for the total field $A^\mu=\delta^{\mu -}\calA^-+a^\mu$ (color
indices will be often kept implicit in what follows)
 \beq\label{YM}
 D_\nu F^{\nu\mu} \,=\,\delta^{\mu -}\rho + J^\mu_{\rm dip}\,.\eeq
The dipole color current $J^\mu_{\rm dip}=J^\mu_q+J^\mu_\bq$ involves
contributions from the quark and the antiquark and also depends upon the
background field, because it obeys a covariant conservation law: ${\cal
D}_\mu J^\mu_{\rm dip}=0$, where ${\cal D}^\mu=\del^\mu + \delta^{\mu -}
ig\calA^-$.

In the vacuum ($\calA^-=0$), the color current associated with a pair of
classical particles with constant velocities can be written as
$j^\mu_{\rm dip}=j^\mu_q+j^\mu_\bq$, with
 \beq\label{jvac}
 j^\mu_{q,a}(x)&\,=\,&gu^\mu\,\theta(x^+)\,\delta(x^--u^-x^+)\,
 \,\delta^{(2)}(\bxt-{\bm u}_\perp x^+)\,\mcal{C}_a\,,\nn
 j^\mu_{\bq,a}(x)&=&-g\bar u^\mu\,\theta(x^+)\,
 \delta(x^--\bar u^-x^+)\,
 \,\delta^{(2)}(\bxt-\bbu x^+)\,\mcal{C}_a\,.
 \eeq
We have used here LC notations, with $u^\mu\equiv p_q^\mu/p^+_q =
(1,u^-,\bu)$ and $\bar u^\mu\equiv p_\bq^\mu/p^+_\bq = (1,\bar
u^-,\bbu)$. Note that for the right--moving dipole, $x^+$ plays the role
of `time' while $x^-$ is the `longitudinal coordinate'. The `color
charges' $\mcal{C}_a$ are the components of a color vector in the adjoint
representation describing the orientation of the quark current in the
internal SU$(N_c)$ space; for the antiquark,
$\bar{\mcal{C}}_a=-\mcal{C}_a$. The current is conserved, $\del_\mu
j^\mu_{\rm dip}=0$, since $\del_\mu
j^\mu_{q,a}=g\mcal{C}_a\delta^{(4)}(x)=-\del_\mu j^\mu_{\bq,a}$.

In the eikonal approximation, the effect of the medium on the dipole
consists merely in color rotations, separately for the quark and the
antiquark:
 \beq\label{jmed}
 J^\mu_{q,a}(x)&=&\mcal{U}_q^{ab}(x^+,0)\,j^\mu_{q,a}(x),\qquad
 J^\mu_{\bq,a}(x)\,=\,\mcal{U}_\bq^{ab}(x^+,0)\,j^\mu_{\bq,a}(x).\eeq
$\mcal{U}_q(x^+,0)$ is a Wilson line in the adjoint representation, which
is the special case of
 \beq\label{Uq}
 \mcal{U}(x^+, y^+; [\bm{r}_\perp])\,=\,{\rm P}\,\exp\left\{
 -ig\int_{y^+}^{x^+}\rmd z^+\,\calA^-(z^+,\bm{r}_\perp(z^+))\right\}\eeq
for $y^+=0$ and the transverse path $\bm{r}_\perp(z^+)=\bu z^+$ (the
quark trajectory in the transverse plane). In \eqn{Uq},
$\calA^-=\calA^-_aT^a$ is a color matrix in the adjoint representation
and the symbol P denotes time--ordering in $z^+$. Using $(\del/\del
x^+)\mcal{U}_q^{ab}=-ig\calA^-_{ac}(x)\mcal{U}_q^{cb}$, it is easy to
check that ${\cal D}_\mu J^\mu_{q,a}=g\mcal{C}_a\delta^{(4)}(x)=-{\cal
D}_\mu J^\mu_{\bq,a}$, so the dipole current is {\em covariantly}
conserved, as it should. Note that, although the wavefunction of a
physical quark is known to transform according to the {\em fundamental}
representation of the color group, the corresponding color current
\eqref{jmed} involves a Wilson line in the {\em adjoint} representation,
since this current is a vector in the color space SU$(N_c)$. The only
trace of the underlying fundamental representation lies in the
normalization of the color vector $\mcal{C}_a$, namely
$\mcal{C}_a\mcal{C}_a=C_F=(N_c^2-1)/2N_c$.

In the LC gauge $A^+=0$, only the transverse components $a^i$ (with
$i=1,2$) of the radiated field contribute to the matrix element for gluon
emission (see below). The corresponding, linearized, equation of motion
is readily obtained from \eqn{YM} and reads
 \beq\label{EOM}
 \big(2\del^+{\cal D}^-\,-\,\grad^2\big)a^i\,=\,J^i_{\rm dip}
 -\,\frac{\del^i}{\del^+}J^+_{\rm dip}\,.\eeq
This equation can be formally solved in terms of the background field
Klein--Gordon propagator, {\em i.e.} the Green's function for the
differential operator in the left hand side. The corresponding solution
is well known in the literature (see e.g. \cite{MehtarTani:2006xq}) and
will be succinctly described here. Given that the background field is
independent of $x^-$, it is convenient to first perform a Fourier
transform to the $k^+$ representation. Then the solution to \eqn{EOM} can
be written as
 \beq\label{sol0}
 a^i_a(x^+,\bxt; k^+)\,=\,\frac{i}{2k^+}\int \rmd y^+ \rmd {\bm y}_\perp\,
 {\mcal G}_{ab}(x^+,\bxt; y^+,{\bm y}_\perp; k^+)\,{\mcal J}^i_b
 (y^+,{\bm y}_\perp; k^+)\,,\eeq
where ${\mcal J}^i$ refers to the total current in the r.h.s. of
\eqn{EOM} and the Green's function ${\mcal G}$ obeys
 \beq\label{eqG}
 \left( i{\cal D}^-+\,\frac{\grad^2}{2k^+}\right)
 {\mcal G}(x^+,\bxt; y^+,{\bm y}_\perp; k^+)\,=\,
 i\delta(x^+-y^+)\,\delta^{(2)}(\bxt-{\bm y}_\perp)\,.\eeq
${\mcal G}$ is formally the same as the $D=2+1$ Schr{\"o}dinger evolution
operator for a quantum--mechanical particle with mass $k^+$ and time
$x^+$ propagating in a time--dependent potential $g\calA^-$. As well
known, this propagator admits the following representation as a path
integral:
 \beq\label{GPI}
 {\mcal G}(x^+,\bxt; y^+,{\bm y}_\perp; k^+)\,=\,
 \int [D\bm{r}_\perp(z^+)]\,\exp\left\{
 i\frac{k^+}{2}\!\int_{y^+}^{x^+}\rmd z^+\dot{\bm r}^2_\perp(z^+)\right\}
 \mcal{U}(x^+, y^+; [\bm{r}_\perp])\,,\eeq
with the paths $\bm{r}_\perp(z^+)$ obeying the boundary conditions
$\bm{r}_\perp(y^+)={\bm y}_\perp$ and $\bm{r}_\perp(x^+)=\bxt$. The
corresponding vacuum propagator ($\calA^-=0$) will be also needed :
 \beq\label{G0}
 {\mcal G}_0(x^+,\bxt; y^+,{\bm y}_\perp; k^+)\,=\,\Theta(x^+-y^+)
 \,\frac{k^+}
 {2\pi\,i(x^+-y^+)}\,\exp\left\{
 i\,\frac{k^+(\bxt-{\bm y}_\perp)^2}{2(x^+-y^+)}\right\}\,.\eeq
Note that \eqn{GPI} goes beyond the eikonal approximation in the sense
that the gluon trajectory is not {\em a priori} imposed (as we did for
the quark and the antiquark), but rather is determined by the gluon
interactions with the background field.

Given the solution $a^i$, the gluon emission amplitude is obtained as
(for an on--shell gluon with 4--momentum $k^\mu$, color $a$, and
polarization $\lambda$)
 \beq\label{M0}
  {\mcal M}^a_\lambda(k^+,\bk)= -\lim_{k^2\to 0} k^2
  a_\mu^a(k)\epsilon^\mu_\lambda(k)\,,\qquad
  \epsilon^\mu_\lambda(k^+,\bk)=\Big(0,\frac{{\bm \epsilon}_\perp\cdot
  \bk}{k^+},{\bm \epsilon}_\perp\Big).\eeq
In the LC gauge $a^+=0$, this involves only the transverse components
$a^i$, as anticipated. Using \eqn{sol0} for $x^+\to\infty$ together with
the following composition law (valid for $x^+>z^+>y^+$)
 \beq\label{compG}
  {\mcal G}(x^+,\bxt; y^+,{\bm y}_\perp; k^+)=\int
   \rmd {\bm z}_\perp\,
   {\mcal G}(x^+,\bxt; z^+,{\bm z}_\perp; k^+)\,
   {\mcal G}(z^+,{\bm z}_\perp; y^+,{\bm y}_\perp; k^+),\eeq
applied to $z^+=L^+ \equiv \sqrt{2}L$ (notice that for $x^+>L^+$, one has
$\calA^-=0$ and then ${\mcal G}={\mcal G}_0$), one obtains the amplitude
as a sum of two pieces,
 \beq\label{M1}
  {\mcal M}^i_a(k^+,\bk)\equiv -\lim_{k^2\to 0} k^2
  a^i_a(k)={\mcal M}^{i,\,{\rm in}}_a + {\mcal M}^{i,\,{\rm out}}_a,\eeq
describing emissions inside the medium ($0<x^+ <L^+$) and outside the
medium ($x^+>L^+$), respectively. Each of these pieces is a sum of quark
and antiquark contributions and below we only show the respective quark
contributions. The `out' piece is the simplest:
 \beq\label{Mqout}
  {\mcal M}^{i,\,{\rm out}}_{a,q}(k^+,\bk)&=&
  \int\rmd^4x\, \rme^{ik\,\cdot \,x}\,\Theta(x^+-L^+) \,{\mcal J}^i_{a,q}(x)
  \nn &=& g(u^i-v^i)\,\mcal{U}_q^{ab}(L^+,0)\,\mcal{C}_b\int_{L^+}^\infty
  \rmd x^+\, \rme^{i(k\,\cdot\, u)x^+}
 \eeq
where $k^-=\bk^2/2k^+$, ${\mcal J}^i_{q}= J^i_{q} - ({\del^i}/{\del^+})
J^+_{q}$, and in the second line we have used the $\delta$--functions in
\eqn{jvac} to perform the integrations over $x^-$ and $\bxt$ and denoted
$v^\mu\equiv k^\mu/k^+= (1,v^-,\bv)$. The `in' piece of the
quark--emission amplitude reads
 \beq\label{Mqin}
  {\mcal M}^{i,\,{\rm in}}_{a,q}(k^+,\bk)
  &=& g\int_0^{L^+}
  \rmd x^+\, \rme^{ik^-L^+ + i(k^+ u^-)x^+}\int
   \rmd {\bm z}_\perp\,\rme^{-i\bk\,\cdot\,{\bm z}_\perp} \nn
   &{}&\quad(u^i+ i\del^i_x/k^+)\,
   {\mcal G}_{ab}(L^+, {\bm z}_\perp; x^+, \bxt; k^+)\Big|_{\bxt
   =\bu x^+}\,
  \mcal{U}_q^{bc}(x^+,0)\,\mcal{C}_c\,.
 \eeq
The corresponding formul\ae{} for the antiquark are obtained by replacing
$u^\mu\to \bar u^\mu$ and $\mcal{C}_a\to - \mcal{C}_a$.

Given the amplitude, the emission probability ${\mcal P}$ and the gluon
spectrum are obtained by taking the modulus squared and then summing over
colors and polarizations:
 \beq\label{spectrum}
 \omega\frac{\rmd N}{\rmd^3{\bm k}}\,=\,\frac{1}{16\pi^3}\,
 {\mcal P}({\bm k})\,,\qquad {\mcal P}({\bm k})\,\equiv
 \sum_{a,i}\,\langle |{\mcal M}^i_a|^2\rangle,\eeq
where we have written the gluon momentum in normal coordinates as
$k^\mu=(\omega,{\bm k})$ with $\omega=|{\bm k}|$ and performed the
polarization sum by using $ \sum_\lambda
\epsilon^i_\lambda(k)\epsilon^{j\,*}_\lambda(k)=\delta^{ij}$. The
brackets in \eqn{spectrum} refer to the medium average according to
\eqn{2pA}.

%(throughout this work, averages below denotes a medium average)

%\section{The `out--out' terms as a warm up}
%\label{out}

The amplitude ${\mcal M}^i_a$ is truly a sum of four terms:
$(q,\,\mbox{in})$, $(q,\,\mbox{out})$, $(\bq,\,\mbox{in})$, and
$(\bq,\,\mbox{out})$. Hence the emission probability in \eqn{spectrum}
involves 16 terms: 8 of them describe direct emissions by either the
quark or the antiquark, and 8 represent $q\bar q$ interference terms.
Each of these types of contributions --- direct or interference ---
involves three types of pieces: (in, in), (in, out), or (out, out). For
the (in, in) contributions, the gluon is emitted inside the medium in
both the direct amplitude and the complex conjugate one\footnote{It would
be perhaps more appropriate to say that the gluon is {\em emitted} in the
direct amplitude and {\em reabsorbed} in the complex conjugate amplitude.
For brevity, we shall refer to all such processes as `emissions'.};
denoting the respective emission times as $x^+$ and $y^+$, we have $0 <
x^+,\,y^+ < L^+$. For the (in, out) terms, one has $0 < x^+ < L^+$ and
$y^+ > L^+$, or vice--versa. Finally, for the (out, out) pieces, both
$x^+$ and $y^+$ are larger than $L^+$.

We anticipate that, for the problem of medium--induced gluon radiation,
the (in, in) contributions will be the most important ones, both for
direct emissions and for the interference terms. Here, however, we shall
start by computing the respective (out, out) pieces, with the purpose of
illustrating the medium averaging and the phenomenon of color decoherence
in the simplest possible setting. This will also allow us to make contact
with the results in Ref.~\cite{MehtarTani:2011tz}.

\subsection{The `out--out' terms as a warm up}
 \label{out}

Consider first the direct gluon emission, say from the quark. \eqn{Mqout}
implies
 \beq\label{Pqout}
 {\mcal P}_q^{\mbox{(out)}}({\bm k})&=&g^2{(\bu-\bv)^2}
  \langle \mcal{U}_q^{ab}(L^+,0)\mcal{C}_b
 \mcal{C}_d\,\mcal{U}_q^{\dagger\,ad}(L^+,0)\rangle\int_{L^+}^\infty
  \rmd x^+\! \int_{L^+}^\infty
  \rmd y^+ \rme^{i(k\,\cdot\, u)(x^+-y^+)}
 \nn
 &=&g^2 C_F\frac{(\bu-\bv)^2}
 {(k\cdot u)^2}\,=\,\frac{2g^2 C_F}{(k^+)^2}\,\frac{1}{v\cdot u}
 \,,
 \eeq
where the second line follows after using $\mcal{C}_b\, \mcal{C}_d=
\delta_{bd} C_F/(N_c^2-1)$ --- the condition that prior to the emission
the $q\bar q$ pair be in a color singlet state. We have also used (recall
that e.g. $v^\mu= (1,v^-,\bv)$ with $v^2=0$ and hence $2v^-= \bv^2$)
 \beq\label{buv}
 (\bu-\bv)^2 = \bu^2 + \bv^2 - 2\bu\cdot \bv =
 2(u^- + v^- - \bu\cdot \bv)= 2 v\cdot u\,.\eeq
An expression similar to \eqn{Pqvac} but with $v\cdot u\to v\cdot {\bar
u}$ holds for the direct emission by the antiquark. Note that there is no
medium dependence in the final result for ${\mcal P}_q^{\mbox{(out)}}$
because \texttt{(1)} the quark Wilson lines in the direct and the complex
conjugate amplitude have compensated each other, and \texttt{(2)} there
was a similar cancelation of the $L$--dependent phases $\rme^{\pm
i(k\,\cdot\, u)L^+}$ generated by the lower limit $L^+$ of the time
integrations. Accordingly, \eqn{Pqvac} is formally identical to the
corresponding probability in the vacuum, ${\mcal P}_q^{\mbox{(vac)}}$.
For later use, it is convenient to rewrite this vacuum probability as
 \beq\label{Pqvac}
 {\mcal P}_q^{\mbox{(vac)}}({\bm k})\,=\,2g^2 C_F\,{(\bu-\bv)^2}
 \,\tau_q^2\,,\eeq
where we have recognized the formation time (for an emission by the
quark) $\tau_q=1/[\sqrt{2}\,(k\cdot u)]$. Indeed, in a frame where the
quark propagates along the longitudinal axis ($\bu=0$), one has $k\cdot
u=k^-=k^2_\perp/2k^+$, which is the inverse formation time adapted to the
LC time variable $x^+$. More generally, in the actual frame where the
quark has a transverse velocity $\bu$, we can write
 \beq\label{tq}
 k\cdot u =\,\frac{\omega E_q}{p^+_q}\,(1-\cos\theta_q)\,\simeq\,
 \frac{\omega\tq^2}{2\sqrt{2}}\ \quad\Longrightarrow\quad
 \tau_q=\,\frac{1}{\sqrt{2}\,(k\cdot u)}\,\simeq\,
 \frac{2}{\omega\tq^2}\,,\eeq
where one recognizes the expression \eqref{tau} for the formation time at
small angles. In writing \eqn{tq} we performed approximations valid when
all the angles in the problem (the dipole angle and the gluon emission
angles) are small; then $p^+_q \simeq\sqrt{2}E_q$ etc. In what follows we
shall systematically perform such approximations which are strictly valid
in the plasma rest frame.

As explained in Sect.~\ref{heur}, throughout this paper we are mostly
interested in emissions at relatively large angles, for which the
formation times are small: $\tau_q\ll L^+$. On the other hand, the (out,
out) piece \eqref{Pqout} of the emission probability is controlled, by
construction, by emission times $x^+$ and $y^+$ within the range $L^+\le
x^+,\,y^+\lesssim L^+ +\tau_q$, which are much larger than $\tau_q$. It
would be very unnatural that gluons be emitted (in the vacuum) at times
much larger than their formation times. But as a matter of facts, when
$\tau_q\ll L^+$ there is no {\em physical} out--of--medium emission; in
that case, \eqn{Pqout} represents merely a piece of the total result
which cancels against other pieces. More precisely, the boundary terms
generated by emission times within an interval $\Delta x^+\simeq \tau_q$
around the medium boundary at $L^+$ cancel among the (in, in), (in, out),
and (out, out) contributions, separately for the direct emissions and for
the interference terms. The net result is that all the {\em vacuum--like}
emissions at large angles are emitted at early times $x^+,\,y^+\lesssim
\tau_q\ll L^+$, as expected on physical grounds.

%But of course, the situation is different for the {\em medium--induced
%emissions}, as already mentioned in Sect.~\ref{heur} (see also
%Sect.~\ref{direct}).

The above argument also explains why the (out, out) pieces play no
physical role for the situation of interest in this paper, which is
characterized by small formation times: $\tau_q,\,\tau_\bq\ll L$. On the
other hand, these pieces become important for the relatively soft and
collinear emissions which have large formation times
$\tau_q,\,\tau_\bq\gtrsim L$. This is the situation considered in
Ref.~\cite{MehtarTani:2011tz}. To make contact with the results in that
paper, we shall now consider the (out, out) contribution to the
interference term. From \eqn{Mqout}, this is obtained as
  \beq\label{Iout}
 {\mcal I}^{\mbox{(out)}}({\bm k})&\,=\,&-\frac{2g^2 C_F}{(k^+)^2}
 \,\frac{(u^i-v^i)(\bar
 u^i-v^i)}{(v\cdot u)(v\cdot \bar u)}\,\cos[L^+ k\cdot(u-\bar u)]\,
 S_{q\bq}(L^+,0)\,,\eeq
where the quark and the antiquark Wilson lines combined in the following
2--point function
 \beq\label{Sqq}
 S_{q\bq}(L^+,0)\,=\,\frac{1}{N_c^2-1}\langle {\rm Tr}\,
 \mcal{U}_q (L^+,0)\,\mcal{U}_\bq^{\dagger}(L^+,0)\rangle\,,\eeq
which describes the residual color coherence between the two fermions
after having crossed the medium --- that is, the probability for the
$q\bar q$ pair to remain in a color singlet state. $S_{q\bq}$ is formally
the same as the scattering $S$--matrix for a dipole made with a pair of
colored particles in the adjoint representation that we shall succinctly
refer to as the `$q\bq$ dipole'. For a background field with a Gaussian
distribution, cf. \eqn{2pA}, the expectation value in \eqn{Sqq} is easily
computed as (see e.g. \cite{Iancu:2003xm})
 \beq\label{Sdip0}
 S_{\rm dip}\big(x^+,y^+; [\bm{r}_\perp]\big)\,=\,\exp\Biggl\{-g^2N_c\int\limits_{y^+}^{x^+}\rmd z^+
 n_0(z^+)\int \frac{\rmd^2\bm{q}_\perp}{(2\pi)^2}\,\frac{1-
 \rme^{i\bm{q}_\perp\,\cdot\,\bm{r}_\perp(z^+)}}{(\bm{q}_\perp^2+
 \mu_D^2)^2}\Biggr\}\,,\eeq
where for later convenience we have kept generic endpoints in time, $y^+$
and $x^+$, and a generic `trajectory' $\bm{r}_\perp(z^+)$ for the dipole
transverse size in the interval $y^+ < z^+ < x^+$. For the cases of
interest in this work, the dipole size is always much smaller than the
medium screening length, $r\equiv |\bm{r}_\perp|\ll\mu_D^{-1}$. Then the
integral over $\bm{q}_\perp$ in \eqn{Sdip0} is controlled by transverse
momenta within the range $\mu_D < q_\perp < 1/r$ and to leading
logarithmic accuracy it can be estimated by expanding
$\rme^{i\bm{q}_\perp\,\cdot\,\bm{r}_\perp}$ to second order. This yields
 \beq\label{Sdip1}
 S_{\rm dip}\big(x^+,y^+; [\bm{r}_\perp]\big)\,\simeq\,
 \exp\Biggl\{-\frac{\alpha_s N_c}{4}\int\limits_{y^+}^{x^+}\rmd z^+
 n_0(z^+)\,r^2(z^+)\,\ln\frac{1}{r^2(z^+)\mu_D^2}\Biggr\}\,,\eeq
where the logarithm $\rho\equiv \ln({1}/{r^2\mu_D^2})$ is assumed to be
relatively large, $\rho\gg 1$. A more compact version of \eqn{Sdip1} can
be obtained by assuming $n_0(z^+)=n_0$ to be constant and neglecting the
variation of the logarithm within the interval of integration; then,
 \beq\label{Sdip}
 S_{\rm dip}\big(x^+,y^+; [\bm{r}_\perp]\big)\,\simeq\,
 \exp\biggl\{-\frac{1}{4}\,\hat q\,\rho\int_{y^+}^{x^+}\rmd z^+
 \,r^2(z^+)\biggr\}\,,\eeq
where (the saturation scale $Q_s$ is introduced for later reference)
 \beq\label{Qs}
 \hat q\,\equiv\,\alpha_s N_c n_0\,\sim\,\alpha_s N_c \mu_D^2 T\,,\qquad
 Q_s^2\equiv \hat q L^+\,,\eeq
and it is understood that the logarithm $\rho$ in \eqn{Sdip} is evaluated
with the maximal dipole size $r_{\rm max}$ within the interval $y^+ < z^+
< x^+$. (Eqs.~\eqref{Qs} and \eqref{dkdt} are consistent with each other
since $\ell\sim 1/(\alpha_s N_c T)$ for a weakly--coupled QGP.)

When applying these formul\ae{} to the $q\bq$ dipole in \eqn{Sqq}, one
has $\bm{r}_\perp(z^+)=(\bu-\bbu)z^+$ with $0<z^+<L^+$ and therefore
 \beq\label{Sqq1}
 S_{q\bq}(L^+,0)\,\simeq\,
 \exp\biggl\{-\frac{1}{12}\,\hat q\,\rho \,(\bu-\bbu)^2 (L^+)^3
 \biggr\}\,\simeq\,
 \exp\biggl\{-\frac{1}{24}\,\big(Q_s\tqq L^+)^2\,\rho
 \biggr\}\,,\eeq
where in writing the second estimate we have used a small--angle
approximation which holds, strictly speaking, in the plasma rest frame
(the product $\tqq L^+$ is boost invariant):
 \beq\label{smalltqq}
 (\bu-\bbu)^2 \,=\, 2 u\cdot\bar u \,=\,2\,\frac{p_q\cdot p_\bq}{p^+_q
p^+_\bq}\,=\,2\,\frac{E_qE_\bq}{p^+_q p^+_\bq}\,\Big(1-\cos\tqq\Big)
 \,\simeq\,\frac{\tqq^2}{2}\,.\eeq

At this point one should recall that we consider a relatively large
dipole angle, $\tqq\gg\theta_c$ or $Q_s\tqq L^+\gg 1$. The exponent in
\eqn{Sqq1} is therefore large, which implies that $S_{q\bq}\ll 1$. We see
that the (out, out) contribution to interference is washed out by the
medium, due to the color decoherence suffered by the $q\bar q$ pair after
passing through the medium: since they move along different directions,
the quark and the antiquark undergo different color precessions, so after
leaving the medium, they do not form a color singlet anymore. Since the
interference term vanishes, it follows that {\em in the regime where the
(out, out) piece yields a physical contribution}, the total dipole
radiation is the incoherent sum of two vacuum--like contributions (cf.
\eqn{Pqout}), by the quark and the antiquark. This is the main conclusion
in Ref.~\cite{MehtarTani:2011tz} and admits interesting consequences: it
implies that, in the presence of the medium, there should be an
enhancement in the radiation at large angles, outside the dipole cone:
$\tq,\,\tbq >\tqq$.

To fully appreciate the impact of this conclusion, it is important to
specify the kinematical region where it applies. From the previous
arguments, it is clear that this relies on two main assumptions:
\texttt{(i)} relatively large formation times\footnote{Formally, this
condition is necessary to avoid the rapid oscillations of the cosine
factor in \eqn{Iout}, whose argument is the same as $L^+ k\cdot(u-\bar
u)=L/{\tau_q}-L/{\tau_\bq}$. Less formally, the regime of large formation
times $\gtrsim L$ is the only one where the (out, out) piece of the
spectrum is physically irrelevant, as already explained.}
$\tau_q,\,\tau_\bq \gtrsim L$, and \texttt{(ii)} a sufficiently large
dipole angle $\tqq\gg\theta_c$. As we shall now argue, these conditions
are satisfied for sufficiently soft gluons and for relatively small, but
not too small, emission ($\tq,\,\tbq$) and dipole ($\tqq$) angles. The
precise conditions read
  \beq\label{outtheta}
 \omega\,\ll\, \omega_c\qquad{\rm and}\qquad \theta_c\,\ll\,\tq,\,\tbq,\,
 \tqq
 \,\lesssim\,\theta_c\left(\frac{\omega_c}{\omega}\right)^{1/2}\,.\eeq
The upper limit on $\omega$ comes up by combining the two conditions
above and focusing on emission angles which are commensurable with the
dipole angle: $\tq\sim\tbq\sim \tqq$ (this is the regime where the
conclusion in Ref.~\cite{MehtarTani:2011tz} have non--trivial
consequences). Then, one can write
\beq
\tau_q\simeq \frac{2}{\omega\tq^2}\,\gtrsim\,L\quad \Longrightarrow\quad
 \omega\,\lesssim\,
\frac{2}{L\tqq^2}\,\ll\,\frac{2}{L\theta_c^2}\,
 =\,\omega_c\,,\eeq
where we have used $\tq\sim\tqq \gg \theta_c$ and
$\theta_c^2={2/(\omega_c L)}$, cf. \eqn{omegac}. Then the upper limit on
the values of the angles follows by rewriting the condition
$\tau_q\gtrsim L$ in the form
 \beq
 \tq\,\lesssim\,\sqrt{\frac{2}{\omega L}}\,=\,\theta_c
 \sqrt{\frac{\omega_c}{\omega}}\,.
 \eeq
It should be also clear from the above that for a very small dipole angle
$\tqq\lesssim\theta_c$, the medium effects become irrelevant (since the
$q\bq$ pair preserves its color coherence throughout the medium), so the
soft emissions with $\tau_q\gtrsim L$ proceed exactly as in the vacuum
--- in particular, the dipole antenna shows the characteristic
angular ordering.

Note that the region \eqref{outtheta} has some overlap with the
`small--angle regime' for medium--induced gluon radiation as defined in
Sect.~\ref{intphys} --- that is, the regime characterized by $\theta_c\ll
\tqq\ll\tf$. There is, however, an important difference: the respective
range in Sect.~\ref{intphys} refers to the {\em dipole} angle $\tqq$
alone; while this angle can be as small as $\theta_c$, the actual {\em
emission} angles for the BDMPS--Z gluons are much larger:
$\tq,\,\tbq\gtrsim\tf\gg \theta_c$. On the other hand, the upper limit in
\eqn{outtheta} is much smaller than $\tf$, as it can be easily checked
using \eqn{theta}. So, even for dipole angles $\tqq$ as small as shown in
\eqn{outtheta}, the out--of--medium emissions discussed in
Ref.~\cite{MehtarTani:2011tz} and the BDMPS--Z--like emissions that we
presently focus on are geometrically separated, with the latter being
distributed at significantly larger angles than the former.

\section{Medium--induced gluon radiation: direct emission}
\label{direct}

Starting with this section, we shall concentrate on the in--medium, or
(in, in), pieces, which are the dominant contributions to medium--induced
gluon radiation in the kinematical range of interest (relatively small
frequencies, $\omega \ll\omega_c$, or large emission angles $\tq \gtrsim
\theta_f(\omega)\gg \theta_c$, cf. \eqn{theta}). Although we are
ultimately interested in the quark--antiquark interference terms, for
which we shall present original results in the next section, here we
shall start our analysis with the direct emission terms, from which we
shall extract the BDMPS--Z spectrum. This will give us the opportunity to
develop a series of approximations that we shall test on the case of
direct emissions and then apply to the interference terms. These
approximations, which are correct to parametric accuracy, have the virtue
to render the physical interpretation transparent and thus allow us to
pinpoint the subtle mechanisms at work in the interference effects. More
precise calculations will be presented in Appendix \ref{App} and they
will confirm the results in this and the next coming section to the
accuracy of interest.

\begin{figure}[tb]
\centerline{
\includegraphics[width=0.7\textwidth]{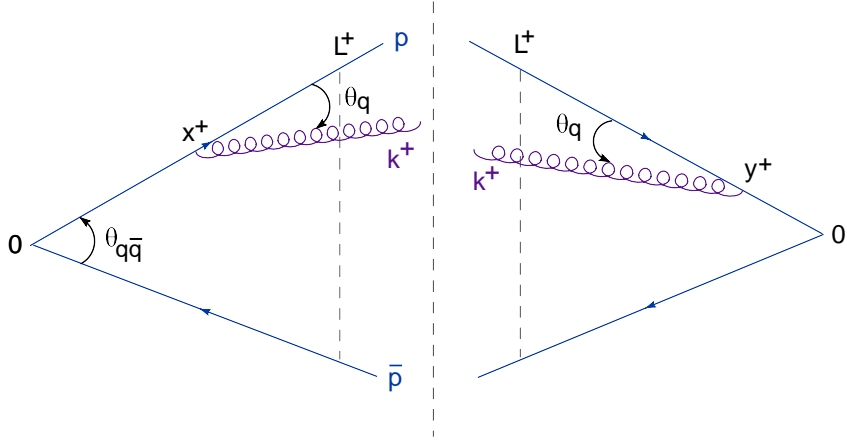}}
\caption{\label{fig:dirst}\sl The standard representation of the
Feynman graph for direct emission by the quark
(amplitude times the complex conjugate amplitude).}
\end{figure}

The probability for in--medium gluon radiation by the quark is obtained
by taking the modulus squared of the amplitude \eqref{Mqin}, summing over
the final color indices, averaging over the initial ones, and performing
the medium average over the background field. This yields
 \beq\label{Pqin}
 {\mcal P}_q^{\mbox{(in)}}({\bm k})&=&2g^2 C_F\,{\rm Re}
 \int_0^{L^+}\! \!
  \rmd x^+\! \int_0^{x^+} \! \!  \rmd y^+\, \rme^{ik^+u^-(x^+-y^+)}
  \nn  &{}&\quad\times\,
  \int \rmd {\bm z}_{1\perp}\!\int \rmd {\bm z}_{2\perp}
  \,\rme^{-i\bk\,\cdot\,({\bm z}_{1\perp}-{\bm z}_{2\perp})}
  \big(u^i+ {i\del^i_x}/{k^+}\big)
   \big(u^i- {i\del^i_y}/{k^+}\big) \\
   &{}&\quad\times\,
   \frac{1}{N_c^2-1}\,\big\langle {\rm Tr}\,
   {\mcal G}(L^+, {\bm z}_{1\perp}; x^+, \bxt; k^+)
   %\Big|_{\bxt =\bu x^+}
   \, \mcal{U}_q(x^+,y^+)\,
   % \mcal{U}_q^{\dagger}(y^+,0)\,
   {\mcal G}^\dagger(L^+, {\bm z}_{2\perp}; y^+,
   {\bm y}_\perp; k^+)
   %\Big|_{{\bm y}_\perp =\bu y^+}
   \big\rangle \,,\nonumber
    \eeq
where $\mcal{U}_q(x^+,y^+)$ is given by \eqn{Uq} with
$\bm{r}_\perp(z^+)=\bu z^+$ and it is understood that after the
performing the transverse derivatives $\del^i_x$ and $\del^i_y$ one sets
$\bxt\!=\bu x^+$ and ${\bm y}_\perp\!=\bu y^+$. In writing \eqn{Pqin} we
have restricted the time integrals to $0<y^+<x^+<L^+$ and multiplied the
result by a factor of 2. The Feynman graph representing this emission is
shown in Fig.~\ref{fig:dirst}.

\begin{figure}[tb]
\centerline{
%\includegraphics[width=0.62\textwidth]{Inter_Standard.eps}
%\hskip 4mm
\includegraphics[width=0.45\textwidth]{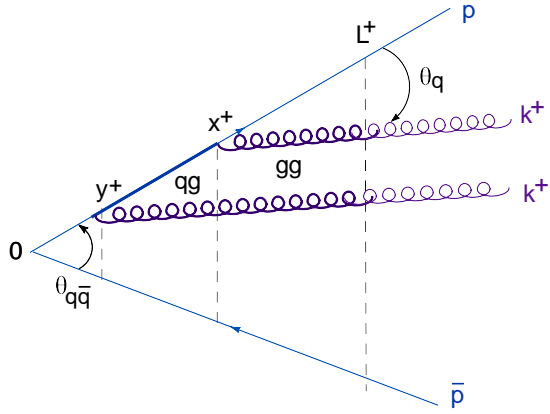}
}
\caption{\label{fig:dirwil}\sl A folded version of the Feynman
graph for direct emission where the amplitude and
the complex conjugate amplitude are represented on top of each
other, to more clearly exhibit the $qg$ and $gg$ dipoles.
The (quark and gluon) Wilson lines are indicated with thick lines.}
\end{figure}

Note that the quark Wilson lines prior to the first emission time $y^+$
have canceled each other between the direct and the complex conjugate
amplitude. The color trace in the last line of \eqn{Pqin} can be further
simplified by using the fact that the background field correlations are
local in time. To that aim one first uses the composition law
\eqref{compG} to break the last gluon propagator in \eqn{Pqin} into two
pieces --- from $y^+$ to $x^+$ and from $x^+$ to $L^+$. Then the medium
average factorizes as (below, the $k^+$ variable is kept implicit, to
simplify writing)
 \beq\label{fact1}
&{}& \int \rmd {\bm z}_{\perp} \frac{1}{N_c^2-1}\,\big\langle {\rm Tr}\,
   {\mcal G}(L^+, {\bm z}_{1\perp}; x^+, \bxt)
   \, \mcal{U}_q(x^+,y^+)\,
   {\mcal G}^\dagger(x^+, {\bm z}_{\perp}; y^+,
   {\bm y}_\perp)
   {\mcal G}^\dagger(L^+, {\bm z}_{2\perp}; x^+,
   {\bm z}_\perp)
   \big\rangle \,\nonumber\\
 &{}& \quad  = \int \rmd {\bm z}_{\perp} \frac{1}{N_c^2-1}\,
 \big\langle {\rm Tr}\,
   {\mcal G}(L^+, {\bm z}_{1\perp}; x^+, \bxt)
   {\mcal G}^\dagger(L^+, {\bm z}_{2\perp}; x^+,
   {\bm z}_\perp) \big\rangle\,\nonumber\\
 &{}& \quad \qquad \times\,
   \frac{1}{N_c^2-1}\,\big\langle {\rm Tr}\,
   \mcal{U}_q(x^+,y^+)\,
   {\mcal G}^\dagger(x^+, {\bm z}_{\perp}; y^+,
   {\bm y}_\perp)\big\rangle\,.
   \eeq
The two color traces in the r.h.s. of \eqn{fact1} are recognized as the
2--body propagators of two effective dipoles --- a quark--gluon ($qg$)
dipole extending from $y^+$ to $x^+$ and a gluon--gluon ($gg$) dipole
from $x^+$ to $L^+$ --- whose interactions in the medium are here
computed beyond the eikonal approximation (cf. the discussion after
\eqn{G0}). These dipoles can be easier visualized by folding the Feynman
graph in Fig.~\ref{fig:dirst} in such a way that the direct and complex
conjugate amplitudes overlap with each other, as shown in
Fig.~\ref{fig:dirwil}.

Using \eqref{GPI}, one obtains the following path--integral
representation for the propagator of the $qg$ dipole:
\beq\label{Kdef}
 {\mcal K}_{qg}(x^+,{\bm z}_{\perp}; y^+,{\bm y}_\perp; k^+)
 &\equiv&\frac{1}{N_c^2-1}\,\big\langle {\rm Tr}\,
   \mcal{U}_q(x^+,y^+)\,
   {\mcal G}^\dagger(x^+, {\bm z}_{\perp}; y^+,
   {\bm y}_\perp)\big\rangle\\
 &=& \int [D\bm{r}_\perp]\,\exp\Bigg\{\!-
 i\,\frac{k^+}{2}\!\int_{y^+}^{x^+}\!\rmd z^+
 \dot{\bm r}^2_\perp\Bigg\}\,
 S_{qg}\big(x^+, y^+; [\bm{r}_\perp-\bu z^+]\big)\,,
 \nonumber\eeq
which features a $qg$ pair with fluctuating size $\bm{r}_\perp(z^+)-\bu
z^+$ and path--dependent $S$--matrix %(cf. \eqn{Sdip})
 \beq\label{Sqg}
 S_{qg}\big(x^+, y^+; [\bm{r}_\perp-\bu z^+]\big)\,\simeq\,
 \exp\biggl\{-\frac{1}{4}\,\hat q \,\rho\,\int_{y^+}^{x^+}\rmd z^+
 \,\big(\bm{r}_\perp(z^+)-\bu z^+\big)^2\biggr\}\,.\eeq
We recall that the boundary conditions for the gluon paths are $
\bm{r}_\perp(y^+)={\bm y}_\perp$ and $\bm{r}_\perp(x^+)={\bm z}_{\perp}$
and that $\rho$ is a slowly varying function of the dipole size (cf.
\eqn{Sdip1}).

As for the $gg$ dipole in \eqn{fact1}, the corresponding mathematics
turns out to be simpler: on the average, the medium is homogeneous in the
transverse plane, as manifest on \eqn{2pA}. Then the medium averaging
also averages out the fluctuations in the dipole transverse size, with
the net effect that the respective $S$--matrix depends only upon the {\em
initial} dipole size at time $x^+$, that is $\bxt-{\bm z}_{\perp}$.
Specifically, the following identity holds (see e.g.
\cite{CasalderreySolana:2007zz,MehtarTani:2006xq} for details) :
  \beq \label{Sggdef}
  &{}& \int \rmd {\bm z}_{1\perp}\!\int \rmd {\bm z}_{2\perp}
  \,\rme^{-i\bk\,\cdot\,({\bm z}_{1\perp}-{\bm z}_{2\perp})}\,
   \frac{1}{N_c^2-1}\,
 \big\langle {\rm Tr}\,
   {\mcal G}(L^+, {\bm z}_{1\perp}; x^+, \bxt)
   {\mcal G}^\dagger(L^+, {\bm z}_{2\perp}; x^+,
   {\bm z}_\perp) \big\rangle\,\nn
 &{}& \qquad =\,\rme^{-i\bk\,\cdot\,({\bm x}_{\perp}-{\bm z}_{\perp})}
 \,S_{gg}(L^+, x^+; \bxt-{\bm z}_{\perp})\,,\eeq
with (compare to \eqn{Sdip})
 \beq \label{Sgg}
 S_{gg}(L^+, x^+; \bxt-{\bm z}_{\perp}) \,\simeq\,
 \exp\biggl\{-\frac{1}{4}\,\hat q\,\rho\, (L^+-x^+)\,\big(\bxt-{\bm z}_{\perp}
 \big)^2 \biggr\}\,.\eeq
Note that in writing Eqs.~\eqref{Sqg} and \eqref{Sgg} above, we have
tacitly assumed that the respective dipole sizes are much smaller than
$\mu_D^{-1}$, so that the approximations leading to
Eqs.~\eqref{Sdip1}--\eqref{Sdip} indeed apply. This will be checked
later, when we shall see that the typical dipole sizes are of order
$1/k_f$ for \eqref{Sqg} and respectively of order $1/Q_s$ for
\eqref{Sgg}.

Putting together the previous results, we deduce the following expression
for the probability for direct emission from the quark
\beq\label{Pqinfin}
 {\mcal P}_q^{\mbox{(in)}}({\bm k})&=&2g^2 C_F \,{\rm Re}
 \int_0^{L^+}\! \!
  \rmd x^+\! \int_0^{x^+} \! \!  \rmd y^+\, \rme^{ik^+u^-(x^+-y^+)}
  \big(u^i+ {i\del^i_x}/{k^+}\big)
   \big(u^i- {i\del^i_y}/{k^+}\big) \nn
   &{}&\quad\times
   \int \rmd {\bm z}_{\perp} \,
   \rme^{-i\bk\,\cdot\,({\bm x}_{\perp}-{\bm z}_{\perp})}
 \, {\mcal K}_{qg}(x^+,{\bm z}_{\perp}; y^+,{\bm y}_\perp; k^+)\,
 S_{gg}(L^+, x^+; \bxt-{\bm z}_{\perp})\,,\eeq
where it is understood that $\bxt\!\to \bu x^+$ and ${\bm y}_\perp\!\to
\bu y^+$ after taking the derivatives. Within the limits of our
calculation, this expression is exact. It is also rather formal, in the
sense of involving a path integral and holding for an arbitrary
kinematics of the emitted gluon. In the `harmonic approximation', which
consists in treating the slowly varying logarithm $\rho$ in
Eqs.~\eqref{Sqg} and \eqref{Sgg} as a fixed quantity, the integrations
become Gaussian and can be performed exactly (see the Appendix). To keep
the discussion as intuitive as possible, in what follows we shall perform
a series of approximations which are valid in the kinematics of interest.

But before we proceed with more formal steps, let us emphasize a point of
physics\footnote{We would like to thank Al Mueller for an illuminating
discussion of this point.}: the gluon formation time $\tau_f$ for
medium--induced radiation is controlled by the intermediate, quark--gluon
dipole, stage of the dynamics in \eqn{Pqinfin} and hence it is of the
order of the typical duration $x^+-y^+$ of that stage. Indeed, the
$S$--matrix \eqref{Sqg} of this {\em effective} dipole, built with the
quark in the direct amplitude and the gluon in the complex conjugate one
(or vice--versa), is a measure of the color coherence between the emitted
gluon and the parent quark. So long as this dipole is relatively small
(meaning for sufficiently small time separations $x^+-y^+$), one has
$S_{qg}\simeq 1$ and then one cannot distinguish the gluon from the
quark: in any process involving color exchanges, the emerging
quark--gluon pair acts in the same way as the original, bare, quark would
do. But with increasing $x^+-y^+$, the dipole size increases (via gluon
diffusion) and then $S_{qg}$ starts to decrease from one, because of the
medium rescattering. One can consider the gluon as being formed when the
$qg$ dipole suffers a first inelastic collision in the medium, {\em i.e.}
when the exponent in $S_{qg}$ becomes of $\order{1}$. The respective
value of $x^+-y^+$ sets the formation time. For even larger values of
$x^+$, one has $S_{qg}\ll 1$ and the emission probability is strongly
suppressed.

The starting point of our approximation is an expression for the
propagator \eqref{Kdef} of the $qg$ dipole valid in the harmonic
approximation. With $\rho\approx$~const. and absorbed into the
normalization of $\hat q$ for convenience\footnote{This is a standard
convention in the literature; the factors of $\rho$ can be recovered, if
needed, by replacing everywhere $\hat q\to\hat q\rho$.}, the path
integral \eqref{Kdef} describes a harmonic oscillator with {\em imaginary
squared frequency}
 \beq\label{Omega}
 \Omega^2\,=\,i\,\frac{\hat q}{2k^+}\quad\Longrightarrow\quad
 \Omega\,=\,\frac{1+i}{\sqrt{2}}\,\sqrt{\frac{\hat q}{2k^+}}\,,
 \eeq
and hence it can be exactly computed (see e.g.
\cite{Baier:1998yf,CasalderreySolana:2007zz} for details and also the
Appendix below). To be specific, let us ignore the transverse derivatives
in \eqn{Pqinfin} for the time being (we shall return to them latter) and
fix $\bxt\!= \bu x^+$ and ${\bm y}_\perp\!= \bu y^+$. Then one obtains
 \beq\label{Kharmu}
 {\mcal K}_{qg}(x^+,{\bm b}_{\perp}\!+\bu x^+; y^+,\bu y^+; k^+)=
 \exp\left\{
 -i{k^+}
 \Big[(x^+-y^+)\frac{\bu^2}{2}+\bu\cdot{\bm b}_{\perp}\Big]\right\}\,\nn
\times\ {\mcal K}_{qg}(x^+,{\bm b}_{\perp}; y^+,{\bm 0}_\perp; k^+)
 \eeq
where we set ${\bm b}_{\perp}\equiv {\bm z}_{\perp} - \bu x^+$ and
 \beq\label{Kharm}
 {\mcal K}_{qg}(x^+,{\bm b}_{\perp}; y^+,{\bm 0}_\perp; k^+)=
 \,\frac{k^+\Omega}
 {2\pi i\sinh \Omega(x^+\!-y^+)} \exp\left\{-
 \frac{k^+\Omega}
 {2i}\,\coth\Omega(x^+\!-y^+)\,
 {\bm b}_{\perp}^2\right\}\,.\eeq
The quantity ${\bm b}_{\perp}$ is the transverse size of the $qg$ dipole
at time $x^+$ and thus also the size of the ensuing $gg$ dipole at any
time $z^+>x^+$. We shall denote
  \beq\label{formt}
 \tau_f\,\equiv\,\frac{1}{|\Omega|}\,=\,\sqrt{\frac{2k^+}{\hat q}}
 \,,\eeq
anticipating that this quantity plays the role of the formation time.

Two limits of \eqn{Kharm} will be useful in what follows:

\texttt{(i)}{\em Small times $|\Omega|(x^+\!-y^+)\ll 1$ or $x^+\!-y^+\ll
\tau_f$ :} then, by expanding the r.h.s. of \eqn{Kharm} to quadratic
order in $|\Omega|(x^+\!-y^+)$ one finds
  \beq\label{Ksmall}
 {\mcal K}_{qg}(x^+,{\bm b}_{\perp}; y^+,{\bm 0}_\perp; k^+)\simeq
 \frac{k^+}
 {2\pi\,i(x^+-y^+)}\,\exp\left\{-
 i\frac{k^+ {b}_\perp^2}{2(x^+-y^+)}
 -\frac{1}{12}\,\hat q\,(x^+-y^+)\,
 b_\perp^2\right\}.\eeq
This is recognized as the saddle point approximation to \eqref{Kdef} with
the saddle point determined by the kinetic piece of the action alone;
that is, ${\mcal K}_{qg}\approx {\mcal G}_0 \,S_{qg}$ where ${\mcal G}_0$
is the free propagator \eqref{G0} and $S_{qg}$ is the $S$--matrix
\eqref{Sqg} evaluated along the classical path, which reads  :
 \beq\label{qgclas}
 % \bm{r}_{\rm class}(z^+)\,=\, \bu y^+ +\,
 % \frac{z^+-y^+}{x^+-y^+}\,\big({\bm z}_{\perp} - \bu y^+\big)
 % \ \Longrightarrow\
 \bm{r}_{\rm class}(z^+)-\bu z^+=\,
 \frac{z^+-y^+}{x^+-y^+}\,
 {\bm b}_{\perp}
 \,.\eeq

\texttt{(ii)}{\em Large times $|\Omega|(x^+\!-y^+)\gg 1$ or $x^+\!-y^+\gg
\tau_f$ :} then, one finds
  \beq\label{Klarge}
 {\mcal K}_{qg}(x^+,{\bm b}_{\perp}; y^+,{\bm 0}_\perp; k^+)\,\propto\,
\frac{k^+}{\tau_f}\ \rme^{-(x^+\!-y^+)/\tau_f}\,\exp\left\{-
 \,\frac{1+i}{4}\,\sqrt{{\hat q}{k^+}}\,
 b_\perp^2\right\}\,.\eeq

\eqn{Ksmall} implies that, at early times $x^+\!-y^+\lesssim \tau_f$, the
size of the $qg$ dipole increases through diffusion, $b_\perp^2\propto
(x^+\!-y^+)/k^+$ (as shown by the first term in the exponent) and this
increase enhances the dipole rescattering off the medium (as described by
the second term in the exponent of \eqn{Ksmall}, coming from $S_{qg}$).
When $x^+\!-y^+\simeq \tau_f$, this second term becomes of order one,
showing that $\tau_f$ is the formation time, as anticipated. For larger
times $x^+\!-y^+\gg \tau_f$, the dipole propagator is exponentially
suppressed, cf. \eqn{Klarge}, meaning that the color coherence of the
$qg$ pair has been destroyed by the medium. The maximal size of the $qg$
dipole, as attained for $x^+\!-y^+\simeq \tau_f$, is\footnote{As usual,
when writing parametric estimates, we ignore numerical factors and
identify $k^+$ with $\omega$.}
 \beq\label{qgmax}
  b_{f}^2 \ \simeq\,\frac{\tau_f}{k^+}
  \,\sim\,\frac{1}{\sqrt{{\hat q}{\omega}}}\,.\eeq
  %\,\sim\,\frac{1}{Q_s^2}\,\sqrt{\frac{\omega_c}{\omega}}
  %\,\gg\,\frac{1}{Q_s^2}\,.\eeq
Via the uncertainty principle, this yields the typical transverse
momentum of the gluon at the formation time as $k_{f}^2 \simeq 1/b_{f}^2
\simeq \sqrt{\omega\hat q}$, in agreement with \eqn{tauf}. This obeys the
scaling law $k_{f}^2\simeq\hat q\tau_f$ showing that this transverse
momentum has been acquired via medium rescattering during a time
$\tau_f$. Since $\tau_f\ll L^+$, this $k_{\perp}$ is much smaller than
the final momentum of the gluon, which can be as large as
$k_\perp\!\sim\! Q_s$, as we shall shortly see.

To compute the final gluon spectrum, one has to also take into account
the medium rescattering {\em after} the time of formation, as encoded in
the $S$--matrix \eqref{Sgg} of the $gg$ dipole. Specifically, the
function $S_{gg}(L^+, x^+; \bxt-{\bm z}_{\perp})$ controls the range of
the integration over ${\bm z}_{\perp}= {\bm b}_{\perp}+\bu x^+$, which in
turn fixes the final transverse momentum ${\bm k}_\perp$ via the Fourier
transform in \eqn{Pqinfin}. As we shall shortly check, the typical values
for ${b}_{\perp}$ allowed by the $gg$ dipole are much smaller than $b_f$.
Hence, in evaluating this Fourier transform, one can replace the
propagator ${\mcal K}_{qg}$ of the $qg$ dipole by the corresponding free
propagator ${\mcal G}_0$ (which carries the whole dependence upon ${\bm
b}_{\perp}$ in the limit where ${b}_{\perp}\ll b_f$). Then the integral
over ${\bm b}_{\perp}$ reduces to
\beq\label{intb2}
\rme^{-i\frac{k^+ \bu^2}{2}(x^+-y^+)} \int \rmd {\bm b}_{\perp} \,
   \rme^{i{\bm b}_{\perp}\,\cdot\,(\bk - k^+\bu)}\,
   \exp\left\{
 \frac{-ik^+{b}_{\perp}^2}{2(x^+-y^+)}\right\}\,
 \exp\biggl\{-\frac{\hat q}{4}\, (L^+-x^+)b_{\perp}^2\biggr\}\,,
 \eeq
where it was important to also include the phase factor in the r.h.s. of
\eqn{Kharmu} (which is a part of the free propagator $ {\mcal
G}_{0}(x^+,{\bm b}_{\perp}\!+\bu x^+; y^+,\bu y^+; k^+)$). The overall
phase in front of the above integral is such that it exactly cancels the
phase $\rme^{ik^+u^-(x^+-y^+)}$ in \eqn{Pqinfin}. This is worth
emphasizing in view of the discussion of the interference terms in
Sect.~\ref{Int}, where the corresponding phase cancelation does {\em not}
hold --- which in turn has important consequences.

For medium--induced radiation, the time variables $x^+$ and $y^+$ can lie
anywhere within the medium, $0< x^+,\,y^+< L^+$ (except very close to the
boundaries\footnote{Very small values $0<
x^+,\,y^+\!<\omega/Q_s^2\ll\tau_f$ corresponds to vacuum--like emissions
with relatively large momenta $k_\perp\gtrsim Q_s$. Values close to
$L^+$, such that $L^+-\tau_f< x^+,\,y^+\!< L^+$, yield boundary terms
which cancel when summing up together the (in, in), (in, out), and (out,
out) contributions.}), so long as $x^+-y^+\!\sim\tau_f\ll L^+$.
Accordingly ${\hat q}(L^+\!-x^+)\!\sim {\hat q}\,L^+=Q_s^2$ is much
larger than $k^+/\tau_f\sim k_f^2$ (recall \eqn{tauf}), so the integral
in \eqn{intb2} is controlled by the last factor inside the integrand
%(the $S$--matrix for the $gg$ dipole)
and yields
 \beq\label{intb}
 %\frac{k^+}{2(x^+-y^+)}
 \int \rmd {\bm b}_{\perp} \,
   \rme^{i{\bm b}_{\perp}\,\cdot\,(\bk - k^+\bu)}\,
 \exp\biggl\{-\frac{1}{4}\,Q_s^2
 b_\perp^2\biggr\}\,\sim\,%\frac{k^+}{Q_s^2(x^+-y^+)}\,
 \frac{1}{Q_s^2}\,
 \exp\biggl\{-\frac{(\bk - k^+\bu)^2}{Q_s^2}\biggr\}.\eeq
Note that $k^+\bu$ is the transverse momentum inherited by the gluon from
its parent quark. Accordingly, $\bk - k^+\bu$ is the additional
transverse momentum acquired by the gluon from the medium and is the same
as the component of the gluon momentum which is transverse to the quark;
indeed, using \eqn{tq} one can write
 \beq\label{kperpq}
 (\bk - k^+\bu)^2\,=\,2k^+(k\cdot u)\,\simeq\,
  (\omega\tq)^2\,.\eeq
Hence, \eqn{intb} shows that the momentum gained by the gluon via medium
rescattering can be as large as $Q_s$, as anticipated. Since $Q_s\gg
k_f$, it is clear that most of this momentum gets accumulated {\em after}
the gluon formation (as also shown by the fact that the integral
\eqref{intb} is controlled by the $S$--matrix of the final $gg$ dipole).

The last ingredient that we need in order to evaluate \eqn{Pqinfin} is
the action of the transverse derivatives like $(u^i+ i\del^i_x/k^+)$.
These will be shortly computed, but their effect can be anticipated on
physical grounds:  from the construction of the amplitude in \eqn{Mqin},
we recall that the derivative $\del^i_x$ acts on the gluon propagator at
the emission point. Hence, the operator $(u^i+ i\del^i_x/k^+)$ mesures
the difference between the transverse orientations of the source and of
the emitted gluon, at the time of formation. Then, clearly, we expect its
magnitude to be of order $\tf\equiv k_f/\omega$ (the formation angle
introduced in \eqn{formation}). To explicitly check that, one needs to
compute
 \beq\label{TDdirect}\big(u^i+
{i\del^i_x}/{k^+}\big)
   \big(u^i- {i\del^i_y}/{k^+}\big)
 \, {\mcal K}_{qg}(x^+,{\bm b}_{\perp}+\bxt; y^+,{\bm y}_\perp; k^+)\,
 \eeq
with the derivatives evaluated at $\bxt\!= \bu x^+$ and ${\bm y}_\perp\!=
\bu y^+$. The general expression for ${\mcal K}_{qg}$ which is required
for that purpose is given in the Appendix. But for a parametric estimate,
one can replace ${\mcal K}_{qg}$ by the free propagator ${\mcal G}_{0}$.
We thus deduce
  \beq
   \bigg(u^i- \frac{i\del^i_y}{k^+}\bigg)\bigg(u^i+\frac
{i\del^i_x}{k^+}\bigg)\, {\mcal G}_{0}&=&\bigg(u^i-
\frac{i\del^i_y}{k^+}\bigg)\bigg[\bigg(u^i-k^+\,\frac{b^i
 +x^i-y^i}{x^+-y^+}\bigg){\mcal G}_{0}\bigg]\nn
 &\,=\,&\bigg[\frac{b_\perp^2}
 {(x^+-y^+)^2}\,+\,\frac{2i}{k^+(x^+-y^+)}\bigg]{\mcal G}_{0}
 \eeq
where the last equality is obtained after setting $x^i-y^i=u^i(x^+-y^+)$.
Using $x^+-y^+\sim \tau_f$ and $b_\perp\sim 1/Q_s$, it is easy to check
that the second term in the square brackets is the dominant one and is of
order $1/(k^+\tau_f)\sim\theta_f^2$, as anticipated.

We are finally in a position to estimate the spectrum \eqref{Pqinfin} for
direct emissions. To that aim, one has to multiply the Gaussian in
\eqn{intb} by a factor $L^+\tau_f$ coming from the integrals over the
time variables $y^+$ and $x^+$ (this factor is the longitudinal
phase--space for medium--induced gluon radiation), by the factor
$\theta_f^2$ which estimates the effects of the transverse derivatives,
by a factor ${k^+}/{(x^+-y^+)}\sim \omega/\tau_f$ coming from the
normalization of $\mcal{G}_0$ in \eqn{Ksmall} and, finally, by the
overall factor $g^2 C_F$ manifest on \eqn{Pqinfin}. Putting all that
together, one finds
 \beq\label{BDMPS}
 {\mcal P}_q^{\mbox{(in)}}(\omega,{\bm k}_\perp)\,\propto\,\alpha_s
 C_F\,\theta_f^2\,L^+\,\frac{\omega}{Q_s^2}\,\exp\biggl\{
 -\frac{(\bk - k^+\bu)^2}{Q_s^2}\biggr\}.\eeq
\eqn{BDMPS} is indeed the expected parametric estimate for the BDMPS-Z
spectrum of the medium--induced radiation by a quark. A perhaps more
familiar form of this spectrum is obtained by using \eqn{formation} for
$\theta_f$ to deduce (for $\bu=0$)
 \beq\label{BDMPSZ}
 \omega \,\frac{\rmd N}{\rmd \omega\rmd k_\perp^2}
 \,\propto\,\frac{\alpha_s
 C_F}{\sqrt{\omega\hat q}}\,\exp\biggl\{
 -\frac{k_\perp^2}{Q_s^2}\biggr\}\,.
 \eeq
It is here understood that $k_\perp\gtrsim k_{f}\simeq (\omega\hat
q)^{1/4}$, since the gluon acquires a transverse momentum of order $
k_{f}$ already by the time of formation. The spectrum \eqref{BDMPSZ} is
roughly flat in the range $k_{f}< k_{\perp} <Q_s$ and it is exponentially
suppressed at larger values $k_{\perp}\gg Q_s$. After integration over
$k_\perp$ and recalling that $Q^2_s=\hat q L^+ \gg k_{f}^2$ and
$\omega_c=\hat q L^2/2$, this yields
 \beq\label{DeltaE}
  \omega \,\frac{\rmd N}{\rmd \omega}
 \,\propto\,\alpha_s
 C_F \sqrt{\frac{\omega_c}{\omega}}\quad\Longrightarrow\quad
 \Delta E\equiv \int_0^{\omega_c}\omega \,\frac{\rmd N}{\rmd \omega}\,
 \propto\,\alpha_s
 C_F\,\omega_c\,,
 \eeq
where the integration has been restricted to the phase--space for
medium--induced radiation, {\em i.e.} $ \hat q^{1/3}<{\omega}\le
{\omega}_c$, cf. \eqn{range} (but the lower limit is irrelevant for
computing the total energy loss, which is dominated by the upper limit
${\omega}_c$).

It is finally interesting to compare the spectrum \eqref{BDMPSZ} for
medium--induced radiation to the bremsstrahlung spectrum in \eqn{brems},
for the same kinematics. By inspection of these equations, it is apparent
that the medium--induced spectrum is formally the same (for any
$k_{\perp}$ within the range $k_{f}< k_{\perp} <Q_s$) as the vacuum
spectrum evaluated at $k_{\perp}=k_f$. Hence, clearly,
 \beq\label{Ratio}
 \frac{{\mcal P}_q^{\mbox{(in)}}}{{\mcal P}_q^{\mbox{(vac)}}}
 \,\sim\, \frac{k_\perp^2}{\sqrt{\omega\hat q}}\,\sim\,
 \frac{k_\perp^2}{k_{f}^2}\,,\eeq
which shows that the medium--induced radiation dominates over
bremsstrahlung for all the relevant momenta. This ratio is largest for
$k_\perp\simeq Q_s$, when it becomes
 \beq\label{Rinout}
 %\quad\Longrightarrow\quad
 \frac{{\mcal P}_q^{\mbox{(in)}}}{{\mcal P}_q^{\mbox{(vac)}}}
 \, %\frac{Q_s^2}{\sqrt{\omega\hat q}}\,
 \sim\,\frac{L^+}{\tau_f}\,\sim\,
 \sqrt{\frac{\omega_c}{\omega}}\,\gg\,1
 \quad{\rm for}\quad k_\perp\simeq Q_s \,.\eeq
Physically, this is so because a gluon which is formed via medium
rescattering can be emitted at any place inside the medium
($x^+,\,y^+\!\lesssim L^+$), in contrast to the vacuum--like emissions,
which are restricted to relatively short distances/times
$x^+,\,y^+\lesssim\tau_q\ll L^+$. Accordingly, the longitudinal
phase--space $L^+\tau_f$ for medium--induced radiation is parametrically
larger than the corresponding phase--space $\sim\tau_q^2$ for
bremsstrahlung.

\section{Medium--induced gluon radiation: interference terms}
\label{Int}

\begin{figure}[tb]
\centerline{
\includegraphics[width=0.7\textwidth]{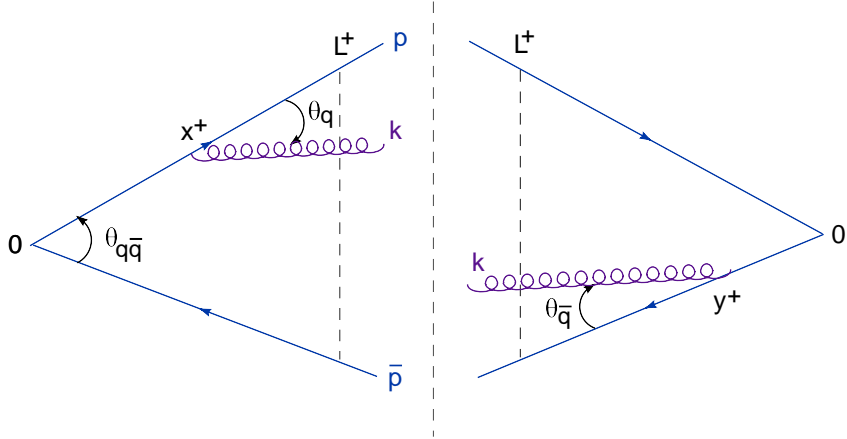}}
\caption{\label{fig:intst}\sl A Feynman graph for
interference (amplitude times the complex conjugate amplitude).}
\end{figure}

\begin{figure}[tb]
\centerline{
\includegraphics[width=0.45\textwidth]{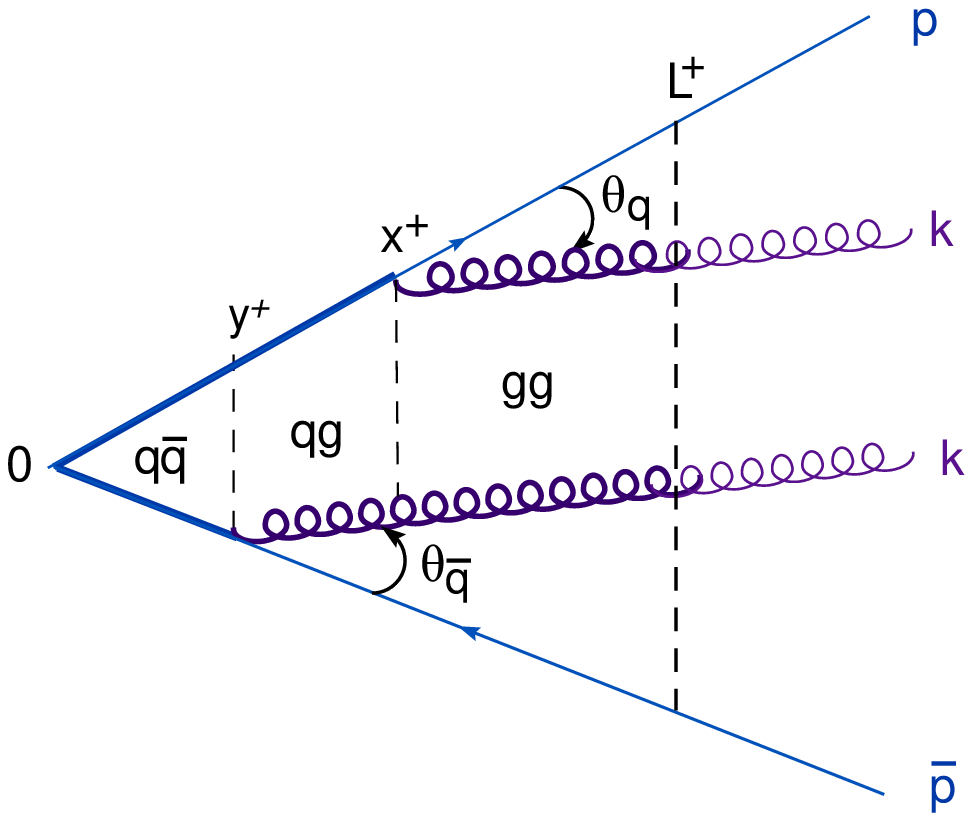}
}
\caption{\label{fig:intwil}\sl A folded version of the Feynman
graph for interference where the amplitude and
the complex conjugate amplitude are represented on top of each
other, to more clearly exhibit the $q\bq$, $qg$ and $gg$ dipoles.
The (quark and gluon) Wilson lines are indicated with thick lines.}
\end{figure}

We now turn to the main problem of interest for us in this paper, namely
the contribution of the quark--antiquark interference to the
medium--induced gluon radiation (see Fig.~\ref{fig:intst}). Once again,
we shall focus on the (in, in) piece, where the gluons are emitted inside
the medium in both the direct and the complex conjugate amplitude. The
respective contribution to the gluon spectrum is obtained by multiplying
the quark amplitude \eqref{Mqin} by the complex conjugate of the
corresponding antiquark amplitude, performing the average over the medium
and the sum (average) over the final (initial) color indices. This yields
\beq\label{Iin}
 {\mcal I}^{\mbox{(in)}}({\bm k})
  &=&-2g^2 C_F \ {\rm Re}
 \int_0^{L^+}\! \!
  \rmd x^+\! \int_0^{x^+} \! \!  \rmd y^+\, \rme^{ik^+(u^-x^+-
  \bar u^- y^+)}
  \nn  &{}&\quad\times\,
  \int \rmd {\bm z}_{1\perp}\!\int \rmd {\bm z}_{2\perp}
  \,\rme^{-i\bk\,\cdot\,({\bm z}_{1\perp}-{\bm z}_{2\perp})}
  \big(u^i+ {i\del^i_x}/{k^+}\big)
   \big(\bar u^i- {i\del^i_y}/{k^+}\big) \nn
   &{}&\quad\times\,
   \frac{1}{N_c^2-1}\,\big\langle {\rm Tr}\,
   {\mcal G}(L^+, {\bm z}_{1\perp}; x^+, \bxt; k^+)
   %\Big|_{\bxt =\bu x^+}
   \, \mcal{U}_q(x^+,0)\,
   \mcal{U}_\bq^{\dagger}(y^+,0)\,
   {\mcal G}^\dagger(L^+, {\bm z}_{2\perp}; y^+,
   {\bm y}_\perp; k^+)
   %\Big|_{{\bm y}_\perp =\bu y^+}
   \big\rangle \,,\nonumber\\[0.1cm]
  &{}&\quad + \ (q\,\to\,\bq),
    \eeq
where the Wilson lines $\mcal{U}_q(x^+,0)$ and
$\mcal{U}_\bq^{\dagger}(y^+,0)$ refer to the quark and the antiquark,
respectively, and it is understood that after performing the transverse
derivatives $\del^i_x$ and $\del^i_y$ one has to identify $\bxt$ and
${\bm y}_\perp$ with the emission points $\bu x^+$ and $\bbu y^+$,
respectively. The explicit integrals in \eqn{Iin} are written for the
situation where the gluon is emitted at $x^+$ by the quark in the direct
amplitude and absorbed by the antiquark at $y^+$ in the complex conjugate
amplitude, with $y^+ < x^+$. The other possible configurations are
obtained by exchanging the quark and the antiquark, as indicated in the
last line of \eqn{Iin}. After `folding' the Feynman graph as shown in
Fig.~\ref{fig:intwil}, in such a way to superpose direct and conjugate
amplitudes, one can view $y^+$ as the `first emission time', for an
emission off the antiquark, and $x^+$ as the `second emission time', for
an emission by the quark. Although somewhat formal, this perspective
allows one to easily visualise the effective `color dipoles' encoded in
\eqn{Iin}, that we now discuss.

The subsequent manipulations are rather similar to those in
Sect.~\ref{direct}. Once again, one splits the quark Wilson line as
$\mcal{U}_q(x^+,0)=\mcal{U}_q(x^+,y^+)\mcal{U}_q(y^+,0)$ and one breaks
the last gluon propagator into two pieces --- from $y^+$ to $x^+$ and
from $x^+$ to $L^+$ ---, by introducing an intermediate integration point
${\bm z}_{\perp}$. Then one uses the locality of the medium correlations
in time to factorize the color trace into effective dipole contributions
(cf. \eqn{fact1}). This procedure now generates three dipole
$S$--matrices: a quark--antiquark ($q\bq$) dipole which extends in time
from 0 up to $y^+$, a quark--gluon ($qg$) dipole from $y^+$ to $x^+$, and
a gluon--gluon ($gg$) dipole from $x^+$ to $L^+$. The integrations over
${\bm z}_{1\perp}$ and ${\bm z}_{2\perp}$ are again performed as in
\eqn{Sggdef} and the outcome can be written as (compare to \eqn{Pqinfin})
 \beq\label{Iinfin}
 {\mcal I}^{\mbox{(in)}}({\bm k})&=&-2g^2 C_F\ {\rm Re}
 \int_0^{L^+}\! \!
  \rmd x^+\! \int_0^{x^+} \! \!  \rmd y^+\, \rme^{ik^+(u^-x^+-
  \bar u^- y^+)}
  \big(u^i+ {i\del^i_x}/{k^+}\big)
   \big(\bar u^i- {i\del^i_y}/{k^+}\big) \nn
   &{}&\ \times\,S_{q\bq}(y^+,0)
   \int \rmd {\bm z}_{\perp} \,
   \rme^{-i\bk\,\cdot\,({\bm x}_{\perp}-{\bm z}_{\perp})}
 \, {\mcal K}_{qg}(x^+,{\bm z}_{\perp}; y^+,{\bm y}_\perp; k^+)\,
 S_{gg}(L^+, x^+; \bxt-{\bm z}_{\perp})\nonumber\\[0.1cm]
  &{}&\ + \ (q\,\to\,\bq)\,,\eeq
where it is understood that $\bxt\!\to \bu x^+$ and ${\bm y}_\perp\!\to
\bbu y^+$ after taking the derivatives. The $q\bq$ dipole is evaluated
similarly to \eqn{Sqq1} :
\beq\label{Sqq2}
 S_{q\bq}(y^+,0)\,\simeq\,
 \exp\biggl\{-\frac{1}{12}\,\hat q\,\rho \,(\bu-\bbu)^2 (y^+)^3
 \biggr\}\,\simeq\,
 \exp\biggl\{-\frac{1}{24}\,\hat q\,\tqq^2\,(y^+)^3\,\rho
 \biggr\}\,.\eeq
(We have also used $(\bu -\bbu)^2 \simeq\tqq^2/2$ for small angles.) The
$qg$ dipole is now built with the {\em quark} line in the direct
amplitude and the gluon emitted by the {\em antiquark} in the complex
conjugate amplitude. The corresponding propagator ${\mcal K}_{qg}$ is
defined as in Eqs.~\eqref{Kdef}--\eqref{Sqg} but with different boundary
conditions for the path integral \eqref{Kdef}, namely
$\bm{r}_\perp(y^+)=\bbu y^+$ and $\bm{r}_\perp(x^+)={\bm z}_{\perp}$.
Finally, the $gg$ dipole is given by \eqn{Sgg}, as before.

There are several important differences between the interference term
\eqref{Iinfin} and the corresponding expression \eqref{Pqinfin} for the
direct emission. Two of them are quite obvious:

\texttt{(a)} The presence of the initial $q\bq$ dipole, which measures
the {\em color coherence} between the quark and the antiquark at the time
$y^+$ of the first emission, which is restricted to
 \beq\label{ymax}
 y^+\,\lesssim\,\tau_{coh}\,\simeq \left(\frac{24}{
 \hat q\rho\,\tqq^2}\right)^{1/3}\sim\,L^+
 \left(\frac{\theta_c}{
 \tqq}\right)^{2/3}.\eeq
For larger values $ y^+\gtrsim \tau_{coh}$, one has $S_{q\bq}(y^+,0)\ll
1$, that is, the color coherence is washed out. The parametric estimate
in the r.h.s. shows that $\tau_{coh}\ll L^+$ so long as
$\tqq\gg\theta_c$. (As before, in writing parametric estimates we neglect
numerical factors and treat $\rho$ as a constant of $\order{1}$.)

\texttt{(b)} The fact that the $qg$ dipole starts at $y^+$ with a
non--zero transverse size $r_0$ equal to the separation between the quark
and the antiquark at that time (the maximal size of the $q\bq$ dipole):
$r_0=|\bu-\bbu|y^+\sim\tqq\, y^+$.

A third difference between direct and interference terms, which is
perhaps less obvious at this stage but will play a major role for the
final results, is the following:

\texttt{(c)} The vacuum--like phase $\rme^{ik^+(u^-x^+- \bar u^- y^+)}$
in \eqn{Iinfin} is not compensated in the calculation of medium--induced
radiation, in contrast to what happened for the direct emissions (recall
the discussion after \eqn{intb2}). Rather, there is a left--over phase
which controls the {\em quantum coherence} between the two emitters (see
\eqn{Phi} below).

By itself, the constraint \eqref{ymax} represents a strong limitation on
the longitudinal phase--space for interference and shows that the
interference terms are suppressed with respect to the direct emissions.
However, it turns out that the limitation introduced by the quantum
coherence, cf. point \texttt{(c)} above, can be even stronger, depending
upon the value of $\tqq$. To understand the interplay between the
different types of coherence, we shall now perform a more detailed
analysis of \eqn{Iinfin}.

The first step consists in clarifying the formation time. From
Sect.~\ref{direct}, we recall that this is controlled by the propagator
${\mcal K}_{qg}$ of the quark--gluon dipole. In the present case, this
propagator measures the (quantum and color) coherence between the gluon
emitted by one of the emitters and the {\em other} emitter. The `exact'
expression of ${\mcal K}_{qg}$ valid in the harmonic approximation will
be given in the Appendix. Here we shall merely use a combination of
small--time and large--time approximations, like in
Eqs.~\eqref{Ksmall}--\eqref{Klarge}. The time scale separating between
the two regimes is, once again, $\tau_f= 1/|\Omega|$, cf. \eqn{formt}.

For small $x^+\!-y^+\ll \tau_f$, ${\mcal K}_{qg}$ can be approximated by
the saddle point approximation to the path integral in \eqn{Kdef}, with
the saddle point determined by the kinetic term alone. The corresponding
classical path is readily determined as
  \beq\label{intqgclas}
 % \bm{r}_{\rm class}(z^+)\,=\, \bu y^+ +\,
 % \frac{z^+-y^+}{x^+-y^+}\,\big({\bm z}_{\perp} - \bu y^+\big)
 % \ \Longrightarrow\
 \bm{r}_{\rm class}(z^+)-\bu z^+=\br_0+
 \frac{z^+-y^+}{x^+-y^+}\big(
 {\bm z}_{\perp} - \bu x^+ +\br_0\big)\,,\qquad \br_0\equiv(\bbu-\bu)y^+
 \,.\eeq
As in Sect.~\ref{direct}, it is convenient to change variables from ${\bm
z}_{\perp}$ (the gluon transverse position at time $x^+$) to ${\bm
b}_{\perp}\equiv {\bm z}_{\perp} - \bu x^+$ (the final size of the $qg$
dipole and hence also the size of the $gg$ dipole at any time $z^+\ge
x^+$). Then the saddle point \eqref{intqgclas} yields ${\mcal
K}_{qg}\approx {\mcal G}_0 S_{qg}$, with
 \beq\label{G0int}
 {\mcal G}_0 (x^+, {\bm z}_{\perp};
 y^+,\bbu y^+; k^+)\,=\,\frac{k^+}
 {2\pi\,i(x^+-y^+)}\,\exp\left\{-
 i\,\frac{k^+({\bm b}_{\perp}+\bu x^+ -\bbu y^+)^2}{2(x^+-y^+)}\right\},
 \eeq
 \beq\label{intSqg}
S_{qg}\big(x^+, y^+; {\bm b}_{\perp}\big)&\,\approx\,&
\exp\biggl\{-\frac{1}{12}\,\hat q\,(x^+-y^+)(b_\perp^2+r_0^2+ {\bm
b}_{\perp}\cdot\br_0)
 \biggr\}.\eeq
For larger time difference, $x^+\!-y^+\gg \tau_f$, the $qg$ dipole is
exponentially suppressed, as manifest on \eqn{Klarge}~: ${\mcal
K}_{qg}\propto\rme^{-(x^+\!-y^+)/\tau_f}$.

So, clearly, the actual formation time cannot be larger than $\tau_f$.
However, depending upon the value of $r_0\sim\tqq\, y^+$, this time could
be {\em shorter} --- that would be the case if the exponent in
\eqn{intSqg} could become of order one already for $x^+\!-y^+\ll \tau_f$.
To find out what is the actual scenario, one needs to consider ${\mcal
K}_{qg}$ simultaneously with the other constraints on the time
integrations in \eqn{Iinfin}, which are specific to the interference
problem. The first one is the condition for color coherence between the
two emitters, as expressed by \eqn{ymax}. The second one is the condition
for their quantum coherence, as encoded in the phase $\rme^{ik^+(u^-x^+-
\bar u^- y^+)}$ manifest in \eqn{Iinfin} together with a similar phase
encoded in ${\mcal K}_{qg}$.

To better appreciate the role of these phases, let us first consider the
vacuum limit of the present calculation. In the vacuum, all the dipole
$S$--matrices are set to one, the function ${\mcal K}_{qg}$ reduces to
the free gluon propagator ${\mcal G}_0$, and then the integral over ${\bm
z}_{\perp}$ in \eqn{Iinfin} is straightforward. Using the integration
variable ${\bm b}_{\perp}\equiv {\bm z}_{\perp} - \bu x^+$, one finds
 \beq\label{intG0}
 \rme^{ik^+(u^-x^+-
  \bar u^- y^+)}\int \rmd {\bm b}_{\perp} \,
   \rme^{i{\bm b}_{\perp}\,\cdot\,\bk}\,{\mcal G}_0(x^+-y^+;
   {\bm b}_{\perp}+\bu x^+ -\bbu y^+)=\,
   \rme^{i(k\,\cdot\, u)x^+-i(k\,\cdot\, \bar u)y^+}
   \eeq
where the r.h.s. is recognized as the product of phases controlling the
in--vacuum emission times, from the quark and the antiquark. These phases
imply $x^+\lesssim \tau_q$ and $y^+\lesssim \tau_\bq$, where we recall
that $\tau_q= 1/(k\cdot u)\sim 1/\omega\tq^2$ and similarly for
$\tau_\bq$. Then the time integrations generate the expected longitudinal
phase--space $\tau_q\tau_\bq$ for interference in the vacuum.

In the case of the medium, the integral over ${\bm b}_{\perp}$ is
controlled by the $S$--matrix $S_{gg}(L^+, x^+; {\bm b}_{\perp})$ of the
$gg$ dipole, which enforces a rather small value $b_\perp\sim 1/Q_s$. (A
similar argument applied to direct emissions; recall the discussion after
\eqn{intb2}.) For such small values of $b_\perp$, we can replace ${\mcal
K}_{qg}\approx {\mcal G}_0$ for the purposes of the ${\bm
b}_{\perp}$--integration, which then reduces to
 \beq\label{intG01}
 \rme^{ik^+(u^-x^+-
  \bar u^- y^+)}&{}&\int \rmd {\bm b}_{\perp} \,
  \rme^{i{\bm b}_{\perp}\,\cdot\,\bk}\,
  \,\exp\left\{-
 i\,\frac{k^+({\bm b}_{\perp}+\bu x^+ -\bbu y^+)^2}{2(x^+-y^+)}\right\}
   \exp\biggl\{-\frac{1}{4}\,Q_s^2 b_\perp^2\biggr\}\nn
 &{}&\sim\,\rme^{i\Phi}\ \frac{1}{Q_s^2}\,
 \exp\biggl\{-\frac{1}{Q_s^2}\left(\bk\! - k^+\frac{\bu x^+ -\bbu y^+}
 {x^+-y^+}
 \right)^2
 \biggr\},
   \eeq
with the phase
 \beq\label{Phi}
 \Phi\equiv\,k^+(u^-x^+- \bar u^- y^+)\,-\,
 \frac{k^+(\bu x^+ -\bbu y^+)^2}{2(x^+-y^+)}\,=\,-
 \frac{k^+(\bu -\bbu)^2 x^+ y^+}{2(x^+-y^+)}\,.
 \eeq
In these manipulations, we have anticipated that $k^+/(x^+-y^+)\sim
k_f^2\ll Q_s^2$ and we have used $\bu^2=2u^-$. At this point, one should
recall that in the corresponding calculation for the direct emission,
Eqs.~\eqref{intb2}--\eqref{intb}, the analog of this phase $\Phi$ has
exactly canceled. The phase \eqref{Phi} is not quite the same as it would
be in the vacuum, cf. \eqn{intG0}~: it does not constrain the $x^+$ and
$y^+$ variables {\em individually}, but a particular combination of them.

To summarize, the integrations over $x^+$ and $y^+$ are controlled by the
following product
 \beq\label{factor}
 \exp\left\{-
 i\,\frac{k^+\tqq^2\, x^+ y^+}{4(x^+-y^+)}\right\}\,\exp\biggl\{
 -\frac{1}{24}\,\hat q\, (x^+-y^+) \big(\tqq y^+\big)^2
 \biggr\}\,\exp\biggl\{-\frac{x^+\!-y^+}{\tau_f}\biggr\},
 \eeq
together with the constraint \eqref{ymax} coming from color coherence.
The first factor in \eqn{factor} is the `vacuum--like' phase $\Phi$. The
second factor comes from the $S$--matrix \eqref{intSqg} of the $qg$
dipole, where we have neglected $b_\perp\sim 1/Q_s$ next to $r_0\sim\tqq
y^+$. The third factor is of course the large--time decay of the $qg$
propagator.

The four constraints introduced by the three factors in \eqn{factor}
together with \eqn{ymax} have to be simultaneously considered. Their
analysis becomes streamlined if one first identifies the characteristic
times scales associated with each of them. Let us enumerate these scales
here (they have already appeared in the qualitative discussion in
Sect.~\ref{intphys}):

\texttt{(i)} {\em the color coherence time} $\tau_{coh}$ : this is the
maximal value of the first emission time $y^+$ at which the quark and the
antiquark do still form a color singlet. This scale has been already
shown in \eqn{ymax};

\texttt{(ii)} {\em the in--medium formation time} (here in the context of
interference) : this is the typical time interval $x^+-y^+$ during which
the $qg$ dipole looses quantum and color coherence. As we shall shortly
argue, this scale is still determined by the last factor in \eqn{factor},
like for direct emissions, and thus is equal to $\tau_f$, cf. \eqn{tauf}.

\texttt{(iii)} {\em the transverse resolution time} $\tau_\lambda$ : this
represents the characteristic time scale for quantum coherence between
the two emitters during the process of gluon formation. This scale is
determined by the phase in the first factor in \eqn{factor}: the
condition $\Phi\lesssim 1$ together with the fact that $x^+-y^+\!\sim
\tau_f$ implies the following constraint on the emission times $x^+$ and
$y^+$ :
 \beq\label{tlambda0}
 x^+ y^+\,\lesssim\,\frac{4\tau_f}{k^+\tqq^2}\ \Longrightarrow\
 \sqrt{x^+ y^+}\,\lesssim\,\frac{\lambda_f}{\tqq}\,\equiv\,
 \tau_\lambda,\eeq
where $\lambda_f=1/k_f$ is the transverse wavelength of the gluon at the
time of formation (we have used $\tau_f\sim k^+/k_f^2$). Some useful
estimates for $\tau_\lambda$ have been given in \eqn{tlambda}.

\eqn{tlambda0} represents in an average way the condition that the gluon
overlap with both sources during the formation time. Since $x^+\!\simeq
y^+\!+ \tau_f$, it is clear that this condition must be viewed as a
constraint on the first emission time $y^+$.

\texttt{(iv)} {\em the interference time} $\tau_{int}$ : as discussed in
Sect.~\ref{intphys}, this is the upper limit on $y^+$ which follows from
\eqn{tlambda0} in the large angle regime where $\tqq\gg\tf$ (and hence
${\tlambda}\ll {\tau_f}$) :
  \beq\label{tint}
 y^+\,\lesssim\, \tau_{int}\,\equiv\,\frac{\tlambda^2}{\tau_f}
  =\,\frac{2}{\omega\tqq^2}\,.\eeq
Some useful estimates for $\tau_{int}$ are shown in \eqn{tint}. In the
other interesting regime at small angles $\tqq\ll\tf$ (or ${\tlambda}\gg
{\tau_f}$), the upper limit on $y^+$ is essentially $\tlambda$ (see
Sect.~\ref{intphys} for details).

So far, we have not discussed the time scale introduced by the original
size $r_0\sim\tqq y^+$ of the $qg$ dipole. Moreover, we have implicitly
assumed this scale not to influence the formation time. Let us check that
this is indeed the case. The exponent in the middle factor in
\eqn{factor} becomes of order one when $x^+-y^+\!\sim\tau_f$ and
$y^+\!\sim \tau_\lambda$ (we have used $\hat q\tau_f\simeq k_f^2$, cf.
\eqn{tauf}). This shows that the characteristic time scale associated
with $r_0$ is the same as the scale $\tau_\lambda$ for quantum coherence.
Since $y^+$ cannot become larger than $\tau_\lambda$, as clear from
\eqn{tlambda0}, we conclude that the original dipole size $r_0$ plays at
most a marginal role in the formation process and thus it cannot modify
the formation time to parametric accuracy.

We have thus recognized in our calculation all the characteristic time
scales for color and quantum coherence that were previously introduced,
via physical considerations, in Sect.~\ref{intphys}. The interplay
between these scales leads to the various regimes for interference
identified in Sect.~\ref{intphys}, that we shall not repeat here. Rather,
we shall now explicitly check the arguments in Sect.~\ref{intphys}
concerning the suppression of the interference effects relative to the
direct emissions.

To that aim, we shall estimate the contribution of the interference terms
to the spectrum for medium--induced radiation for dipole angles
$\tqq\gg\theta_c$. This contribution is obtained by multiplying the
Gaussian in \eqn{intG01} by the corresponding longitudinal phase--space
$\tau_{min}\tau_f$, by the normalization ${k^+}/{(x^+-y^+)}\sim
\omega/\tau_f$ of the gluon propagator \eqref{G0int}, and by a factor
$\theta_f^2$ which estimates the transverse derivatives in \eqn{Iinfin}.
As in Sect.~\ref{intphys}, $\tau_{min}\equiv {\rm
min}(\tau_{int},\tau_{coh})$ is the smallest among the coherence scales
which limit $y^+$ in the context of interference: $\tau_{min}
=\tau_{int}$ when $\tqq \gtrsim \tf$ and, respectively, $\tau_{min}
=\tau_{coh}$ when $\theta_c\ll \tqq\lesssim\tf$. The angular factor
$\theta_f^2$ is the same as for direct emissions. In the present context,
this factor is not {\em a priori} obvious and will be later checked via
explicit calculations. But before doing that, let us first exhibit the
interference contribution to the spectrum, which reads (for
$k_\perp\gtrsim k_f$)
 \beq\label{Ispec}
 {\mcal I}^{\mbox{(in)}}(\omega,\bk)\propto -\alpha_s
 C_F\,\theta_f^2\,\tau_{min}\,\frac{\omega}{Q_s^2}\,\exp\left\{
 -\frac{\left(\bk\! - \Delta\bk
 \right)^2}{Q_s^2}
 \right\}.\eeq
The off--set $\Delta\bk$ in the Gaussian is obtained from the respective
quantity in \eqn{G0int} after averaging over the emission times:
 \beq\label{ymtf}
 \Delta\bk\,=\, k^+
\left\langle\frac{\bu x^+ -\bbu y^+}
 {x^+-y^+}\right\rangle\,\simeq\,
 k^+\bu\,+\,k^+(\bu-\bbu)\,\frac{\tau_{min}}{\tau_f}
 \,.\eeq
As rather obvious from the fact ${\tau_{min}}/{\tau_f}\ll 1$, the second
term in the r.h.s. is negligible in all the interesting cases. Hence the
ensuing off--set $\Delta\bk\simeq k^+\bu$ is the same as for direct
emissions by the {\em quark}, cf. \eqn{BDMPS}, although we have
considered here a contribution to interference where the gluon is truly
emitted by the {\em antiquark}. This interplay is in agrement with our
physical picture that, in order to allow for interferences, the gluon
emitted by the antiquark must be co--moving with the quark. Clearly, for
the reversed situation, where the gluon is emitted by the quark (and thus
is co--moving with the antiquark), one would obtain $\Delta\bk\simeq
k^+\bbu$.

By taking the ratio between the interference term \eqref{Ispec} and the
spectrum \eqref{BDMPS} for a direct emission by the quark, one finds, for
$k_f\lesssim k_\perp\lesssim Q_s$,
  \beq\label{IDratio}
 %\quad\Longrightarrow\quad
 {\mcal R}(\omega,\bk)\,\equiv\,
 \frac{\big|{\mcal I}^{\mbox{(in)}}\big|}{{\mcal P}_q^{\mbox{(in)}}}
 \ \simeq \ \frac{\tau_{min}}{L}\,,
 \eeq
which in turn implies
 \beq\label{IDratio1}
 \left(\frac{\omega}{
 \omega_c}\right)^{1/2}\gtrsim\, {\mcal R}(\omega,k_{\perp})\,\gtrsim\,
\frac{\omega}{
 \omega_c}
  \qquad\mbox{when}\qquad \tf\,\lesssim\, \tqq \,\lesssim\,
  \ts\,,\eeq
for relatively large dipole angles $\tqq\gtrsim \tf$ where
$\tau_{min}=\tau_{int}$, and respectively
 \beq\label{IDratio2}
 1 \,\gg\, {\mcal R}(\omega,k_{\perp})\,\gtrsim\,
\left(\frac{\omega}{
 \omega_c}\right)^{1/2}  \qquad\mbox{when}\qquad \theta_c\,\ll\,
 \tqq \,\lesssim\,
  \tf\,,\eeq
for smaller angles, $\theta_c\lesssim \tf$, where
$\tau_{min}=\tau_{coh}$. Eqs.~\eqref{IDratio}--\eqref{IDratio2}
explicitly show the suppression of the interference effects relative to
the direct emissions for the medium--induced radiation of the dipole and
confirm the respective estimates in Sect.~\ref{intphys}.

To complete the argument, one still needs to evaluate the transverse
derivatives appearing in \eqn{Iinfin}, that is
 \beq\big(u^i+
{i\del^i_x}/{k^+}\big)
   \big(\bar u^i- {i\del^i_y}/{k^+}\big)
 \, {\mcal K}_{qg}(x^+,{\bm b}_{\perp}+\bxt; y^+,{\bm y}_\perp; k^+)\,
 \eeq
with the derivatives evaluated at $\bxt\!= \bu x^+$ and ${\bm y}_\perp\!=
\bbu y^+$. Like in the corresponding calculation for direct emission,
\eqn{TDdirect}, we can replace ${\mcal K}_{qg}\to {\mcal G}_{0}$ to
obtain a parametric estimate. This yields
\beq\label{trint}
\big(u^i+ {i\del^i_x}/{k^+}\big)
   \big(\bar u^i- {i\del^i_y}/{k^+}\big)\,{\mcal
G}_{0}&\to&   \left(u^i-\frac{b^i
 +u^ix^+-\bar u^iy^+}{x^+-y^+}\right)
  \left(\bar u^i-\frac{b^i
 +u^ix^+-\bar u^iy^+}{x^+-y^+}\right)\nn
 &{} &\quad+\,\frac{2i}{k^+(x^+-y^+)}\,.
 % \,\simeq\,\frac{{\bm b}_\perp\cdot(\bu-\bbu)}
 %{\tau_f}\,+\,\frac{2i}{k^+\tau_f}
 \eeq
The last term in the r.h.s. if of order $1/(k^+\tau_f)\sim \tf^2$. For
small dipole angles $\tqq\ll\tf$ it is easy to check that this is the
dominant term. So, in what follows we focus on the less trivial case
where $\tqq\gg \tf$. Then one can use $y^+\!\simeq\tau_{int}\!\ll x^+\!
\simeq\tau_f$ and $b_\perp\sim 1/Q_s$ to simplify the algebra. The terms
within the brackets in the r.h.s. of \eqnum{trint} thus yield
 \beq
 \frac{[b^i-(\bar u^i-u^i)y^+][b^i-(\bar u^i-u^i)x^+]}{(x^+-y^+)^2}
 \,\simeq\,\frac{{\bm b}_\perp\cdot(\bu-\bbu)+\tau_{int}(\bu-\bbu)^2 }
 {\tau_f}\,.\eeq
Using $(\bu-\bbu)^2\sim\tqq^2$ together with \eqn{taucomp}, it becomes
clear that the last term above is of order $\tf^2$. As for the first
term, this is estimated as (after performing the ${\bm
b}_{\perp}$--integration, cf. \eqn{intG01})
 \beq\label{polariz}
 \frac{({\bm k}_\perp-k^+\bu)\cdot(\bu-\bbu)}{Q^2_s\,\tau_f}
 \,\lesssim\,\frac{\omega\tq\tqq}{Q^2_s\,\tau_f}\,\sim\,
 \tf^2\,\frac{\tq\tqq}{\ts^2}\,\lesssim\,\tf^2\,.\eeq
We have also used here \eqn{kperpq} together with the relations
$k_f^2\sim\omega/\tau_f$ and $k_f/Q_s=\tf/\ts$. To conclude, the dominant
effect of the transverse derivatives in the interference terms is a
factor $\tf^2$, so like for the direct emissions.

\section{Discussion and outlook}
\label{outlook}

Throughout this work, we have mostly focussed on medium--induced
radiation of the BDMPS--Z type, whose main characteristic is that the
gluons are emitted {\em inside} the medium, as a result of multiple
scattering. However, we have also noticed at several places that this is
not the only type of medium--induced radiation for the case of a dense
medium. Indeed, as found in
Refs.~\cite{MehtarTani:2010ma,MehtarTani:2011tz} (and reviewed in our
Sect.~\ref{out}), the color decoherence of the $\qq$ antenna leads to
additional radiation {\em outside} of the medium, which is localized in a
region of (angular) phase space that would be forbidden
--- by destructive interference --- in the vacuum. The essential reason
why this new type of radiation exists is because the characteristic time
scale $\tau_{coh}$ beyond which the $q\bq$ pair looses its color
coherence becomes much smaller than the vacuum--like formation time for a
gluon radiated outside the dipole cone, which is typically
$\tau_{int}=2/(\omega \tqq^2)$. However, according to \eqn{taucomp}, the
inequality $\tau_{coh}\ll\tau_{int}$ holds whenever $\tqq\ll\tf$, which
allows for (vacuum) formation times $\tau_{int}$ that are both larger
{\em and} smaller than the medium size $L$. Hence, the same mechanism
could in principle lead to additional gluon radiation {\it inside} the
medium. This possibility has not been mentioned in the previous
literature, so we shall succinctly explore it here, via qualitative
considerations based on our previous results.

Specifically, one has $\tau_{int}\simeq L$ when $\tqq\simeq
\theta_{out}$, where
\beq \label{thetaoutdef} \theta_{out} \,\equiv\,
\sqrt{\frac{2}{\omega L}}\,=\,\theta_c
 \sqrt{\frac{\omega_c}{\omega}}\,=\,\tf
\left(\frac{\omega}{\omega_c}\right)^{1/4} \, ,
\eeq
is the same as the upper limit in our \eqn{outtheta}. So, {\em a priori}
there are two angular regions where the mechanism proposed in
Refs.~\cite{MehtarTani:2010ma,MehtarTani:2011tz} could
operate\footnote{We recall that the lower limit $\theta_c$ on $\tqq$
comes from the condition that $\tau_{coh}\ll L$, cf. \eqn{cohL}.} :
\texttt{(1)} $\theta_c\ll \tqq\lesssim\theta_{out}$, where
$\tau_{int}\gtrsim L$, so the respective emissions take place {\em
outside} the medium; this is the situation considered in
\cite{MehtarTani:2010ma,MehtarTani:2011tz} and \texttt{(2)}
$\theta_{out}\ll \tqq\ll\theta_{f}$ where $\tau_{int}\ll L$, so the
gluons are emitted (via vacuum--like processes) {\em inside} the medium;
this is the new possibility that we would like to explore. Note that,
together, these two angular domains completely overlap with our region of
`relatively small dipole angles' for BDMPS--Z radiation, as defined in
Sect.~\ref{intphys}. This observation naturally leads to the following
two questions: \texttt{(a)} what is the dominant mechanism for
medium--induced radiation for dipole angles within this common range at
$\theta_c\ll \tqq\ll\tf$, and \texttt{(b)} what are the most interesting
values of $\tqq$ for applications to the phenomenology ? These are the
main questions that we would like to address in this section.

With respect to the first question above, its answer depends upon the
ration $\tqq/\theta_{out}$, as we argue now. When
$\theta_c\ll\tqq\lesssim\theta_{out}$, that is, in region \texttt{(1)}
above, the radiation due to the new mechanism of
Refs.~\cite{MehtarTani:2010ma,MehtarTani:2011tz} is concentrated {\em
outside}  the dipole cone, but relatively {\em close to it}~: indeed,
this radiation has the angular distribution of the usual bremsstrahlung
spectrum, that is, it is strongly peaked around the sources and it decays
as $1/\theta$ at large emission angles $\theta\gg\tqq$. By contrast, the
BDMPS--Z gluons are emitted at relatively large angles
$\tf\gg\theta_{out}$ w.r.t. their sources, meaning far outside the dipole
cone. Hence in range \texttt{(1)} for $\tqq$, both types of
medium--induced radiation exist, but they are widely separated in angle
from each other: the out--of--medium emissions dominate the spectrum for
$\tq\, ,\tbq \sim\tqq$, while the in--medium emissions \`a la BDMPS--Z
dominate for $\tq\, ,\tbq \sim\tf$.

Consider now larger dipole angles, $\theta_{out}\ll \tqq\ll\theta_{f}$
(the angular region \texttt{(2)}). Then the previous discussion of the
BDMPS--Z gluons goes unchanged, whereas the mechanism proposed in
Refs.~\cite{MehtarTani:2010ma,MehtarTani:2011tz} is expected to become
ineffective: indeed, in--medium radiation with small emissions angles
$\tqq\ll\tf$ and hence relatively large formation time
$\tau_{int}\gg\tau_f$ is strongly suppressed as compared to the BDMPS--Z
radiation, since the soft gluons cannot avoid accumulating transverse
momenta of order $k_f$, via medium rescattering; as a consequence, they
are liberated from the parent quark after a relatively short time
$\tau_f$ and at an angle $\sim\tf$. Hence, for dipole angles within
region \texttt{(2)}, the medium--induced radiation is predominantly of
the BDMPS--Z type and is distributed at large angles
$\theta\gtrsim\tf\gg\tqq$, far outside the dipole cone.

%In other words, the probability that the transverse momentum acquired
%during an in--medium formation process be much smaller than $k_f$ is
%suppressed.

These considerations show that the physical consequences of the two
mechanisms for medium--induced radiation should be quite different:
whereas the BDMPS--Z gluons are more effective in broadening the energy
distribution of a jet in the transverse plane, in qualitative agreement
with the observations at the LHC \cite{Chatrchyan:2011sx,Ploskon:2009zd},
the mechanism proposed in \cite{MehtarTani:2010ma,MehtarTani:2011tz} is
probably more important for the softening of the intra--jet radiation and
its redistribution at small angles. But a more detailed phenomenological
analysis is still needed before drawing firm conclusions on the last
point.

Since the in--medium antenna pattern is so sensitive to the value of the
dipole angle, it is important to estimate what are the relevant values in
the context of heavy ion collisions. A physical process where in--medium,
color--singlet, antennas like the one that we have discussed are
naturally generated  is the hadronic decay of a heavy vector boson like
the Z or the W. In this scenario, the dipole angle of the pair depends
upon the boson kinematics, in particular, on its boost:
$\tqq\sim1/\gamma$. However, while such bosons are copiously produced in
Pb+Pb collisions at the LHC, their identification via hadronic decays is
complicated, if at all possible, by the large QCD background\footnote{We
thank P. Quiroga, S. Sapeta and G. Soyez for useful discussions on this
topic.}.

Another source of in--medium antennas, but typically in non--singlet
color representations, is the evolution of jets produced via hard
processes in heavy collisions. Although our calculations have been
restricted to the color singlet case, our arguments are sufficiently
general to be applicable to a qualitative discussion of antennas in other
representations.  We expect no modification in the physical regimes for
interference and the associated angular ranges summarized in
Sect.~\ref{intphys}. In particular, for relatively large angles
$\tqq\gg\theta_c$ our main conclusion remains unchanged: the interference
effects are suppressed and the overall antenna pattern is the sum of two
independent BDMPS--Z spectra produced by the two legs of the antenna. For
smaller angles, on the other hand, the interference effects {\em are}
important and their result is such that, when $\tqq\ll\theta_c$, the
total in--medium radiation by the antenna coincides with that by a single
source which carries the global color charge of the antenna (e.g., a
source in the adjoint representation if the antenna was produced by a
gluon decay). It would be interesting to check this conclusion
explicitly. (The calculation of the octet channel in
\cite{MehtarTani:2011tz} provides a partial check in that sense.)

Within the in--medium jet evolution we can distinguish two types of
antennas: those generated via hard, vacuum--like, parton splittings and
those arising via medium--induced emissions. Addressing the relevant
angles in either case will ultimately resort on in--medium Monte--Carlo
generators, such as
\cite{Lokhtin:2005px,Zapp:2008af,Armesto:2009qg,Armesto:2009ab,Schenke:2009gb,Renk:2010zx},
which can keep track of all the kinematical and probabilistic effects.
Here we will provide some simple estimates based on physical arguments,
to be ultimately confronted to explicit calculations. For simplicity, we
shall treat $\hat q$ as a fixed parameter in these estimates, in lines
with our general strategy throughout this paper.

Consider first an antenna resulting from medium--induced radiation. The
main question is, what is the typical angle between the emitted gluon and
the parent parton {\em by the time of a subsequent gluon emission}. This
angle starts with a value $\sim\tf$ at the time of formation but it can
be enlarged by additional multiple scattering occurring after the
formation. So, we need to estimate the typical time interval $\tau_{rad}$
between two successive emissions. The probability for emitting a new
gluon can be estimated as $\mcal{P}\sim \alpha_s C_R\, n_{\rm eff}$ where
$n_{\rm eff}\equiv \tau_{rad}/\tau_f$ is the number of effective
scattering centers along $\tau_{rad}$. This probability becomes of
$\mathcal{O}(1)$ when
\beq
\tau_{rad}\,\sim\, \frac{\tau_f}{\alpha_s C_R}\,.
\eeq
This estimate is a bit simplistic, since emissions can happen at
different frequencies and the relevant probability is the inclusive one.
Since the number of emitted gluons grows with decreasing $\omega$, cf.
\eqn{spectrumf}, a more realistic estimate (or, at least, a strict lower
limit) reads
\beq
\label{tauRest} \tau_{rad}\,>\, \frac{\tau_f(\omega_{min})} {\alpha_s
C_R}\,\sim\, \frac{1} {\alpha_s C_R\,\omega_{min}} \,,
\eeq
where $\omega_{min}\sim \hat q^{1/3}$ is the lowest energy for BDMPS--Z
gluons. Thus, for gluon frequencies which are not parametrically larger
(in $\alpha_s$) that $\omega_{min}$,  the typical time between successive
emissions is much larger than the formation time and the partons that
form the antenna acquire significant momentum by the time of secondary
emissions, increasing the angle of the effective dipole.  Thus,
medium--induced gluons lead, typically, to relatively large dipoles,
following the classification of Sect.~\ref{intphys}. Note also that for
the medium--induced parton cascade, this radiation time $\tau_{rad}$
plays the role of an effective medium length. Hence that fact that
$\tau_{rad}\gg\tau_f$ (at least for not too large frequencies) guarantees
the validity of our central argument for the suppression of the
interference terms (the suppression factor being $\tau_f/\tau_{rad}$ in
this case).

A different source of antennas propagating in the medium is the
vacuum--like evolution of hard partons. This refers to the emission of
gluons with large transverse momenta $k_\perp \gg Q_s$ (which cannot be
produced via in--medium interactions) and hence very short formation
times $\tau_q\ll\tau_f$ (for a given frequency). The precise values of
the dipole opening depends on the kinematics of the intervening hard
processes, in particular on their energy and virtuality.  Indeed, for a
hard parton of energy $E$ and virtuality $Q$ which emits a hard gluon
carrying a fraction $z$ of its energy, the angle of emission is
\beq
\theta_{hard}^2\approx\frac{1}{ z (1-z)}\,\frac{Q^2}{E^2} \,,
\eeq
and the emission time is estimated via the uncertainty principle as
\beq\label{thard}
\tau_{hard} \sim\, \frac{E}{Q^2}\sim\, \frac{1}{z (1-z)}
\frac{1}{\theta_{hard}^2\, E}\,.
\eeq
(Using $\omega\simeq z E$ for a small--$z$ emission and $\omega
\theta_{hard}\simeq k_\perp$, it becomes clear that \eqn{thard} is
consistent with our basic formula \eqref{tau} for the formation time.)
Thus, unless the branchings are very asymmetric, the vacuum--like
evolution can produce antennas with very small angles and at very early
times $\tau_{hard}\ll L$. We conclude that light jets (those with jet
mass much smaller than their total energy) can be sources for all the
different types of dipoles discussed in Sect.~\ref{intphys}, with a
predilection though for small and very small dipoles in the sense of that
discussion. Depending upon the precise relation between the emission
angle $\theta_{hard}$ and the characteristic medium angle $\theta_c$, the
antenna created via such a hard branching can either act as a set of two
independent sources of BDMPS--Z gluons (if $\theta_{hard}\gg\theta_c$),
or radiate such gluons in the same way as the parent parton would do (if
$\theta_{hard}\lesssim\theta_c$).

So far, we have discussed the in--medium hard branchings only as a
mechanism for generating antennas, but we have not addressed the medium
effects on such a branching by itself. As a matter of facts, we do not
expect such effects to be significant: the in--medium emissions of
relatively hard gluons with transverse momenta $k_\perp \gg Q_s$ should
proceed exactly as in the vacuum, for both direct emissions and the
corresponding interference phenomena leading to angular ordering. This is
quite clear from the fact that the respective formation time
$\tau_{hard}$ is much shorter than the time scale $\tau_{coh}$ for the
color decoherence of the sources. Since there was some confusion on this
point in the recent literature \cite{Leonidov:2010he}, we would like to
take this opportunity and fully clarify this issue. The analysis in
Ref.~\cite{Leonidov:2010he} was based on the assumption that the color
decorrelation time (the analog of our $\tau_{coh}$) is to be identified
with the mean free path $\ell$ of a colored parton propagating through
the medium (as introduced in the discussion in Sect.~\ref{bdmps0}). That
assumption would be correct if and only if the two emitters which form
the antenna would undergo {\em independent} color rotations in the
medium, which in turn requires their transverse separation $r_\perp\sim
\tau_q\tqq$  at the time of emission to be larger the Debye screening
length $\mu_D^{-1}$. Clearly, this would be the case only for extremely
soft radiated gluons, with transverse momenta $k_\perp \simeq \omega
\tqq\lesssim \mu_D$. For the medium created in heavy ion collisions at
the current energies, this scale $\mu_D$ is of the order of a few hundred
MeV. Gluons with such momenta are truly soft and do not significantly
contribute to the in--medium evolution of a hard jet, which rather
proceeds via hard, vacuum--like, emissions and semi--hard,
medium--induced, ones.

\section*{Acknowledgments}
We would like to thank Al Mueller for insightful and patient explanations
on the BDMPS--Z physics, and Andrei Leonidov, Cyrille Marquet, Guilherme
Milhano, Carlos Salgado, and Urs Wiedemann for many related discussions
and useful comments on the manuscript.

%\newpage
\appendix
\section{Momentum space analysis of the gluon spectrum}
\label{App}

The total radiation probability  from the dipole, which includes the
direct emissions from the quark, \eqn{Pqinfin}, and the antiquark (as
obtained by replacing $u\to\bar u$ within \eqn{Pqinfin}) and the
quark--antiquark interference terms, \eqn{Iinfin}, can be compactly
expressed as
\beq\label{Ptot} \Ptotin&=&
 \Prf  \,{\rm Re} \sum_{F,\,L= q,\, \bar q} {\rm Sign}(F,L) \,\times
%\left(-1\right)^{(u_L-u_F)^2/(1-\cos \tqq)}
\\
&{}& \Intx \Inty \Intb \rme^{ i \bk \cdot\,\bperp} S_{gg}\left(L^+,x^+;
\bperp\right) \Sqqgen\, \Igen \nonumber
 \eeq
(Note a slight change in the notations for the quark 4--velocities as
compared to the main text: we identify $u_q\equiv u$ and  $u_{\bar
q}\equiv \bar u$. To avoid cumbersome notations, we shall indicate the
transverse components of $u_F$ and $u_L$ by boldface symbols without the
`$\perp$' subscript: ${\bm u}_F$ and ${\bm u}_L$.) The quark--quark
dipole is trivial when both indexes are the same ($S_{qq}=S_{\bar q\bar
q}=1$) and the function ${\rm Sign}(F,L)=1$ if $F=L$ and ${\rm
Sign}(F,L)=-1$ otherwise\footnote{ \eqn{Pqinfin} corresponds to the term
$u_L=u_F=u$ and \eqn{Iinfin} is obtained by setting $u_L=u$ and $u_F=\bar
u$. A change of variables $\bperp={\bm z}_\perp-\bxt$ has been also
performed.}. These four terms summarize the four possible combinations
appearing in the total emission probability which are the direct
emissions from either the quark or the antiquark (Fig~\ref{fig:dirwil})
and the two interference terms in which the gluon is first emitted, at
time $y^+$, by the fermion (quark or antiquark) which has velocity $u_F$
and then absorbed, at time $x^+$, by the other fermion (antiquark or
quark), with velocity $u_L$ (Fig.~\ref{fig:intwil}). The function
$\Igenxy$ encodes the quark--gluon dipole together with its transverse
derivatives:
\beq \label{Igendef}
\Igen&=&\rme^{i k ^+ (u^-_L x^+-u^-_F y^+)}
\left(u_L^i + i \del^i_x / k^+\right)\left(u_F^i - i \del^i_y / k^+\right) \\
  &{}&
   \left . {\mcal K}_{qg}(x^+,{\bm x}_{\perp}+\bperp;
   y^+,{\bm y}_\perp; k^+)\right |_{{\bm x}_{\perp}=
   {\bm u}_L x^+ \, , \, {\bm y}_{\perp}={\bm u}_F y^+ }
\nonumber
 \eeq
For the case of direct emissions, where $u_L$ and $u_F$ coincide with
each other, the $qg$ dipole is made with the gluon and the quark which
has emitted that gluon (the {\em parent} quark). For the interference
terms, this dipole is made with the gluon emitted by the quark with
velocity $u_F$ and the {\em other} quark, which has a velocity $u_L$.

We restrict ourselves to the case the `harmonic approximation', in which
the slowly varying logarithm $\rho$ which enters the various dipole
$S$--matrices (see e.g. Eqs.~\eqref{Sqg} and \eqref{Sgg}) is treated as a
fixed quantity, which is moreover absorbed into the normalization of
$\hat q$. In this approximation, the quark--gluon path integral
(\eqn{Kdef}) is exactly known
\cite{Baier:1998yf,CasalderreySolana:2007zz} for the case of a single
emitter with vanishing transverse velocity. The generalization to the
present case, where the quark which enters the quark--gluon dipole
possesses a non--zero transverse velocity ${\bm u}_L$, is easily to find
and reads
\beq {\mcal K}_{qg}(x^+,{\bm x}_{\perp}; y^+,{\bm y}_\perp; k^+)&=&
\rme^{-i k^+ u^-_L(x^+-y^+) + i k^+ {\bm u}_L \cdot\,({\bm x}- {\bm
y})_{\perp}}
 \\
&{}& \times
  \left .
   \frac{A}{2\pi}\, \exp\left\{-\frac{A}{2}\left(B( \bar {\bm x}^2_\perp
   + \bar{\bm  y}^2_\perp)
-2 \bar {\bm x}_\perp\cdot \bar{\bm  y}_\perp \right)\right\}
 \right|_{\bar{\bm x}_{\perp}=
{\bm x}_{\perp}-\bm{u}_L x^+ \,, \, \bar{\bm y}_{\perp}={\bm
y}_{\perp}-\bm{u}_L y^+ } \nonumber
 \eeq
where $\Omega$ has been already defined in \eqn{Omega} and we have
introduced
\beq A\equiv \,\frac{k^+ \Omega}{i \sinh \Omega\left(x^+-y^+\right)} \, ,
\quad \quad B\equiv \cosh \Omega\left(x^+-y^+\right)\, .
\eeq

\subsection{The gluon spectrum at the time of formation}

The quark--gluon dipole encodes the process of in--medium gluon
formation. Right after formation, the gluon spectrum is obtained via the
Fourier transform of \eqn{Igendef}. After  some lengthy but
straightforward algebra, we obtain
%the spectrum is given by the $x^+$ and $y^+$ integration of
\beq \label{dIdqgen}
\Igenq&=& \frac{1}{B}\left[
 \left({\bm v}_\perp-\ulperp  \right)^2\frac{1}{B}
  +\left({\bm v}_\perp-\ulperp\right)\cdot
  \left(\ulperp-\ufperp\right)
 \big(1+ C \,\Omega y^+ \big)
 \right]   \nonumber
 \\
 &{}&\ \times\,
 \exp\bigg\{-iC\, \frac{k^+\left({\bm v}_\perp-
\ulperp\right)^2}{2\Omega}\, + i\,\frac{k^+}{2}\, (\ulperp-\ufperp)^2
 \big(1+ C \,\Omega y^+ \big)y^+  \nn &{}&
  \qquad\qquad
+ \,i\,\frac{k^+}{B}\,({\bm v}_\perp- \ulperp)
\cdot\left(\ulperp-\ufperp\right) y^+ \bigg\},
 \eeq
where ${\bm v}_\perp\equiv{\bqv }/{k^+}$ is the gluon transverse velocity
when it is formed and $C\equiv \tanh \Omega\left(x^+-y^+\right)$ has the
properties that $C/\Omega\simeq x^+-y^+$ when $x^+-y^+\ll \tau_f$ and
$C\rightarrow 1$ when $x^+-y^+\gg \tau_f$. (Recall that $\tau_f\equiv
1/\left| \Omega\right|$, cf. \eqn{formt}.)

The analysis of this expression shows the main features of the gluon
spectrum {\it at the time of formation}. For $x^+-y^+\!\simeq\tau_f$,
which is the typical value fixed by the subsequent integrations over
$x^+$ and $y^+$, one can write $C\Omega\approx 1/\tau_f$ to parametric
accuracy, and then the first term in the exponent is parametrically the
same as
 \beq\label{Gaussform}
 \exp\bigg\{-i\, \frac{k^+\left({\bm v}_\perp-
\ulperp\right)^2}{2\tau_f}\bigg\}\,=\,
 \exp\bigg\{-i\, \frac{\left({\bm q}_\perp-
k^+\ulperp\right)^2}{2k_f^2}\bigg\}.
 \eeq
Hence, for both the interference and the direct terms, the {\em
transverse momentum distribution} is a Gaussian with width $k^2_f\sim
\sqrt{\omega \hat q}$ centered around the direction $\ulperp$ of the
quark which enters the quark--gluon dipole. (Note that, in the
interference term, this is {\em not} the quark which emitted the gluon,
but the other quark.)

Concerning the {\em angular structure} of the spectrum, as encoded in the
first line of \eqn{dIdqgen}, this is a natural generalization of the
corresponding result in the vacuum, to which it reduces, as it should, in
the limit $|\Omega|\to 0$. Indeed, in that limit, $B\to 1$, $C\to 0$, so
the expression within the square brackets becomes $\left({\bm
v}_\perp-\ulperp\right)\cdot \left({\bm v}_\perp -\ufperp\right)$; this
is the expected result for both the direct terms, cf. \eqn{Pqvac}, and
the interference ones, cf. \eqn{Iout}. (Of course, in the vacuum, the
gluon velocity at the time of formation is the same as its {\em final}
velocity.) One can similarly check that, when $|\Omega|\to 0$, the
exponent in \eqn{dIdqgen} reduces to the respective vacuum result, {\em
i.e.}, to the formation--time phases visible e.g. in \eqn{intG0}.

In addition to the angle and momentum distributions, \eqn{dIdqgen} also
shows what are the {\em time scales} involved in the radiation process.
The overall prefactor $1/B$, which at long times behaves as $1/B\sim
\exp\left\{-\Omega (x^+-y^+)\right\}$, sets the {\em formation} time of
the gluons as $x^+-y^+\!\simeq\tau_f$, in agreement with \eqn{tauf}. The
other relevant time scale is the typical value of the {\em first
emission} time $y^+$ (more properly, this is the time at which the gluon
formation is initiated). For direct emissions $(\ulperp=\ufperp)$, there
is no characteristic value of $y^+$ and emissions happen all along the
medium length with equal probability. By contrast, for the interference
terms $(\ulperp\neq\ufperp)$, the values of $y^+$ are constrained by two
new time scales, $\tlambda$ and $\tint$, which are generated by the
middle term in the exponent in \eqn{dIdqgen} and its interplay with the
other terms.

Specifically, for $x^+-y^+\!\simeq\tau_f$, one can write $1+ C \,\Omega
y^+\approx (\tau_f+y^+)/\tau_f \approx x^+/\tau_f$ to parametric
accuracy, and hence
 \beq\label{Philam}
 \frac{i k^+}{2}\, (\ulperp-\ufperp)^2
 \big(1+ C \,\Omega y^+ \big)\,y^+\,\approx\,
  i\,\frac{k^+\tqq^2\, x^+ y^+}{4\tau_f}\,.\eeq
This is clearly equivalent with the first factor in \eqn{factor}. As
explained in Sects.~\ref{intphys} and \ref{Int}, this term encodes {\em
two} time scales: $\tlambda$, which is an upper limit on
$\sqrt{y^+(y^++\tau_f)}$ for generic values of $\tqq$, cf.
\eqn{tlambda0}, and $\tint$, which is the ensuing limit on $y^+$  for
relatively large angles $\tqq\gg\tf$, cf. \eqn{tint}. The last term in
the exponent in \eqn{dIdqgen}, which involves the momentum acquired by
the gluon during formation, leads to the same time scale $\tau_\lambda$,
as clear from the fact that $k^+|{\bm v}_\perp- \ulperp|\sim\omega\tf $
for the typical gluon velocity ${\bm v}_\perp$. Finally, in addition to
\eqn{dIdqgen}, the time dependence of the interference term is also
sensitive to the overall suppression due to the initial $q\bq$ dipole,
\eqn{Sqq2}, which introduces the additional time scale $\tau_{coh}$ for
color decoherence.

At this level, it is straightforward to make contact between
\eqn{dIdqgen} and the expression \eqref{spectrumf} for the `formation'
spectrum deduced in Sect.~\ref{intphys} via qualitative arguments: for
direct emissions, the only surviving term in the first line of
\eqn{dIdqgen} is $({\bm v}_\perp-\ulperp)^2\simeq\tq^2$. To compute the
spectrum at the formation time, one needs to perform the time
integrations in \eqn{Ptot} {\em without} the factor $S_{gg}$ there, which
describes multiple scattering {\em after} formation. For $\tau_f\ll L$
the integral over the time difference $x^+-y^+$ is cut off by the factor
$1/B^2$ and yields a factor $\tau_f$, while the subsequent integral over
$y^+$ is unrestricted and yields a factor $L$. Altogether, we have a
factor $\tf^2\tau_f L$ multiplying the Gaussian in \eqn{Gaussform} plus,
of course, the overall factor $\alpha_s C_F$. This reproduces the
parametric estimate in \eqn{spectrumf}.

\subsection{The final gluon spectrum}

The previous arguments also show that, in order to compute the {\em
final} spectrum, as it would be measured by a detector, one needs to also
take into account the $S$--matrix $S_{gg}$ of $gg$ dipole. Working in the
momentum representation,  the final spectrum is obtained by convoluting
the spectrum at the time of formation with the Fourier transform of
$S_{gg}$. Within the `harmonic approximation', this amounts to an
additional Gaussian broadening:
 \beq  \Fgen=\int \frac{\rmd^2 \bqv}{(2 \pi)^2}\,\Igenq
 \ \frac{4 \pi}{Q^2_s} \ \rme^{-(\bk -\bqv )^2/Q^2_s}
 \eeq
where $Q^2_s\equiv\hat q (L^+ -x^+)$ depends upon the final formation
time $x^+$. Since the typical gluons are produced within the bulk of the
medium ($x^+\ll L^+$), one can neglect the $x^+$--dependence of $Q^2_s$
to parametric accuracy. After also using $\tau_f\ll L$, we find (with
$Q^2_s=\hat q L^+$ from now on)
  \beq  \label{Ffinal}
 \Fgen&\approx&
\\
&{}& \hspace{-3cm} \approx\,\frac{2 k^+ \Omega}{B\, Q^2_s}  \left[
\frac{2}{B} \frac{\Omega}{i k^+} +\frac{ y^+ \Omega
\left(\ulperp-\ufperp\right)^2}{\sinh \Omega(x^+-y^+)}
		\left(
		1+ y^+ \Omega \left(1+ \frac{2}{B}\right)
		\right) - \right. \nonumber
\\
&{}& \quad \quad \quad \quad \left.		 -
		2 i\frac{\Omega}{Q^2_s}\,
(\bk-k^+\ulperp)\cdot(\ulperp-\ufperp)\,
		\left(1+ y^+\Omega \left(1+\frac{4}{B}\right)\right)
		\right]\, \times
		\nonumber
\\
&{}& \quad \hspace{-3cm} \exp\left\{ i\,\frac{k^+}{2}\,
(\ulperp-\ufperp)^2
 \big(1+ \Omega y^+ \big)y^+
 							-\frac{1}{Q^2_s} \left(\bk-k^+ \ulperp-
							                  \frac{k^+ \Omega\,
y^+(\ulperp-\ufperp)}{\sinh \Omega(x^+-y^+)}\right)^2
 							\right\}
							\, , \nonumber
 \eeq
where we have further approximated $C\approx 1$ since we anticipate that
$x^+-y^+\!\simeq\tau_f$. To understand \eqn{Ffinal} to parametric
accuracy, one can also set $B\approx 1$ and $\sinh
\Omega(x^+-y^+)/\Omega\approx\tau_f$.

As before, the first line of \eqn{Ffinal} specifies the angular
dependence of the final spectrum. Unlike in \eqn{dIdqgen}, there is not a
term proportional to the square of the final angle formed by the gluon
and the quark. This is so since the final gluon spectrum receives
contributions from the entire gluon spectrum at formation. The final
distribution is  characterized by $\theta_f$, the typical angle at
formation, which can be identified in the first term of this line:
$|\Omega|/k^+=1/(\tau_f k^+)\simeq\tf^2$. For the interference term there
is, however, a residual dependence upon the final direction of the gluon:
this enters via the middle line of \eqn{Ffinal}, which is due to the fact
that there is some correlation between the final direction of the gluon
and its direction at the time of emission (as encoded in the off--set
$k^+ \ulperp$ in the final gluon momentum). This term is essentially the
same as that in \eqn{polariz} from the main text and, as shown there, it
is generally subleading.

The last line in \eqn{Ffinal} encode both the time and momentum
dependence. The first term  in the exponent was already present in
\eqn{dIdqgen} (the middle term in the exponent there) and, as already
explained, it encodes the condition of quantum coherence --- that is, the
two time scales $\tau_\lambda$ and $\tau_{int}$. The second term in the
exponent shows the transverse momentum spectrum, which is a Gaussian of
width $Q^2_s$ around the direction $\ulperp$ of the quark which
participates in the formation process ({\em i.e.}, the quark from the
$qg$ dipole). The $\bqv$--dependence of the exponent in \eqn{Ffinal}
leads to a shift in the center of the transverse momentum gaussian; in
fact, by using $\sinh \Omega(x^+-y^+)/\Omega\approx\tau_f$ to parametric
accuracy, one sees that this additional shift is the same as the second
term in the r.h.s. of \eqn{ymtf}. As will be later verified, this
additional shift is negligible in all the interesting cases.

Note finally that, as in the case of the spectrum at formation,
\eqn{Ffinal} must be supplemented with the $qq$ dipole \eqn{Sqq2}, which
introduces the coherence time. We will now specify the parametric
dependence of the radiation spectrum for the different terms in the gluon
amplitude.

 \subsection{Direct emission: the BDMPS--Z spectrum}
In the case of direct emission by either the quark or the antiquark, the
final spectrum is obtained by integrating the following expression
 \beq  \label{FDirect}
    F_q(x^+,y^+,\bk)\,\simeq\,
   \frac{4 \Omega^2}{i Q^2_s} \frac{1}{B^2}
  \exp\left\{
  	-\frac{\left(\bk-k^+\bu\right)^2}{Q^2_s}
          \right\}
 \eeq
over the time variables $x^+$ and $y^+$. For definiteness, we have shown
here the direct emission by the quark but there is of course a similar
contribution by the antiquark. As expected, at this level of
approximation the spectrum is a Gaussian centered around the transverse
momentum $k^+\bu$ inherited from the parent parton. As already explained,
the subsequent integrations over the time variables introduce a factor
$\tau_f L$. By also using $\tau_f\sim 1/\left|\Omega\right|$ and
$\tf^2\sim |\Omega|/ k^+$ we recover the parametric dependencies shown in
\eqn{BDMPS} of the main text.

\subsection{The interference terms for relatively large dipoles: $\tf\lesssim \tqq \lesssim
\ts$.}

We shall now provide estimates for the interference contributions to the
gluon spectrum, as obtained by integrating the expression in \eqn{Ffinal}
with $\ulperp\neq\ufperp$ over the time variables $x^+$ and $y^+$. We
first consider dipole angles within the range $\tf\lesssim \tqq \lesssim
\ts$. As before, the integration over $x^+-y^+$ is dominated by $\tau_f$.
But unlike the previous case of direct emissions, the $y^+$--integration
is now more complicated since there are several competing time scales. As
we have extensively discussed in Sect.~\ref{intphys}, within the present
range for dipole angles, the relevant time scales are strictly ordered,
$\tau_{int}\lesssim \tau_\lambda \lesssim \tau_{coh} \lesssim \tau_f$,
and the integral over $y^+$ is controlled by the smallest time scale,
$\tau_{int}$. In addition, since $\tau_{int}/\tau_f\ll 1$, all terms
proportional to $y^+\Omega$ in \eqn{Ffinal} can be neglected. Then the
only dependence upon $y^+$ which survives in the exponent is that encoded
in the first term there, $\simeq i k^+\tqq^2 y^+$, which after
integration yields a factor $\tau_{int}$, as expected. The same
approximations allow us to simplify the angular dependence of the final
spectrum (the first line of \eqn{Ffinal}) which contains terms
proportional to $\tf^2$, $\tqq^2$ and $\theta_L \tqq$, where
$\theta_L=\theta_q \,\, {\rm or} \,\, \tbq$ is the gluon angle with
respect to the quark which enters the $qg$ dipole. By also using $
|\Omega|/ k^+\sim\tf^2$, $|\Omega| k^+\sim k_f^2$ and the following
estimates,
 \beq   \tau_{int} |\Omega| \tqq^2 & \sim & \tf^2 \, ,\nn
\frac{k^2_f}{Q^2_s}\, \theta_L \tqq
  &\sim& \tf^2 \, \frac{\tqq^2}{\ts^2} \frac{\theta_L}{\tqq} \,,
\eeq
one eventually finds that the total interference term is parametrically
given by
 \beq  \label{Ildf}
 {\mcal I}^{\mbox{(in)}}(\omega,k_\perp)\,\propto\,-\alpha_s  C_F\,\theta_f^2
 \left(1-c_1 \frac{\tqq^2}{\ts^2} \frac{\theta_L}{\tqq}\right) \tau_{int} \frac{\omega}{Q^2_s}
 \,\exp\biggl\{
 -\frac{(k_\perp-k^+\ulperp)^2}{Q_s^2}\biggr\}
 \eeq where $c_1$ is a number of order one. We  see that, unless $\theta_L$ is
arbitrary large, $\theta_L\gg \tqq$, the interference term, \eqn{Ildf},
is suppressed with respect to the direct term, \eqn{BDMPS}, by
 \beq  \label{smdshift}
 \frac{\tau_{int}}{L^+}\ll \frac{\tau_f}{L^+}\ll,1\,.
 \eeq
Note that the term proportional to $c_1$ within the parentheses in
\eqn{Ildf} is the same as that appearing in \eqn{polariz} of the main
text.

%\vspace*{0.2cm}
\subsection{The interference terms for relatively small
dipoles: $\theta_c\ll \tqq\ll\tf$.}

Consider similarly the other relevant range for the dipole angles, at
$\theta_c\ll \tqq\ll\tf$. Then, as discussed in Sect.~\ref{intphys}, the
ordering of time scales gets now reverted, $\tau_f \ll \tau_{coh} \ll
\tau_\lambda \ll \tau_{int}$ and the $y^+$ integration is restricted by
the smallest of the coherence time scales, that is $\tau_{coh}$. In this
case, one we can safely neglect the $y^+$ dependence of the exponential
in \eqn{Ffinal}. (The integral over $y^+$ is rather controlled by the
$q\bq$ dipole and yields a factor $\tau_{coh}$.) In addition, since
\beq \label{neglectshift}
\frac{(k^+)^2 \tqq^2}{k_f^2} \,\frac{\tau^2_{coh}}{\tau_f^2}\,\sim
 \left(\frac{\tqq}{\tf}\right)^{2/3} \ll 1 \,,
 \eeq
we can neglect the shift in the transverse momentum Gaussian in
\eqn{Ffinal} for any final momentum $k_\perp\gtrsim k_f$. The overall
magnitude of the interference term is controlled by the prefactor
encoding the angular dependence (the first line in \eqn{Ffinal}). In the
present case $\tau_{coh}|\Omega|\gg 1$ and we need to determine the
relative value of the three different contributions to the spectrum.
Simple manipulations show that
\beq (\tau_{coh} \tqq |\Omega|)^2 &\sim&
\tf^2 \left(\frac{\tqq}{\tf}\right)^{2/3}\ll \tf^2 \nn
\frac{k^2_f}{Q^2_s} \frac{\tau_{coh}}{\tau_f} \theta_L \tqq   &\sim&
\tf^2 \frac{\theta_L \tf}{\theta_s^2}
\left(\frac{\tqq}{\tf}\right)^{1/3}\, .
\eeq Using this expression and taking into account that the $x^+$ and $y^+$ integration lead to an overall
factor of $\tau_f \tau_{coh}$ we can estimate the parametric dependence
of the interference spectrum as
 \beq  {\mcal I}^{\mbox{(in)}}(\omega,k_\perp)\,\propto\,-\alpha_s  C_F\,\theta_f^2
 \left(1-c_2 \frac{\theta_L \tf}{\theta_s^2} \left(\frac{\tqq}{\tf}\right)^{1/3}\right) \tau_{coh}\frac{\omega}{Q^2_s}
 \,\exp\biggl\{
 -\frac{(k_\perp-k^+\ulperp)^2}{Q_s^2}\biggr\}
 \eeq where $c_2$ is a number of order 1. It is then
clear that, unless one considers gluons emitted at very large angles
$\theta_L\gg \theta_s$, the interference term is proportional to $\tf^2$
and it is magnitude is suppressed as compared to the direct term by
\beq \frac{\tau_{coh}}{L}\sim \left(\frac{\theta_c}{\tqq}\right)^{2/3}
\ll \,1\,.
\eeq

\providecommand{\href}[2]{#2}\begingroup\raggedright\endgroup

\end{document}